\documentclass[a4paper,11pt]{article}
\pdfoutput=1
\usepackage{jcappub}

\usepackage{amsmath}
\usepackage{amsthm}
\usepackage{amssymb}
\usepackage{amsfonts}
\usepackage{mathtools}

 \usepackage[english]{babel}
\usepackage{lineno}
\usepackage{lscape}
\usepackage{soul}
\usepackage{graphicx}
\usepackage{bm}
\usepackage{epstopdf}
\usepackage{url}
\usepackage{booktabs}
\usepackage{mdframed}
\usepackage[colorinlistoftodos]{todonotes}
\usepackage{booktabs}
\usepackage{graphicx}
\usepackage{booktabs}
\usepackage[super]{nth}
\usepackage[math]{cellspace}

\usepackage{fontawesome}
\usepackage{microtype}
\usepackage{xspace}
\usepackage[capitalise]{cleveref}
\crefname{section}{Sec.}{Secs.} \Crefname{section}{Section}{Sections.}
\usepackage{verbatim}
\usepackage{enumitem}
\usepackage{siunitx}
\usepackage{hyperref}
\usepackage{caption}
\usepackage{booktabs,tabularx}
\usepackage{subcaption}
\DeclareSIUnit\parsec{pc}
\DeclareSIUnit\year{yr}
\DeclareSIUnit\solarmass{\textit{M}_\odot}

\newcommand{\IFCA}{%
Instituto de F\'isica de Cantabria (IFCA), University of Cantabria (UC)-CSIC,\\
 Avenida de los Castros, s/n E-39005 Santander, Spain}

\preprint{}

\begin{document}

\title{Impact of dark matter spikes on the merger rates of Primordial Black Holes}

\author[]{Pratibha Jangra,}
\author[]{Bradley J. Kavanagh,}
\author[]{J. M. Diego}

\affiliation[]{\IFCA}

\emailAdd{pratibha@ifca.unican.es}
\emailAdd{kavanagh@ifca.unican.es}
\emailAdd{jdiego@ifca.unican.es}

\abstract{Mergers of Primordial Black Holes (PBHs) may contribute to the gravitational wave mergers detected by the LIGO-Virgo-KAGRA (LVK) Collaboration. We study the dynamics of PBH binaries dressed with dark matter (DM) spikes, for PBHs with extended mass functions. We analyze the impact of DM spikes on the orbital parameters of the PBH binaries formed in the early Universe and calculate their merger rates at the age of the Universe today. We consider two possible scenarios for the dynamics of the dressed binaries: assuming that either the DM spikes are completely evaporated from the binaries before merger or they remain static until the merger. Contrary to previous studies, we find that the presence of spikes may increase or decrease the present-day PBH merger rates, in some cases dramatically. Comparing with merger rates reported by the LVK Collaboration in the third Gravitational Wave Transient Catalog (GWTC-3), we derive approximate constraints on the fraction of Solar-mass PBHs in cold dark matter as $f_\mathrm{pbh}\leq \mathcal{O}(10^{-5} - 10^{-3})$, depending on the mass function. Our calculations are valid only for the idealized scenarios in which the DM spikes are either evaporated or static. However, they suggest that the impact of DM spikes on PBH merger rates may be more complicated than previously thought and motivate the development of a more general description of the merger dynamics, including feedback of the DM spikes in highly eccentric PBH binaries.}

\maketitle

\section{Introduction}
\label{sec:introduction}
Primordial Black Holes (PBHs) may form shortly after the Big Bang due to the collapse of very large density fluctuations on horizon scales~\cite{Hawking:1971ei,Zeldovich}. Other possible channels of PBH formation include the collapse of domain walls, the collapse of cosmic string loops, and bubble collisions~\cite{Carr:2020xqk,Green:2020jor}. The mass spectrum of PBHs is dependent on the scale of perturbations leading to their origin~\cite{Zeldovich}. 
PBHs lighter than $\sim10^{-19}\, \mathrm{M_{\odot}}$ would have already evaporated due to Hawking radiation but PBHs heavier than $10^{-19}\, \mathrm{M_{\odot}}$ could in principle be very interesting as they can be a possible non-particle candidate for cold dark matter ~\cite{1975Natur.253..251C,Carr:2016drx,Green:2020jor,Villanueva-Domingo:2021spv}. PBHs of mass $\gtrsim  1\, \mathrm{M_{\odot}}$ may contribute to the binary black hole mergers detected by LIGO-Virgo-KAGRA (LVK) Collaboration~\cite{Bird:2016dcv,Clesse:2016vqa,Sasaki:2016jop}.
Solar-mass PBHs might also explain the origin of supermassive black holes, which without lighter seed black holes, may struggle to grow rapidly enough to reach their present observed masses~\cite{1984MNRAS.206..801C,Bean:2002kx,2012Sci...337..544V,Inayoshi:2019fun}. The existence of bright galaxies observed by the James Web Space Telescope (JWST) at high redshift of $z\gtrsim 10$ are also difficult to explain as per the standard $\Lambda$CDM model~\cite{Liu:2022bvr,Boylan-Kolchin:2022kae,Inayoshi_2022,2023MNRAS.518.2511L}. But if  PBHs with mass around $ \gtrsim 100\, \mathrm{M_{\odot}}$ are formed in the early Universe then they can seed the formation of massive structures at such high redshift~\cite{Carr:2018rid}.

As per the Gravitational Wave Transient Catalog-3 (GWTC-3), the LVK Collaboration has reported the detection of $\sim 90$ compact binary black hole (BBH) mergers in the component mass range of $5-105 \,\mathrm{M_{\odot}}$~\cite{LIGOScientific:2021djp, PhysRevX.11.021053}. It has been proposed that some of the mergers detected by LVK Collaboration could originate from the collisions of black holes that might be primordial in nature~\cite{Bird:2016dcv,Clesse:2016vqa,Sasaki:2016jop}. It is predicted that PBHs can decouple from the Hubble expansion very early and form bound systems such as binaries which can ultimately merge to produce ripples in the fabric of space-time. The detection of gravitational waves (GWs) from such mergers can lead to constrain the abundance of PBHs which in turn can impose constraints on various models of inflation leading to their origin~(e.g.~\cite{Josan:2010cj,Inomata:2018cht,Biagetti:2018pjj,Frolovsky:2022qpg}).  

In the mass range probed by the LVK Collaboration, we know that the fractional abundance $f_\mathrm{pbh}$ of PBHs in cold dark matter (CDM) should not be greater than $\sim 10^{-3}$~\cite{Sasaki:2016jop,Raidal:2018bbj,DeLuca:2020qqa,Hall:2020daa,Hutsi:2020sol,Wong:2020yig,DeLuca:2020jug,Bhagwat:2020bzh,Franciolini:2021tla,DeLuca:2021wjr,Deng:2021ezy,Chen:2021nxo,Chen:2022fda,Liu:2022iuf,Postnov:2023ntu}. This implies the existence of other candidates contributing to the dark matter, which may have a particle nature. The presence of such dark particles can lead to the formation of dense spikes around all PBHs~\cite{Mack:2006gz,Ricotti:2007jk}. PBHs with DM spikes around them are often referred to as `dressed' PBHs. In principle, spikes around the PBHs can be formed by any particle-like DM. However, gamma ray observations exclude this possibility if DM particles have a large annihilation cross section, as in the case of classical WIMPs~\cite{Lacki:2010zf,Cerdeno:2010jj,Boucenna:2017ghj,Bertone:2019vsk,Adamek:2019gns,Carr:2020mqm,Cooley:2021rws}. We therefore consider the scenario in which the dominant particle DM is not WIMP-like or is WIMP-like with a very small annihilation cross section~\cite{Bertone:2019vsk,Carr:2020mqm}.
The merger rate of PBH binaries \textit{without} DM spikes has been extensively studied, for monochromatic as well as extended PBH mass functions (MF) (see e.g.~\cite{Ali-Haimoud:2017rtz,Raidal:2017mfl,Kocsis:2017yty,Raidal:2018bbj,Chen:2018czv,Jedamzik:2020ypm,Jedamzik:2020omx,DeLuca:2020fpg}). The effects of DM spikes on the orbital parameters of PBH binaries with monochromatic PBH mass functions was taken in to account by Ref.~\cite{Kavanagh:2018ggo} and it was shown that the merger rates of PBH binaries dressed with DM spikes may be around twice that of binaries without spikes. 
The formation of a common DM spike around the whole of the PBH binary was neglected in Ref.~\cite{Kavanagh:2018ggo}. Taking this into account, Ref.~\cite{Pilipenko:2022emp} showed that the merger rate of PBH binaries with DM spikes may be as much as $6-8$ times larger than PBH binaries without DM spikes.

Since the BBH mergers observed by LVK show a range of component masses, so it becomes important to extend these studies to \textit{dressed PBHs with extended mass functions}. This is crucial in order to quantify how frequently the mergers of PBHs with realistic mass functions could take place in the Universe. In addition, dressed PBHs with unequal masses can lead to intermediate mass ratio inspirals (IMRIs) or extreme mass ratio inspirals (EMRIs) in which the lighter PBH in the binary is hundreds to millions of times less massive than its companion. In such systems, it may be possible to see the effects of DM spikes directly on the GW signal~\cite{Macedo:2013qea,Eda:2013gg,Eda:2014kra,Kavanagh:2020cfn,Becker:2021ivq,Coogan:2021uqv,Cole:2022fir}. The possibility to distinguish the gravitational waves originating from the merger of PBH binaries with DM spikes and without DM spikes was recently explored in Ref.~\cite{Cole:2022ucw}. In order to accurately study the GW signals from these PBH binaries, however, it is important to understand their formation mechanism and time evolution up to the merger. 

In this paper, then, we consider PBH binaries with DM spikes having a wide range of PBH masses and estimate their merger rates by extending the approach of Ref.~\cite{Kavanagh:2018ggo}.
The simultaneous evolution of PBHs and DM spikes in the binaries can be a complex problem in full generality~\cite{Kavanagh:2020cfn,Coogan:2021uqv,Cole:2022ucw}. We therefore calculate the merger rates of these binaries assuming that either the DM spikes are completely evaporated before the merger (as in Ref.~\cite{Kavanagh:2018ggo}) or that they remain static until the merger. These assumptions provide a starting point for understanding how DM spikes affect the orbital properties and evolution of unequal-mass PBH binaries. We compare our results with the merger rates reported by LVK in GWTC-3, in order to derive approximate constraints on the abundance of PBHs in dark matter $f_\mathrm{pbh}$.

 The paper is organized in the following manner. In Sec.~\ref{sec:Formation of DM spikes around PBHs}, we review the accretion process of DM spikes around isolated PBHs by closely following the formalism of Ref.~\cite{Adamek:2019gns}. In this section, which can be skipped by the expert reader, we describe the density profile of dark matter spikes expected around PBHs. Then, in Sec.~\ref{sec:Formation of PBH Binaries}, we describe the formation mechanism of PBH binaries with and without DM spikes taking into account the impact of DM spikes. This set up is an improved and generalized description of binary formation given in Refs.~\cite{Chen:2018czv, Kavanagh:2018ggo}, for extended mass functions with the addition of DM spikes into the picture. In Sec.~\ref{sec:Merger time of the binaries}, we analyze the variation of the final merger time of PBH binaries with and without DM spikes as a function of the initial semi-major axis and eccentricity, which also includes an updated review Ref.~\cite{Kavanagh:2018ggo}. After that in Sec.~\ref{sec:Merger rates of PBH binaries},  we calculate the present-day merger rates of PBH binaries surrounded by DM spikes by convolving the final merger time with the probability distributions of initial semi-major axis and eccentricity of the systems. We present these merger rates for three different extended PBH mass functions as illustrative examples. Then, in Sec.~\ref{subsec:Effect of fpbh on merger rates}, we discuss how the gap between the merger rates of PBH binaries with and without DM spikes varies with the abundance of PBHs in CDM $f_\mathrm{pbh}$, a novel aspect in comparison to previous studies. We also highlight the mass ratio distribution of PBHs merging today, as described in Appendix~\ref{sec:Mass ratio Distribution of merging PBHs}.  Finally in Sec.~\ref{sec:conclusions}, we conclude and point out the possible future directions.

\section{Formation of DM spikes around PBHs}
\label{sec:Formation of DM spikes around PBHs}
In this section, we review the formation mechanism and density profile of DM spikes around isolated PBHs proposed originally in Ref.~\cite{Mack:2006gz,Ricotti:2007jk}, following closely the treatment in Ref.~\cite{Adamek:2019gns}.
We start with the assumption that in the early Universe, non-annihilating CDM particles (such as axions or WIMPs with almost negligible annihilation cross-sections) can encounter isolated PBHs in their vicinity, due to which their motion becomes influenced by the combined effect of the gravitational attraction of the PBHs and Hubble expansion. Then, the equation of motion of a DM shell of radius $r$ around a PBH of mass $m_{\rm{pbh}}$ in the FLRW metric is given by:
\begin{equation}\label{eq:DMshell}
\frac{\mathrm{d}^{2} r} {\mathrm{d} t^{2}}  = -\frac{G \,m_{\mathrm{pbh}}}{r^{2}} + \left(\dot{H} + H^{2}\right)r   \,,
\end{equation}
where the first term corresponds to the gravitational attraction of the PBH and the second term denotes the decelerating Hubble expansion of the Universe. The evolution of these DM shells starts deep in the radiation era, corresponding to the initial conditions $r = r_{i}$ and  $\dot{r} = H_{i}\, r_{i} = r_{i}/\left(2\,t_{i}\right)$.

\begin{figure}[tb!] 
\centering
\begin{minipage}[c]{0.42\linewidth}
\includegraphics[width=\linewidth]{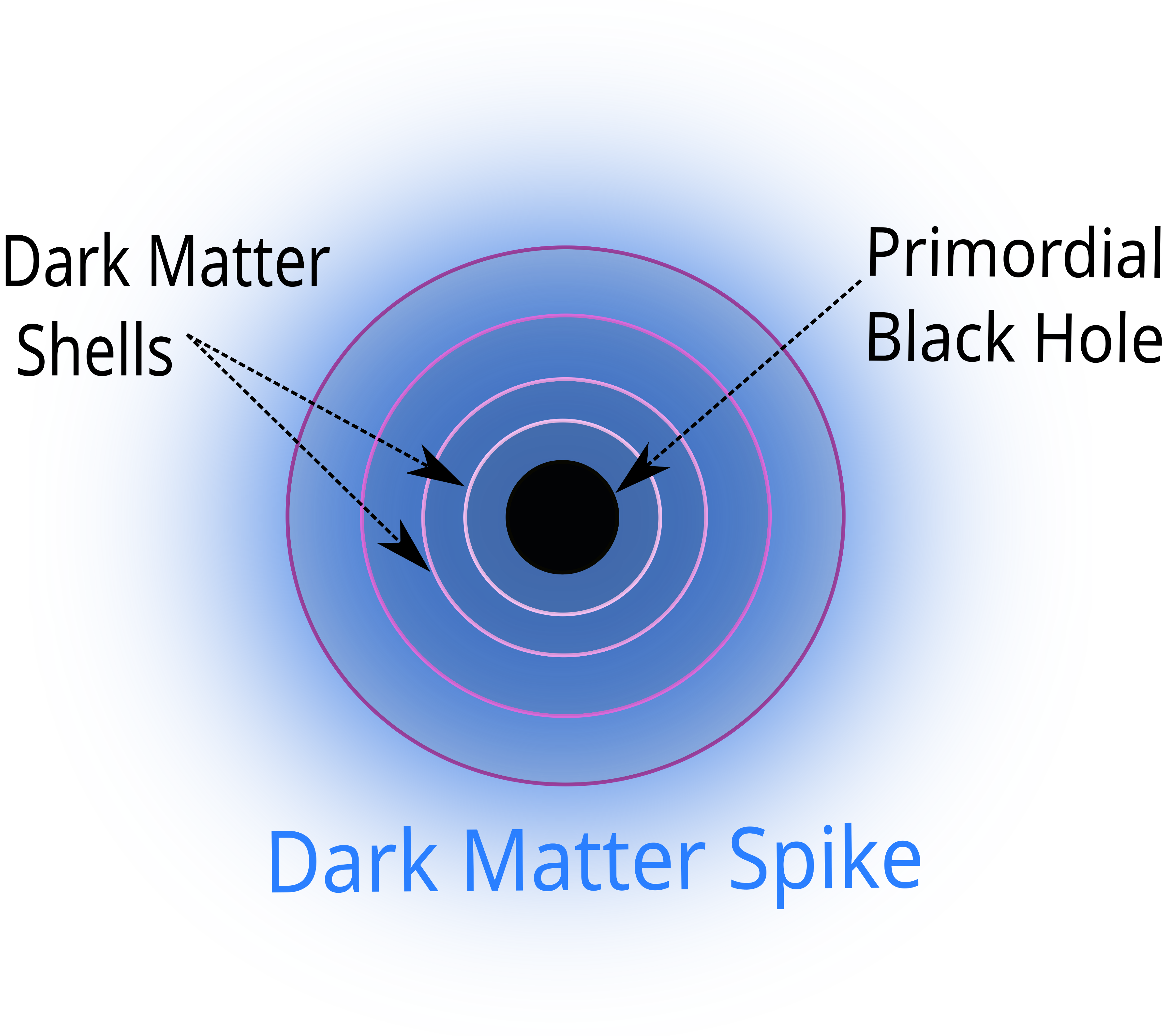}
\end{minipage}
\hspace*{\fill}
\begin{minipage}[c]{0.56\linewidth}
\includegraphics[width=\linewidth]{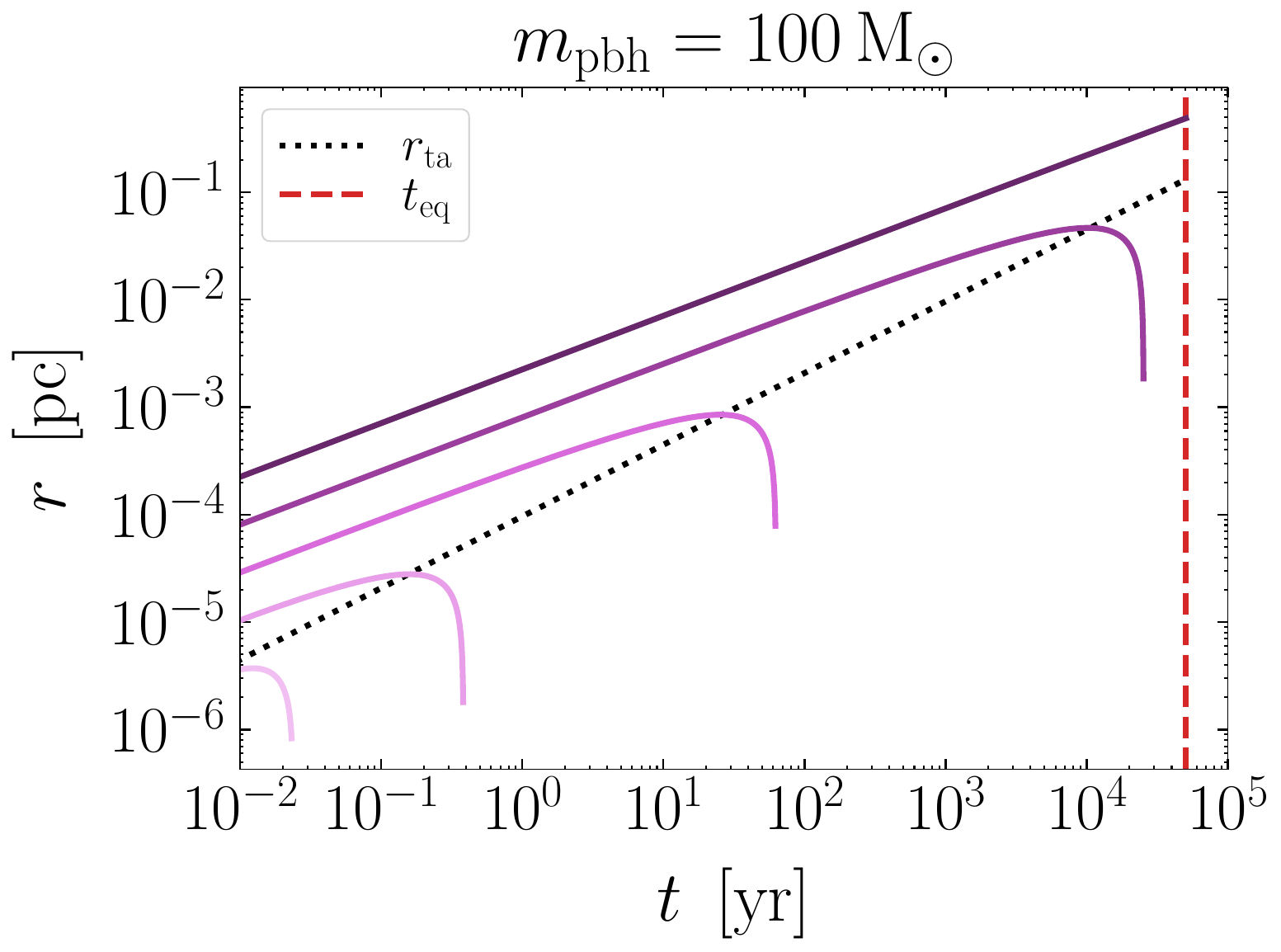}
\end{minipage}
\caption{Dynamics of different DM shells around a PBH of mass $m_{\mathrm{pbh}} = 100\:\mathrm{M_\odot}$ from deep radiation domination up to matter-radiation equality, $t_{\mathrm{eq}}$. The distance $r$ of all the DM shells are measured from the center of the PBH. In the left panel, the blue colored region represents the DM spike around the PBH consisting of various DM shells shown by colored concentric circles. In the right panel, the colored lines show the evolution of different DM shells around the PBH, in accordance with Eq.~\eqref{eq:DMshell}. The black dotted line shows the turnaround radii, $r_{\mathrm{ta}}$ of DM shells given by Eq.~\eqref{eq:r_ta} which grows with time. }
 \label{fig:DMshells}
\end{figure}

At some point, the gravitational attraction term in Eq.~\eqref{eq:DMshell} starts to dominate over the Hubble expansion, i.e.\ $\dot{r}$ becomes zero, due to which the DM shell starts to move towards the PBH and becomes gravitationally bound to it. This is known as the turnaround of the DM shell and occurs at a time $t_{\mathrm{ta}}$. The size of the DM shell at turnaround is known as the turnaround radius $r_\mathrm{ta}$ of the shell. Shells within the turnaround radius are considered bound to the PBH. The turnaround of subsequent DM shells leads to the formation of a DM spike around the PBH  which is shown pictorially in the left panel of Fig.~\ref{fig:DMshells}. During radiation domination, the turnaround radius of the DM shell can be estimated analytically by substituting $\dot{r} = 0$ and $H = 1/(2\,t)$ in Eq.~\eqref{eq:DMshell}.
We have verified that in radiation domination, the numerical estimate of the turnaround radius, 
\begin{equation}\label{eq:r_ta}
 r_{\mathrm{ta}} \simeq \left(2\, G\, m_{\mathrm{pbh}}\, t_{\mathrm{ta}}^{2}\right)^{1/3}\,,
\end{equation}
holds better than its analytic estimate, in agreement with Ref.~\cite{Adamek:2019gns}. So, we shall use the numerical estimate given by Eq.~\eqref{eq:r_ta} as the turnaround radius of DM shells in the rest of the paper.

As per Eq.~\eqref{eq:DMshell}, the variation of the distance $r$ of different DM shells from the center of a PBH of mass $m_{\mathrm{pbh}} = 100\:\mathrm{M_\odot}$ are depicted in the right panel of Fig.~\ref{fig:DMshells}. 
This figure shows that at an initial distance $r_i$ (defined at some arbitrary early time) smaller than $10^{-5}\, \mathrm{pc}$ , the gravitational attraction of the PBH strongly dominates over the Hubble expansion and the shells of the dark matter become quickly gravitationally bound to it. If $r_i$ lies in the range $10^{-5}\,-\,10^{-4} \,\mathrm{pc}$, DM shells initially evolve under the simultaneous effect of gravitational attraction of the PBH and Hubble expansion. After the turnaround point (where gravitational attraction of the PBH becomes stronger than the Hubble expansion), they start moving towards the PBH and become bound to it. For $r_i$ greater than $10^{-4}\,\mathrm{pc}$, the Hubble expansion dominates over the gravitational attraction of the PBH and the DM shells do not get gravitationally bound to the PBH before matter-radiation equality. 

\subsection{Density profile of DM spikes}
\label{subsec:Density profile of DM spikes}
Based on the PBH mass and time scale at which the kinetic decoupling of the DM particles occurs~\cite{Hofmann:2001bi}, the density profile of the DM spikes around the PBHs can be a power law with different radial indices, as discussed in detail in Ref.~\cite{Boudaud:2021irr}. Here, we assume that the formation of the PBHs takes place after the kinetic decoupling of the DM particles such that there is no previous accretion of DM particles around the PBHs~\cite{Eroshenko:2016yve}. Since, in radiation domination, the density of matter varies as $\rho \propto a^{-3}$ and $a \propto t^{1/2}$ then the density of dark matter can be written as~\cite{Adamek:2019gns}:
\begin{equation}\label{eq:totalDMdensity}
    \rho_\mathrm{DM} = \frac{\Omega_\mathrm{cdm}}{\Omega_\mathrm{m}} \, \frac{\rho_{\mathrm{eq}}}{2}\,\left(\frac{t}{t_\mathrm{eq}}\right)^{-3/2}\,.
\end{equation}
In the above equation, the factor of one half comes from the fact that at matter-radiation equality, density of matter is half of the total energy density. We know that the size of the DM spike at any time is given by the turnaround radius $r_{\mathrm{ta}}$ of the DM shells at that time. Then, substituting Eq.~\eqref{eq:r_ta} in Eq.~\eqref{eq:totalDMdensity}, the density of the DM spike in radiation domination at a distance $r$ from the center of a PBH of mass $m_{\mathrm{pbh}}$ can be written as:
\begin{equation} \label{eq:DMdensityprofile}
  \rho_{\mathrm{sp}}(r)  \simeq   \frac{\Omega_\mathrm{cdm}}{\Omega_\mathrm{m}} \, \frac{\rho_{\mathrm{eq}}}{2}\, \left(2\,G \,m_{\mathrm{pbh}}t_{\mathrm{eq}}^{2}\right)^{3/4} r^{-9/4}\,, 
\end{equation} 
with $\rho_{\mathrm{eq}}$ being the total energy density of the Universe at matter-radiation equality ($t_\mathrm{eq}$) and $\Omega_{\mathrm{cdm}}/\Omega_{\mathrm{m}}\approx 0.85$ as the fraction of CDM in the matter density of the Universe~\cite{Planck:2015fie}. Here, the subscript `$\mathrm{sp}$' denotes the DM spike accreted around an isolated PBH. The density profile of $\rho(r)\propto r^{-9/4}$ given by Eq.~\eqref{eq:DMdensityprofile} was originally introduced for the gravitational collapse of collisionless non-relativistic matter in an Einstein-de Sitter universe~\cite{1984ApJ...281....1F,1985ApJS...58...39B}.We note that this profile is steeper than the density profile of $\rho(r)\propto r^{-3/2}$ for the accretion of DM particles around PBHs studied in Ref.~\cite{Kavanagh:2018ggo}. Moreover, Eq.~\eqref{eq:DMdensityprofile} slightly differs from the density profiles of DM spikes calculated in Ref.~\cite{Adamek:2019gns} by a factor of $\Omega_{\mathrm{cdm}}/\Omega_{\mathrm{m}}$, which we have included here to account for the fact that DM does not comprise all of the matter content of the Universe.

Assuming spherical symmetry, the mass of the DM spike around the PBH can be estimated as:
\begin{equation} \label{eq:massofDMspikes}
 m_{\mathrm{sp}}(r) = 4\pi \int_{0}^{r} \rho_{\mathrm{sp}}(r) \, r^{2} \, \mathrm{d}r   =  \frac{8 \pi}{3} \Omega_\mathrm{cdm}\, \rho_{\mathrm{eq}}     \left(2\, G\,  m_{\mathrm{pbh}}\, t_{\mathrm{eq}}^{2}\right)^{3/4} \, 
 \begin{cases}
       \text{$r^{3/4}$} &\quad\text{for $r$} < r_{\mathrm{ta}}\\
       \text{$r_{\mathrm{ta}}^{3/4}$} &\quad\text{for $r$}\ge r_{\mathrm{ta}}\,.
\end{cases}        
\end{equation}
The total mass of the DM spike at any moment is given by the mass enclosed within all the DM shells within the turnaround radius. 

Now, let us define $s = a/a_\mathrm{eq}$ as the scale factor normalized to unity at matter-radiation equality. Then using $H(s) =  \sqrt{4\,\pi\,G\,\rho_{\mathrm{eq}} /3}\,h(s)$ with $h(s)=\sqrt{s^{-3}+s^{-4}}$, the turnaround time of DM shells in terms of turnaround scale factor $s_\mathrm{ta}$ is\footnote{Here, the Hubble parameter does not appear to depend explicitly on $\Omega_\mathrm{m}$ and $\Omega_\mathrm{r}$. Because $s$ is defined with reference to matter-radiation equality, i.e.\ $s = a/a_\mathrm{eq}$, these terms have been absorbed in factors of $a_\mathrm{eq}$.}:
\begin{equation}  \label{eq:t_ta}
      t_\mathrm{ta} =\left(\frac{3}{4G\pi\rho_\mathrm{eq}}\right)^{1/2}\,\left[\frac{2}{3}\left(s_\mathrm{ta}-2\right)\left(s_\mathrm{ta}+1\right)^{1/2} + \frac{4}{3}\right]\,.
 \end{equation}
Substituting the values of the turnaround radius given by Eq.~\eqref{eq:r_ta} and turnaround time given by Eq.~\eqref{eq:t_ta} in Eq.~\eqref{eq:massofDMspikes}, one can easily verify that the mass of the DM spike being accreted around the PBH up to matter-radiation equality is:
\begin{equation}  \label{eq:spikemass}
      m_{\mathrm{sp}}\left(s_\mathrm{eq}\right) = 0.52\,m_\mathrm{pbh}\,.
 \end{equation}
This value is around half of the mass of the corresponding PBH and is in rough agreement with previous calculations of~\cite{Mack:2006gz,Kavanagh:2018ggo} which suggest that at matter-radiation equality, the mass of the DM spike becomes comparable to the mass of the PBH itself. Small differences compared to previous calculations arise from minor numerical factors, including the extra factor of $\Omega_\mathrm{cdm}/\Omega_\mathrm{m}$ in the density profile of the DM spike given by Eq.\eqref{eq:DMdensityprofile}.

\section{Formation of PBH Binaries}
\label{sec:Formation of PBH Binaries}
In this section, we set up a model for the formation of PBH binaries in the presence of DM spikes around each PBH. In doing so, we first provide an improvement over Ref.~\cite{Kavanagh:2018ggo} by self-consistently taking into account the growth of DM spikes during the process of binary formation. Then we extend the formalisms of Refs.~\cite{Ali-Haimoud:2017rtz,Chen:2018czv} for the initial angular momentum of binaries by including the presence of DM spikes. The relevant technical details are given in the following sub-sections, but we here briefly summarize them. In Sec.~\ref{subsec:Initial semi-major axis of PBH binaries}, we show that the impact of DM spikes on the initial semi-major axis of the PBH binaries depends on the scale factor at which the formation of the binaries takes place. As the scale factor of binary formation increases, the initial separation of dressed PBH binaries becomes increasingly smaller than the same PBH binaries without DM spikes. In Sec.~\ref{subsec:Initial angular momentum of PBH binaries} we show that the change in initial angular momentum due the presence of DM spikes is negligible for PBH binaries decoupling early in radiation domination and its value is similar to Ref.~\cite{Ali-Haimoud:2017rtz}.

Similar to Ref.~\cite{Chen:2018czv}, we write $f$ as the total abundance of PBHs in non-relativistic matter such that the fraction of PBHs in CDM is  $f_{\mathrm{pbh}} =  \Omega_{\mathrm{pbh}}/\Omega_{\mathrm{cdm}} \approx f/0.85$. Depending on the formation mechanism, PBHs will possess a mass function $P(m)$, which can be described by a normalized probability distribution function (PDF). Common parametrizations for the mass function include Power Law, Broken Power Law and Lognormal~\cite{Green:2020jor}. In our calculations, we adopt a few specific extended PBH mass functions discussed later in Sec.~\ref{sec:Merger rates of PBH binaries}. We assume that the PBHs of masses $m_{i}$ and $m_{j}$ have the discrete resolution of $\Delta_{i}$ and $\Delta_{j}$ respectively~\cite{Chen:2018czv}\footnote{To keep the notation simple and consistent with Ref.~\cite{Chen:2018czv}, we have written the discrete PBH mass resolution as $\Delta m_{i} \equiv \Delta_{i}$.} such that the equation of normalization for any PDF in the mass range from $m_\mathrm{min}$ to $m_\mathrm{max}$ can be written as:
\begin{equation} 
\int_{m_\mathrm{min}}^{m_\mathrm{max}} P(m_{i}) \,\mathrm{d}m_{i} \simeq 1 \quad \equiv \quad  \sum_{ m_{i} = m_{\mathrm{min}}}^{m_{\mathrm{max}}} P(m_{i}) \Delta_{i} \simeq 1 \,.  
\end{equation}
Since $P(m_i)$ describes the probabilistic distribution of PBHs having mass $m_i$ then the fraction of matter in the forms of PBHs with mass $m_i$ becomes~\cite{Chen:2018czv}:
\begin{equation}
 f P(m_{i}) \Delta_{i}\equiv f_{i} \Delta_{i}\,.   
\end{equation}
We assume that the PBHs are uniformly distributed in the Universe such that the average comoving separation\footnote{We define comoving quantities with respect to matter-radiation equality.} between the PBHs of mass $m_i$ can be written as:
\begin{equation}  \label{eq:x_i}
 \bar{x}_{i}=\left(\frac{3}{4\pi} \frac{m_{i}}{\rho_\mathrm{eq}f_{i}\Delta_{i}}\right)^{1/3}\,.    
\end{equation}
Similarly, PBHs of mass $m_j$ have an average comoving separation $\bar{x}_{j}$ in between them. Then, using the values of comoving separations $\bar{x}_{i}$ and $\bar{x}_{j}$, we can calculate the average comoving separation between the PBHs of masses $m_{i}$ and $m_{j}$ as~\cite{Chen:2018czv}:
\begin{equation}   \label{eq:xij}
 \langle x_{ij}\rangle =  \left(\bar{x}_{i}^{-3} + \bar{x}_{j}^{-3}\right)^{-1/3}
    = \, \mu_{ij}^{1/3} \bar{x}_{ij}\,,    
\end{equation} 
with 
\begin{equation}   \label{eq:mu}
\mu_{ij} = \frac{m_{i}\, m_{j} \,f_{b}\left(\Delta_{i}\Delta_{j}\right)^{1/2}}{ \left(m_{i} + m_{j}\right)\left(f_{j}\Delta_{j}m_{i} + f_{i}\Delta_{i}m_{j}\right)}       \quad; \quad f_{b}=f_{i} + f_{j}\,,   
\end{equation} 
and
\begin{equation}   \label{eq:comovingseparation}
\bar{x}_{ij}^3 = \frac{3\left(m_{i} + m_{j}\right)}{4\pi\rho_{\mathrm{eq}} \,f_{b}\, \left(\Delta_{i}\Delta_{j}\right)^{1/2}}\,.
\end{equation} 
 In the rest of the paper, the subscript `${ij}$' is omitted in the definition of $\mu_{ij}$ and the comoving separation $\langle x_{ij}\rangle$ or $\bar{x}_{ij}$. Note that, unlike in Ref.~\cite{Chen:2018czv}, we have allowed for different discrete spacing i.e.\ $\Delta_{i} \neq \Delta_{j}$ for PBHs of mass $m_{i}$ and $m_{j}$.  In Sec.~\ref{subsec:Initial semi-major axis of PBH binaries} and Sec.~\ref{subsec:Initial angular momentum of PBH binaries}, we mention all relevant technical details as an extension of Ref.~\cite{Chen:2018czv}, which can be skipped by the knowledgeable reader.

\subsection{Initial semi-major axis of PBH binaries}
\label{subsec:Initial semi-major axis of PBH binaries}
 Now, we estimate how the presence of DM spikes with density profile given by Eq.~\eqref{eq:DMdensityprofile} influences the formation of PBH binaries. Since all the neighbouring PBHs and matter-density fluctuations of the Universe also affect the dynamics of the binaries formation, so we include their impact too, which is mentioned in Section~\ref{subsec:Initial angular momentum of PBH binaries}. The formation of a binary takes place between the two nearest neighbouring PBHs of masses $m_{i}$ and $m_{j}$ with $x$ being the comoving separation between them. We assume that no other PBH lies inside a spherical volume of radius equal to the comoving separation $x$. The two PBHs decouple from the Hubble flow  due to their mutual gravitational attraction and form a bound system. However, every PBH in the Universe may have multiple PBHs in its neighbourhood which also might decouple at a similar time, leading to hierarchical interactions  of systems comprising more than two PBHs. These N-body interactions of PBHs have been studied in Ref.~\cite{Raidal:2018bbj} where they conclude that if PBHs are a very tiny fraction of the total DM i.e. $f_{\mathrm{pbh}} \ll 1$ then the possibility of hierarchical PBHs interactions becomes very low. In this work, we consider small values of $f_\mathrm{pbh} \lesssim 10^{-2}$ and we therefore neglect the  disruption of individual binaries via multiple PBH interactions in their vicinity.  

Now, similar to Eq.~\eqref{eq:DMshell}, in a binary, the equation of motion of the proper separation $r_{b}$ between the PBHs with masses $m_i$ and $m_j$ having DM spikes of masses $m_{\mathrm{sp},i}(t)$ and $m_{\mathrm{sp},j}(t)$ respectively is given as:
\begin{equation}  \label{eq:binaryevolution}
  \frac{\mathrm{d}^{2} r_{b}}{\mathrm{d} t^{2}}  = -\frac{G \left[ m_{i} + m_{j} + m_{\mathrm{sp},i}\left(t\right) + m_{\mathrm{sp},j}\left(t\right)\right]}{r_{b}^{2}} \,\frac{r_{b}}{|r_{b}|}  + \left(\dot{H} + H^{2}\right)r_{b}\,,
\end{equation}    
where the dot represents the derivative with respect to proper time $t$. Here, we assume that each PBH and its DM spike can be treated as a point mass which depends on time. As the dressed PBHs evolve under the simultaneous effect of their mutual gravitational attraction and Hubble expansion, mass of each DM spike increases continuously with time $t$. Eventually at some point, the gravitational attraction term in Eq.~\eqref{eq:binaryevolution} dominates over the expansion term and the pair of PBHs reach a maximum separation before turning around to approach each other again. This point is referred to as the point of binary decoupling and we consider this to be the time of formation of the binary. After binary decoupling, we no longer consider the growth of DM spikes around the isolated PBHs or around the binary as a whole\footnote{The accretion of DM spikes around the binary as a whole has been considered in Ref.~\cite{Pilipenko:2022emp}}. So, we assign the initial parameters of the binary at the point of decoupling.

Then, similar to~\cite{Ali-Haimoud:2017rtz} with the addition of DM spikes around PBHs in picture, we define a dimensionless separation, $\chi = r_{b}/x$ and express Eq.~\eqref{eq:binaryevolution} as:
\begin{equation}   \label{eq:binarywithspikes}
 \chi'' + \frac{\left(s h' + h\right)}{s^{2} h} \left(s \chi' - \chi \right) + \frac{1}{\lambda(s)} \frac{1}{\left(sh\right)^2}
   \frac{1}{\chi^{2}} \frac{\chi}{|\chi|} = 0 \,,
   \end{equation} 
where primes represent the derivative with respect to scale factor $s$ and $h \equiv h(s) = \sqrt{s^{-3} + s^{-4}}$. 
Also,
\begin{equation}   
  \lambda(s) = \frac{4 \,\pi \, \rho_\mathrm{eq} x^{3}}{3 \left[m_{i} + m_{j} +  m_{\mathrm{sp},i}(s) +  m_{\mathrm{sp},j}(s)  \right]} = \lambda \cdot n(s)\,, 
\end{equation}
with a dimensionless parameter
\begin{equation}   \label{eq:lambda}
 \lambda = \frac{4 \,\pi \, \rho_\mathrm{eq} x^{3}}{3 \left(m_{i} + m_{j} \right)} = \frac{X}{f_{b} \left(\Delta_{i}\Delta_{j}\right)^{1/2}}     \quad;\quad   X\equiv \frac{x^{3}}{\bar{x}^{3}}\,, \end{equation} 
 which quantitatively describes how the mass of the PBHs present in the volume occupied by the binary differs from the mass of the background matter lying in the same volume.\footnote{The value of $\lambda$ given by Eq.~\eqref{eq:lambda} appears to differ from Ref.~\cite{Ali-Haimoud:2017rtz,Chen:2018czv} by a factor of $2$ because we have considered $\rho_\mathrm{eq}$ as the total energy density of the Universe at matter-radiation equality rather than the energy density of matter only. But our calculations are still consistent with Ref.~\cite{Ali-Haimoud:2017rtz,Chen:2018czv}.} Large values of $\lambda$ correspond to PBH pairs having a large initial separation, due to which the PBHs decouple late as the background density dominates over the mutual density of the two PBHs for a longer time. Here, the only term signifying the effect of DM spikes in the PBH binary is: 
\begin{equation}
\begin{aligned}
n(s) &= \left[1 + \frac{\left( m_{\mathrm{sp},i}(s) +  m_{\mathrm{sp},j}(s)\right)}{\left(m_i + m_j\right)}\right]^{-1}\\
&= \left[1 + 2^{5/4}\,\frac{\Omega_\mathrm{cdm}}{\Omega_\mathrm{m}}
\,\left(\frac{8\pi G\rho_{\mathrm{eq}} \,t_{\mathrm{eq}}^{2}}{3}\right)^{3/4}\left(\frac{2}{3}(s-2)(s+1)^{1/2} + \frac{4}{3}\right)^{1/2}\right]^{-1} \,.   \end{aligned}
\end{equation} 
This explicit expression shows that the impact of DM spikes on the dynamics of binary formation is independent of the PBHs masses in the binary. This is applicable to any PBH binary dressed with DM spikes having the density profile of $\rho(r)\propto r^{-9/4}$. 

\begin{figure}[tb!]
\centering
\includegraphics[width = 0.47\textwidth]{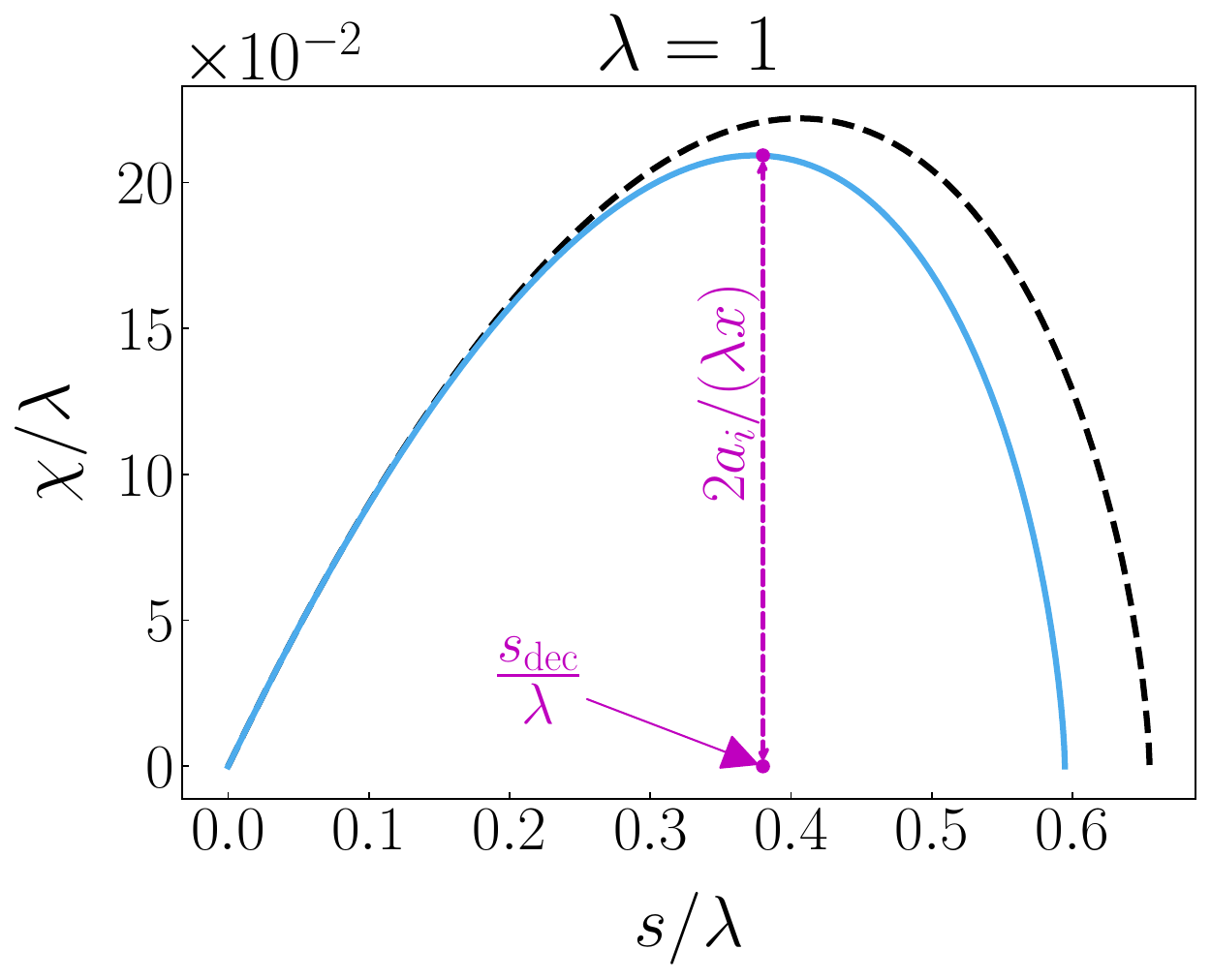}
\includegraphics[width = 0.48\textwidth]{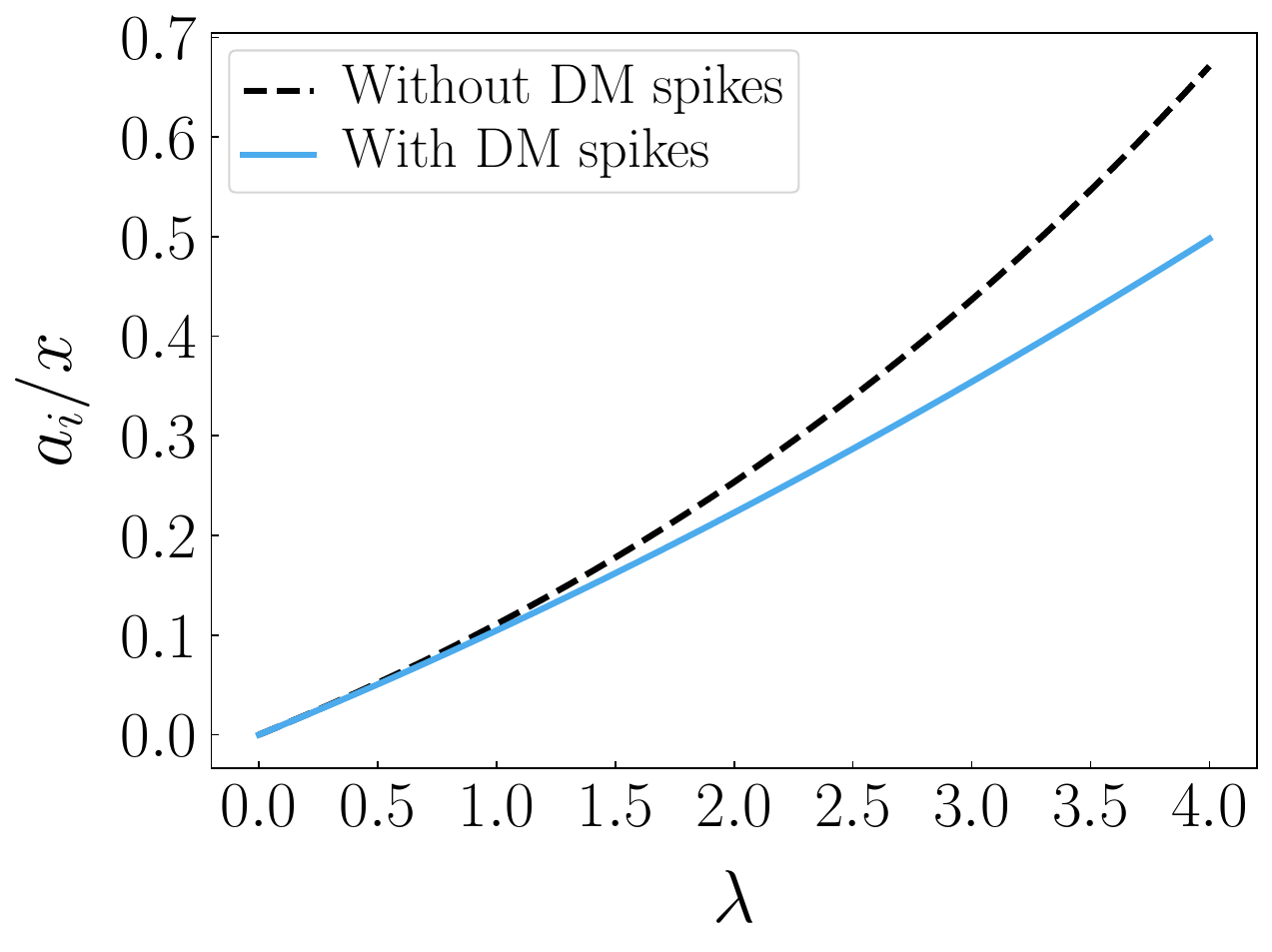}
\caption{The left panel shows the dynamics of PBH binaries with and without DM spikes up to the first cross-over of the PBHs using $\lambda = 1$ in Eq.~\eqref{eq:binarywithspikes}. The right panel shows the variation of the initial semi-major axis $a_i$ of the PBH binaries with and without DM spikes as a function of $\lambda$ as per the numerical solution of Eq.~\eqref{eq:binarywithspikes}. Here, $\lambda$ is the dimensionless parameter given by~\eqref{eq:lambda} which is a proxy for the PBH separation and which quantifies how much larger the background density is compared to the mutual density of the two PBHs. In addition, $x$ is the comoving separation of the PBHs in the binaries and $s_\mathrm{dec}$ is the scale factor of the binary decoupling. }
\label{fig:semimajoraxisvslambda}
\end{figure}

Following Eq.~\eqref{eq:binarywithspikes}, the formation of PBH binaries with and without DM spikes for $\lambda = 1$ is illustrated in the left panel of Fig.~\ref{fig:semimajoraxisvslambda}. This panel shows that PBHs with DM spikes decouple earlier than the PBHs without spikes due to the increased gravitational attraction with the presence of DM spikes\footnote{We write the scale factor at which binary decoupling takes place as $s_\mathrm{dec} = \left(1+z_\mathrm{eq}\right)/\left(1+z_\mathrm{dec}\right)$.}. We estimate the initial semi-major axis $a_{i}$ of the binaries as half of the proper separation of the PBHs at the point of binary decoupling because this is the maximum separation shared by the PBHs of the binaries i.e.\ $r_{b, \mathrm{max}} \approx \left(1 + e_{i}\right) a_{i}$ with $e_{i} \sim 1$. Hence, we take the initial dimensionless separation of the PBHs at binary decoupling i.e.\ when the curve $\chi/\lambda$ first turns over such that $|\chi|$ becomes equal to $2a_{i}/x$. 

Now, for different values of $\lambda$, the impact of DM spikes on the initial semi-major axis of PBH binaries can be seen in the the numerical solution of Eq.~\eqref{eq:binarywithspikes} shown in the right panel of Fig.~\ref{fig:semimajoraxisvslambda}. From this panel, it is clear that for $\lambda \ll 1$ values, the behaviour of a PBH binary having DM spikes is not significantly different from the same binary having no DM spikes. We find that the scale factor $s_\mathrm{dec}$ of binary decoupling is directly proportional to the value of $\lambda$. For $\lambda \ll 1$, binaries decouple very early and hence do not have enough time to accrete substantial DM spikes. But for $\lambda\gtrsim 1$, the initial separation of the binary with DM spikes is comparatively smaller because the binary has enough time to accrete DM spikes. With the accretion of DM spikes, the gravitational attraction between the PBHs in the binary enhances and prevents their further evolution due to Hubble expansion. This ultimately makes the initial separation of the PBH binaries smaller than the PBH binaries without DM spikes. This behaviour of the initial semi-major axis of dressed PBH binaries qualitatively agrees with the initial semi-major axis of PBH binaries having DM spikes with the density profile of $\rho(r)\propto r^{-3/2}$ shown in Fig.~$3$ of Ref.~\cite{Kavanagh:2018ggo}. 
From this panel, the initial semi-major axis of the PBH binaries without DM spikes for a given value of $\lambda$ can be approximated as:
\begin{equation}  \label{eq:initialsemimajoraxisnohalos}
   a_{i,0}\approx\left(0.09 \, \lambda + 0.02 \, \lambda^{2}\right) x\,,
\end{equation}
with the initial semi-major axis for PBH binaries having DM spikes as:
\begin{equation}  \label{eq:initialsemimajoraxishalos}
     a_{i}\approx\left(0.09 \, \lambda+0.01\, \lambda^{2}\right) x \,.
\end{equation}
The subscript `$0$' signifies PBH binaries without DM spikes.

\subsection{Initial angular momentum of PBH binaries}
\label{subsec:Initial angular momentum of PBH binaries}
During the formation of binaries, apart from their mutual gravitational attraction, PBHs also experience tidal torque due the  gravitational force of all other PBHs and large scale density perturbations in the Universe~\cite{Hayasaki:2009ug, Eroshenko:2016hmn}. This tidal force provides an initial angular momentum to the PBHs which avoids their head-on collision, and leads to the formation of very eccentric binaries. Now in Sec.~\ref{subsec:Due to PBHs outside the binaries} and Sec.~\ref{subsec:Due to linear density perturbations}, we calculate the probability distribution of this initial angular momentum of PBH binaries by including the presence of DM spikes around PBHs in the formalism of Ref.~\cite{Ali-Haimoud:2017rtz}.

\subsubsection{Tidal Torque due to all PBHs outside the binaries}
\label{subsec:Due to PBHs outside the binaries} 
The gravitational attraction of all the PBHs outside the binary leads to a net tidal force on the PBHs in the binary. The local tidal field produced by any PBH outside the binary is defined as $T_{ij}=-\partial_i\partial_j\phi$ with $\phi$ being the gravitational potential~\cite{Ali-Haimoud:2017rtz}. Due to this tidal field, the PBHs in the binary experience a force given by~\cite{Ali-Haimoud:2017rtz,Chen:2018czv}:
\begin{equation}
    \textbf{F} = \textbf{T} \cdot \textbf{r}_{b}\,.
\end{equation}
Here, $\textbf{F}$ is the tidal force per unit reduced mass of the binary. This perturbative force provides an angular momentum per unit reduced mass of the PBH binary as:
\begin{equation}  \label{eq:tidalangularmomentum}
  \textbf{\textit{l}} = \int \mathrm{d}t\,  \textbf{r}_{b} \times[\textbf{T} \cdot \textbf{r}_{b}]\,. 
\end{equation}
Now, consider a dressed PBH lying outside the binary, designated as the distant PBH. The total mass of the distant PBH is $m_{\mathrm{total},d}(s) = m_{d} + m_{\mathrm{sp},d}(s)$ with $m_{d}$ being the mass of the PBH itself and $m_{\mathrm{sp},d}$ as the mass of its DM spike at scale factor $s$.
The tidal torque will grow as the total mass of the distant PBH grows with $s$. So, if the comoving separation $y$ of the distant PBH from the center of the PBH binary is approximately constant and much larger than the separation of the binary then $\mathbf{T} \propto m_{\mathrm{total},d}(s)/s^3$. The tidal field exerted on the PBH binary at matter-radiation equality by the distant PBH at comoving separation $y \gg x$ is~\cite{Ali-Haimoud:2017rtz}:
\begin{equation} \label{eq:T_eq}
 \frac{\textbf{T}_{\mathrm{eq}}}{G\,m_{\mathrm{total},d}(s_{\mathrm{eq}})}=\frac{3\hat{\textbf{y}}_{i}
 \hat{\textbf{y}}_{j} - \boldsymbol{\delta}_{ij}} {y^{3}} = \mathcal{T}_\mathrm{eq}\,.
\end{equation}
Then, using $\textbf{T} \propto m_{\mathrm{total},d}(s)/s^{3}$ and the above equation, the tidal field experienced by the PBH binary at any scale factor $s$ can be written as:
\begin{equation}   \label{eq:tidaltorque}
  \textbf{T} \approx s^{-3} \, \textbf{T}_{\mathrm{eq}} \:\frac{m_{\mathrm{total},d}(s)}{m_{\mathrm{total},d}(s_{\mathrm{eq}})}\,.  
\end{equation}

Now substituting the value of the tidal torque given by Eq.~\eqref{eq:tidaltorque} in Eq.~\eqref{eq:tidalangularmomentum}, the angular momentum of the PBH binary becomes:
\begin{equation}  \label{eq:angularmomentumwithspikes}  \textbf{\textit{l}}   =   \left(\frac{3\,G}{4\,\pi\,\rho_\mathrm{eq}}\right)^{1/2} \int_{s_{i}}^{s_\mathrm{dec}} \frac{\mathrm{d}s}{sh(s)}  \, \frac{\chi^{2}\left(s;\lambda\right)}{s^{3}} \, {\left(m_{d} + m_{\mathrm{sp},d}(s)\right) } \: \textbf{\textit{x}} \times [\mathcal{T}_\mathrm{eq} \cdot \textbf{\textit{x}}] \,, 
\end{equation}
with numerical solution
\begin{equation}     \label{eq:angularmomentum}
  \textbf{\textit{l}}   \approx  \left(\frac{3}{4\,\pi\,G\,\rho_\mathrm{eq}}\right)^{1/2} \left(0.26\, \lambda + 0.02\, \lambda^{2}\right)\: \textbf{\textit{x}} \times [\mathcal{T}_\mathrm{eq} \cdot \textbf{\textit{x}}]\,.
\end{equation}
Here, $s_{i} \ll 1$ is some initial scale factor corresponding to radiation domination and $s_\mathrm{dec}$ is the scale factor at which the binary decouples from the Hubble expansion. For $\lambda \ll 1$, the value of the angular momentum given by Eq.~\eqref{eq:angularmomentum} for the PBH binary with DM spikes agrees with the value for PBH binaries without DM spikes calculated in Ref.~\cite{Ali-Haimoud:2017rtz}. And from Eq.~\eqref{eq:angularmomentumwithspikes}, we see that the presence of DM spikes around the distant PBHs impacts the tidal torque substantially only for $s_\mathrm{dec} \geq s_\mathrm{eq}$ corresponding to $\lambda > 1$. This is because the masses of DM spikes are lighter for PBH binaries which decouple much earlier than matter-radiation equality, meaning that the enhanced tidal gravitational force due to the presence of spikes is not enough to substantially change the initial angular momentum.

Now, the dimensionless angular momentum of the PBH binary due to the tidal torques is~\cite{Ali-Haimoud:2017rtz,Chen:2018czv}:
\begin{equation} \label{eq:dimensionlessangularmomentum}
     \textbf{j} = \frac{ \textbf{\textit{l}}}{\sqrt{G\,m_{b}\,a}} \,,
\end{equation}
with $m_{b} = \left[m_{i} + m_{j} +  m_{\mathrm{sp},i}(s) +  m_{\mathrm{sp},j}(s) \right]$ being the total mass of the binary at scale factor $s$. For the tidal field generated by the distant PBH situated at comoving separation $y \gg x$~\cite{Ali-Haimoud:2017rtz} and semi-major axis $a$ given by Eq.~\eqref{eq:initialsemimajoraxishalos}, we can rewrite the dimensionless angular momentum of the binary as:
\begin{equation} 
      \textbf{j} \approx2.45 \, \frac{\mathcal{C}(\lambda)}{\sqrt{m_{b}\,\left(m_{i}+ m_{j}\right)}}\: m_{d}  \: \frac{x^{3}}{y^{3}} \left(\hat{\textbf{\textit{x}}}\cdot\hat{\textbf{\textit{y}}}\right)\left(\hat{\textbf{\textit{x}}}\times\hat{\textbf{\textit{y}}}\right)\,.
\end{equation}
The dimensionless constant
\begin{equation}
 \mathcal{C} (\lambda) = \frac{\left(1 + 0.0861\, \lambda\right)\: }{\sqrt{\left(1 + 0.0696\, \lambda\right)}} \,,   
\end{equation}
characterizes the configuration of the PBH binary based on the PBH masses and their comoving separation $x$. 
Note that in the limit $\lambda \ll 1$, the value of $\mathcal{C} (\lambda)$ becomes unity and matches with the case without DM spikes. 

The total reduced angular momentum of the PBH binary with DM spikes resulting from \textit{all} distant PBHs in the Universe at comoving separation $y\gg x$ is:
\begin{equation}  \label{eq:dimensionlessangularmomentum2}
   \textbf{j}_\mathrm{total}\approx2.45 \: \frac{\mathcal{C}(\lambda)}{\sqrt{m_{b}\,\left(m_{i}+ m_{j}\right)}}\: \sum_{q} m^{q}_{d} \: \left(\hat{\textbf{\textit{x}}}_{q}\cdot\hat{\textbf{\textit{y}}}_{q}\right)\left(\hat{\textbf{\textit{x}}}_{q}\times\hat{\textbf{\textit{y}}}_{q}\right)\,.
\end{equation}
Now, applying the formalism described in Appendix $1$ of Ref.~\cite{Ali-Haimoud:2017rtz}, the probability distribution of the initial angular momentum of PBH binaries is given as:
\begin{equation}    \label{eq:jipdf}
j\frac{dP}{dj}\bigg\vert_{\lambda}  = \frac{\gamma_{\lambda}^{2}}{\left(1+\gamma_{\lambda}^{2}\right)^{3/2}} \,,
\end{equation}
with $\gamma_\lambda = j/j_\lambda$ and 
\begin{equation}     \label{eq:tidalj_Xwithspikes}
j_{\lambda}  = 0.41\: \lambda \cdot \mathcal{C}(\lambda) \,  \sqrt{\frac{\left(m_{i}+ m_{j}\right)}{m_{b}}}  \,.
\end{equation}
 Here, $j_{\lambda}$ is the characteristic value of the initial angular momentum $j$ for a given value of the dimensionless variable $\lambda$ which specifies the comoving separation of the PBHs in the binaries. This distribution of initial angular momentum given by Eq.~\eqref{eq:jipdf} for PBH binaries with DM spikes is valid for $0 < j < 1$ and we set $P(j) = 0$ for $j>1$.

\subsubsection{Tidal torque due to large scale density perturbations}
\label{subsec:Due to linear density perturbations}
In the approximation that PBHs do not make up all of the dark matter, large scale matter density perturbations in the Universe also exert tidal forces on the PBH binaries. Since the dynamics can become very complex if the scale of density perturbations is either smaller or of the order of the separation of PBH binaries, so we are only considering the tidal torque generated by density fluctuations having comparatively larger length scales. Then, the tidal torque exerted on PBH binaries 
by such large scale matter density perturbations $\delta_{m}$ is~\cite{Ali-Haimoud:2017rtz}:
\begin{equation}
 T_{ij}^\mathrm{eq}=-\partial_{i}\partial_{j}\phi_{p}=-4\pi G\rho_\mathrm{eq}\partial_{i}\partial_{j}\partial^{-2}\delta_{m}\,,
 \end{equation}
 with $\phi_{p}$ as the gravitational potential due to the large scale density perturbations. Since the tidal field generated by the perturbations $\delta_{m}$ is Gaussian so the tidal tensor $T_{ij}^\mathrm{eq}$ is Gaussian, resulting in reduced angular momentum of the PBH binaries with DM spikes as
\begin{equation} \label{eq:jofmatterdensity}
     \textbf{j} = 1.61 \,  \:\: \frac{\mathcal{C}(\lambda) }{G\,\sqrt{m_{b}\,\left(m_{i}+ m_{j}\right)}}\:  x^{3}\, \hat{\textit{\textbf{x}}}\times[\textbf{T}_\mathrm{eq}\cdot \hat{\textit{\textbf{x}}}]\,. 
\end{equation}
Now as per the reduced angular momentum given by the above equation and Appendix $2$ of Ref.~\cite{Ali-Haimoud:2017rtz}, the variance of the distribution of $\textbf{j}$ in the plane perpendicular to $\hat{x}$ is:
\begin{equation}  \label{eq:varianceofj}
    \left\langle j^{2}\right\rangle  = \left[0.29 \: \lambda \cdot \mathcal{C}(\lambda)\: \sigma_\mathrm{eq}\: \sqrt{\frac{\left(m_{i}+ m_{j}\right)}{m_{b}}} \right]^{2}\,,
\end{equation}
with $\sigma_\mathrm{eq} = \left\langle \delta^{2}_\mathrm{eq} \right\rangle ^{1/2} \approx 0.005$.

In general, the total initial angular momentum of the PBH binary is the sum of the angular momenta due to all distant PBHs and large scale density perturbations with characteristic value:
\begin{equation}  \label{eq:jXhalo}
j_{\lambda} = \left[ j_{\lambda}^{2} +  \left\langle j^{2}\right\rangle\right]^{1/2}\approx 0.41 \: \lambda \cdot \mathcal{C}(\lambda) \: \left[f^{2} + 0.52\,\sigma_\mathrm{eq}^{2}\right]^{1/2} \: \sqrt{\frac{\left(m_{i}+ m_{j}\right)}{m_{b}}}\,.
\end{equation}
The characteristic angular momentum of PBH binaries without DM spikes can be calculated by substituting $m_{b} = \left(m_i + m_j\right)$ and $\mathcal{C}(\lambda) \rightarrow 1$ (in the limit $\lambda \ll 1$) in the above equation. Then, similar to~\cite{Ali-Haimoud:2017rtz}, we assume that the probability distribution of the total angular momentum has the same expression as given in Eq.~\eqref{eq:jipdf}, but with the characteristic angular momentum given by Eq.~\eqref{eq:jXhalo}. 
We find that the impact of DM spikes is larger for higher values of $\lambda$, since binaries which decouple late have enough time to accrete massive DM spikes. However, comparing with the case without DM spikes, we find that even for large $\lambda$ the impact of DM spikes on the angular momentum is not significant.

\section{Merger time of PBH binaries}
\label{sec:Merger time of the binaries}
Since a lot of detailed analysis has already been done on PBH binaries without DM spikes~\cite{Ali-Haimoud:2017rtz, Kocsis:2017yty,Chen:2018czv,Liu:2018ess,Liu:2019rnx}, our main focus will be on PBH binaries which contain DM spikes. Now, in Sec.~\ref{subsec:merger time Ejected DM spikes} and Sec.~\ref{subsec:merger time Static DM spikes} we calculate the final merger time of PBH binaries by taking into account the impact of DM spikes on their merger. We consider two possible scenarios in which the DM spikes can alter the merger dynamics of PBH binaries, listed as follows:
\begin{enumerate}[]
\item \textbf{Evaporated DM spikes} -  We assume that for very eccentric orbits, every time the PBHs in the binaries come close to each other there is an exchange of energy between the PBHs and the DM spikes which lead to the ejection of DM particles from the binaries. This continuous loss of energy from the PBHs leads to a decrease in the semi-major axis and circularizes the binaries until the DM spikes are completely evaporated before the merger, as studied in Ref.~\cite{Kavanagh:2018ggo}.  
In the rest of this work, PBH binaries merging after complete evaporation of DM spikes are labelled as PBH binaries with \textbf{evaporated} DM spikes. We define the final orbital parameters of these PBH binaries at the stage where the masses of the DM spikes become zero after which we assume that the binaries lose energy only via emission of gravitational waves.

\item \textbf{Static DM spikes} - We consider PBH binaries with DM spikes in analogy to Intermediate Mass Ratio Inspiral (IMRI) systems in which a lighter object inspirals into another object of intermediate mass. The impact of DM spikes in the orbital evolution of IMRIs is an on-going field of study (see e.g.~\cite{Eda:2014kra,Kavanagh:2020cfn,Becker:2021ivq}). We consider highly eccentric PBH binaries in which the primary PBH of mass $m_{i}$ is at the center of its DM spike and the secondary PBH with mass $m_{j} \ll m_i$ inspirals into the primary PBH. Due to the existence of a strong gravitational potential between the massive primary PBH and its DM spike, a lot of energy is required to remove the DM particles from the spike. For a very light secondary PBH spiralling down into the primary PBH, it is therefore difficult to eject DM particles. Hence, under this assumption, the density of the DM spikes can be treated as static up to the merger of the PBH binary. In rest of work, PBH binaries with such type of spikes as denoted as PBH binaries with \textbf{static} DM spikes. 
\end{enumerate}
The actual dynamics of a realistic PBH binary will lie in between these two cases of \textbf{evaporated} DM spikes and \textbf{static} DM spikes. The assumption of \textbf{evaporated} spikes should be an accurate description when the PBH masses are close to equal (as shown in Ref.~\cite{Kavanagh:2018ggo}). The assumption of \textbf{static} spike should be valid in the limit of very large mass ratios. In general, the DM spike of the primary PBH will respond to the inspiral of the secondary PBH but we expect the spike to survive in this process. We will apply these two scenarios regardless of the PBH mass ratio, though we discuss in Sec.~\ref{subsec:Impact of DM spikes on merger time} and Sec.~\ref{sec:conclusions} more precisely when we expect these assumptions to fail and what steps are needed for a more complete treatment.

\subsection{Evaporated DM spikes}
\label{subsec:merger time Ejected DM spikes}
Now, we calculate the final merger time of PBH binaries with \textbf{evaporated} DM spikes by extending the analytical approach of Ref.~\cite{Kavanagh:2018ggo} for extended mass functions and the DM density profile of $\rho_\mathrm{sp}(r)\propto r^{-9/4}$ mentioned in Sec.~\ref{subsec:Density profile of DM spikes}. 
In highly eccentric binaries, the orbit is almost radial, due to which the exchange of angular momenta between the PBHs and the DM spikes is negligible~\cite{Kavanagh:2018ggo}. Hence, we make the assumption that the angular momentum of the PBHs remains conserved irrespective of the angular momentum of the DM spikes. So, the binaries circularize only due to the loss of the binding energy of DM spikes in every close encounter of PBHs. In that case, the angular momentum of the PBHs in the binary is given as:
\begin{equation}  \label{eq:angularmomentumspbhs}
 L = \sqrt{\mu_{0} \, G\, m_{i}\,m_{j}\,a}\,j \,,
\end{equation}
with $\mu_{0} = m_{i} m_{j}/\left(m_{i} + m_{j}\right)$ being the reduced mass of the binary in the absence of DM spikes. Then, applying the conservation of angular momentum before and after the evaporation of the DM spikes, we get:
\begin{equation}  \label{eq:finalangularmomentum}
  j_{f} =  \sqrt{\frac{a_{i}}{a_{f}}}\; j_{i}\quad ; \quad  j = \sqrt{1 - e^{2}}\,,
\end{equation}
with $e$ being the eccentricity of the PBH binary orbit. Now, the binding energy of each DM spike is~\cite{Kavanagh:2018ggo}: 
\begin{equation}   \label{eq:E_bind}
  E^{\mathrm{bind}}_{\mathrm{sp}} = -4\,\pi \,G \int_{r_s}^{\infty} \frac{\left[m_\mathrm{pbh} + m_\mathrm{sp}(r)\right]}{r} \: r^{2}\rho_{\mathrm{sp}}(r)\, \mathrm{d}r\,,
\end{equation} 
where $r_{s}$ is the Schwarzschild radius of the PBH. In the approximation that PBHs with DM spikes can be treated as point objects, the initial orbital energy of the binary with PBH masses $m_{i}$ and $m_{j}$ having DM spikes of masses $m_{\mathrm{sp},i}(r)$ and $m_{\mathrm{sp},j}(r)$ respectively, can be given as:
\begin{equation}   \label{eq:orbitalenergy}
    E_{i} = - \frac{G \, m_{\mathrm{total},i}\, m_{\mathrm{total},j}} {2 \,a_{i}}\,,
\end{equation} 
with $ m_{\mathrm{total},k} = \left(m_{k} + m_{\mathrm{sp},k}(r)\right)$. 
The final orbital energy of the PBH binary after the complete evaporation of DM spikes is:
\begin{equation}   \label{eq:orbitalenergyfinal}
    E_{f} = - \frac{G  m_{i} m_{j}} {2 a_{f}}\,.
\end{equation} 
Then, using the conservation of energy, the final semi-major axis $a_{f}$ of the binary after the DM spikes are totally evaporated~\cite{Kavanagh:2018ggo} can be written as:
\begin{equation}   \label{eq:a_final}
   a_{f}(a_{i}) \equiv \beta \, a_{i} = \frac{G  m_{i}  m_{j} a_{i}} {\left[G m_{\mathrm{total},i}   m_{\mathrm{total},j} - 2  a_{i}\left(E^{\mathrm{bind}}_{\mathrm{sp},i} + E^{\mathrm{bind}}_{\mathrm{sp},j} \right)\right]}  \,.
\end{equation} 
Here, $\beta$ is the ratio of the final to initial semi-major axis of the PBH binary, taking into account the complete evaporation of the DM spikes.
   
\begin{figure}[tb!]
\centering
\includegraphics[width=0.48\textwidth]{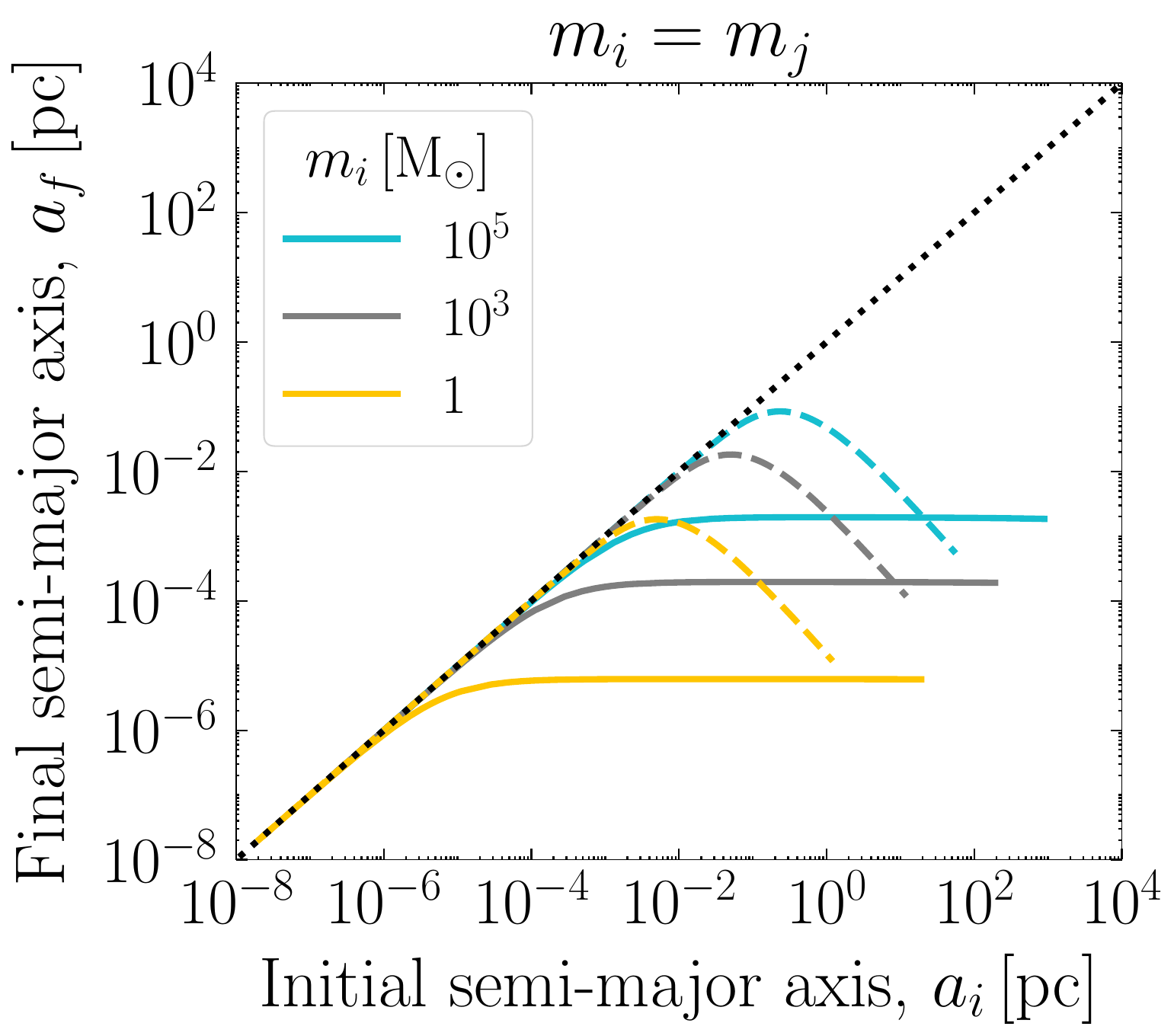}
\hspace*{\fill}
\includegraphics[width=0.48\textwidth]{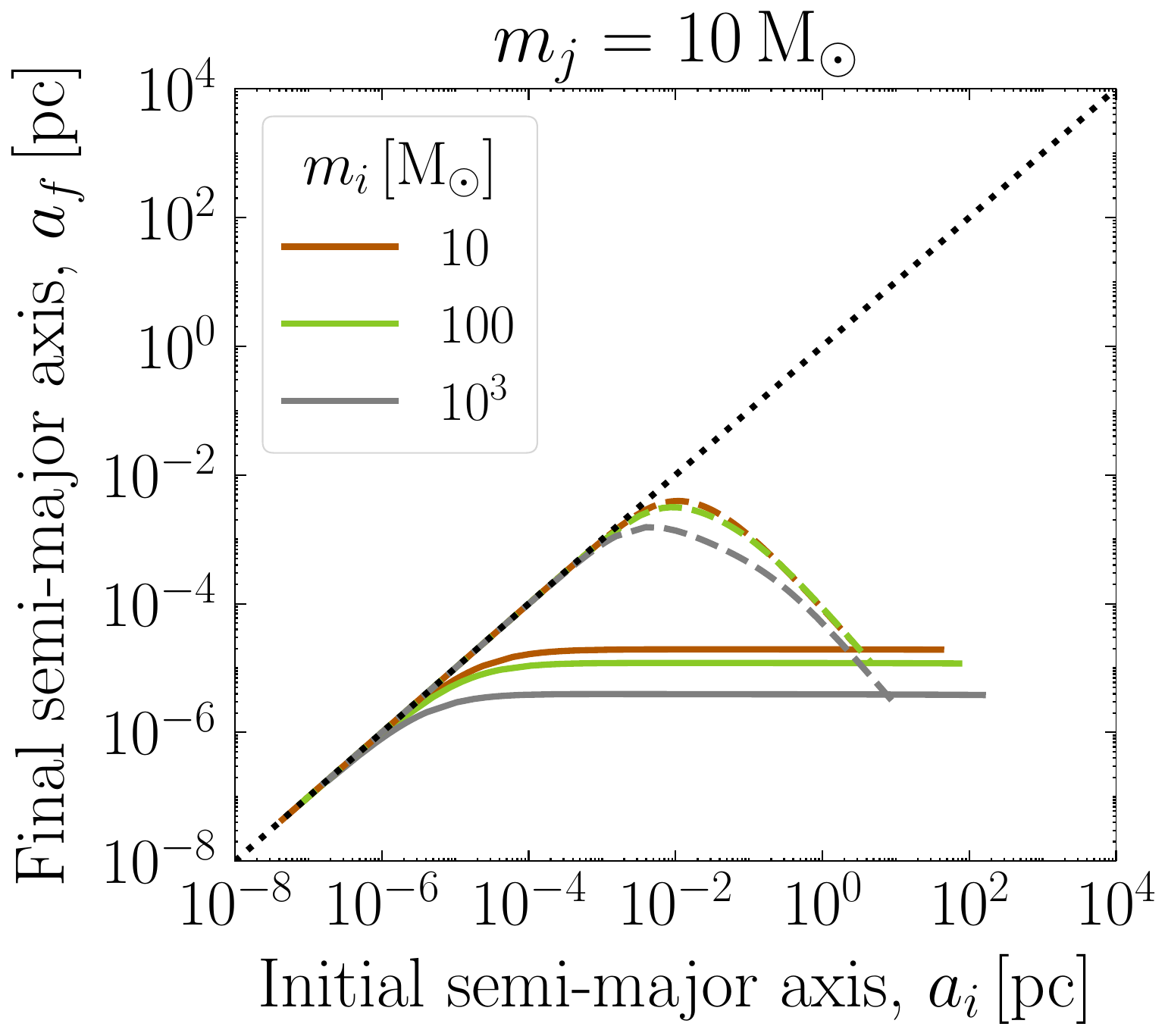}
\caption{Variation of the final semi-major axis $a_{f}$ with initial semi major axis $a_{i}$ for PBH binaries with \textbf{evaporated} DM spikes. In both panels, the black dotted line shows the case in which the final semi-major axis is equal to the initial semi-major axis i.e.~$a_{f} = a_{i}$. The colored solid lines represent results for the density profile of $\rho_{\mathrm{sp}}(r) \propto r^{-9/4}$ given by Eq.~\eqref{eq:a_final} and the colored dashed lines correspond to $\rho_{\mathrm{sp}}(r) \propto r^{-3/2}$~\cite{1985ApJS...58...39B,Kavanagh:2018ggo}.}
 \label{fig:semimajoraxis}
\end{figure}

The variation of the final semi-major axis $a_{f}$ with the initial semi-major axis $a_{i}$ for binaries of dressed PBHs with masses $m_{i}$ and $m_{j}$ is shown in Fig.~\ref{fig:semimajoraxis}. In all cases, we see that $a_f$ remains close to $a_i$ for binaries with small initial sizes. Binaries with small values of $a_i$ decouple early, meaning that the DM spike has not yet had time to grow. The effect of DM spikes comes into play more and more as the semi-major axis increases, because in that case the binary decouples later. The heavier the DM spikes, the more the energy ejected by the PBH binaries during the evaporation of the spike, which ultimately makes the final separation of the binary comparatively smaller.

In the left panel of this figure (for equal mass PBHs), we see that for a density profile of $\rho_\mathrm{sp}(r) \propto r^{-9/4}$, the final semi major axis $a_{f}$ of the binaries becomes constant at large values of the initial semi major axis $a_{i}$.  This is because the binding energy of DM spikes tends to a constant value with the growth of the DM spikes.
In contrast, the final semi-major axis $a_{f}$ decreases with increase in the initial semi major axis $a_{i}$ for a DM density profile of $\rho_\mathrm{sp}(r) \propto r^{-3/2}$~\cite{Kavanagh:2018ggo}, because the binding energies of the DM spikes keep increasing for this shallower density profile as the initial separation of the binaries increases. In the right panel (for un-equal mass PBHs), we see that as the primary PBH mass $m_{i}$ increases, the mass of its DM spike increases which leads to an increase its binding energy. This enhances the loss of energy from the binary during the evaporation of the DM spike, shrinking the binary more in comparison to other binaries with smaller value of $m_{i}$. Also, this decrease in the final separation of binaries is much larger for the DM density profile of $\rho_\mathrm{sp}(r) \propto r^{-9/4}$ compared to $r^{-3/2}$. 

Now, the merger time of highly eccentric ($e \sim 1$) isolated binaries emitting energy only via emission of gravitational waves is~\cite{peters_gravitational_1964}:
\begin{equation}  \label{eq:mergertime}
  t(a, e) =\frac{3}{85}\left[ \frac{a^{4}\, c^{5}\left(1 - e^{2}\right)^{7/2}}{G^{3} m_{i}m_{j}\left(m_{i} + m_{j}\right)} \right]\,.
\end{equation}
Then, substituting the value of the final angular momentum  given by Eq.~\eqref{eq:finalangularmomentum} in the above equation, the final merger time of the PBH binaries with \textbf{evaporated} DM spikes  can be written as:
\begin{equation}  \label{eq:finalmergertime}
  t_{f} =  \sqrt{\frac{a_{f}}{a_{i}}} \, t_{i} \,,
\end{equation}
with $t_{i}$ being the initial merger time predicted by Eq.~\eqref{eq:mergertime} right after the formation of PBH binary. From the left panel of Fig.~\ref{fig:semimajoraxis}, it is evident that the final merger time $t_{f}$ of PBH binaries with \textbf{evaporated} DM spikes is smaller than their initially predicted merger time $t_{i}$. The reason for this behaviour is that while the evaporation of the DM spikes tends to circularize the binary orbit (tending to slow down the merger), it also shrinks the orbit, which tends to speed up the merger, and which dominates over the circularization effect. Hence, PBH binaries with \textbf{evaporated} DM spikes merge faster than the PBH binaries without DM spikes.

\subsection{Static DM spikes}
\label{subsec:merger time Static DM spikes}
To find out the merger time of the PBH binaries with \textbf{static} DM spikes, we consider PBH binaries merging analogous to IMRI systems using the same formalism as that of Ref.~\cite{Becker:2021ivq}. These PBH binaries lose energy via the emission of gravitational waves (GW) and dynamical friction (DF) over timescales which are much larger than the period of the binary orbit. We assume that the path of the secondary PBH around the central PBH can be considered as a classical Keplerian orbit. During the evolution of this orbit the secondary PBH loses energy and angular momentum leading to merger of the binary. 
 The total orbital energy $E$ is given by:
\begin{equation}  \label{eq:orbitalenergystatic}
 E = \frac{ Gm_{i}m_{j}}{2a}\,,
\end{equation}
with angular momentum $L$ of the PBHs given by Eq.~\eqref{eq:angularmomentumspbhs}. Here, we have assumed that the mass of the DM spikes is much less than the reduced mass of the binary. Then, similar to Ref.~\cite{Becker:2021ivq}, using these values of the orbital energy $E$ and angular momentum $L$, the equation of evolution for the semi-major axis $a$ and  eccentricity $e$ of the binary orbit can be respectively written as:
\begin{equation} \label{eq:evolutionofa}
\frac{\mathrm{d} a}{\mathrm{d} t} = \frac{\mathrm{d} E}{\mathrm{d} t}/\frac{\partial E}{\partial a} \,,
\end{equation}
\begin{equation} \label{eq:evolutionofe}
\frac{\mathrm{d} e}{\mathrm{d} t} = -\left(\frac{1-e^{2}}{2\,e}\right) \left(\frac{1}{E}\frac{\mathrm{d} E}{\mathrm{d} t} + \frac{2}{L}\frac{\mathrm{d} L}{\mathrm{d} t} \right)\,.
\end{equation}
 Here, $\mathrm{d}E/\mathrm{d}t$ signifies the sum of the loss of energy due to the emission of gravitational waves $\mathrm{d}E_\mathrm{GW}/\mathrm{d}t$ and due to dynamical friction $\mathrm{d}E_\mathrm{DF}/\mathrm{d} t$ which are further discussed in Appendix~\ref{sec:scaling}. 

\begin{figure}[tb!]
\centering
\includegraphics[width=0.6\textwidth]{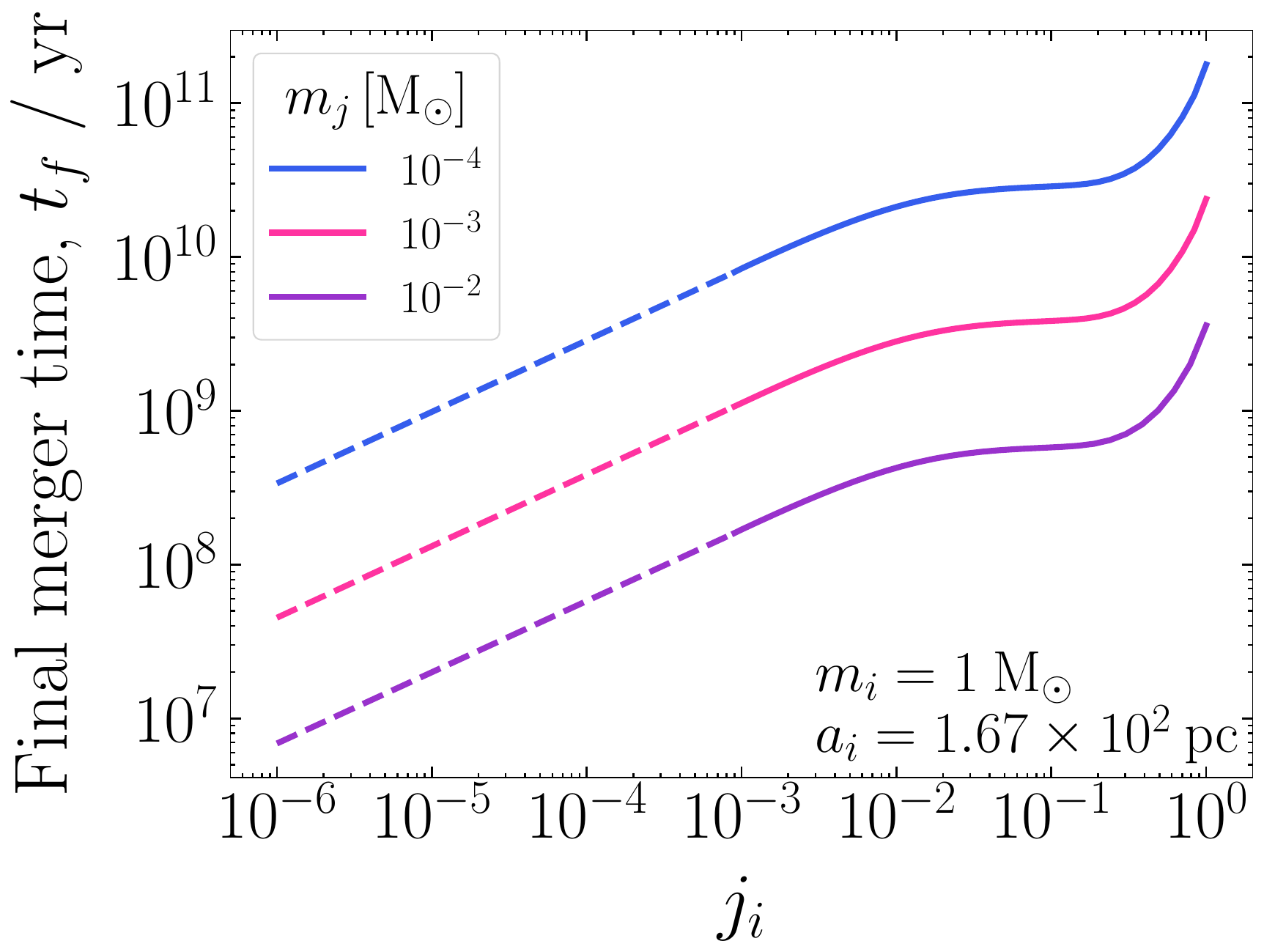}
\caption{Variation of the final merger time $t_{f}$ with initial angular momentum $j_{i}$ for different PBH binaries having \textbf{static} DM spikes. Here, $m_{i}$ is the central/primary PBH and $m_{j}$ is the secondary PBH which spirals down in to $m_{i}$. The solid part of the curves shows the final merger time $t_{f}$ of these binaries obtained numerically from code \href{https://github.com/DMGW-Goethe/imripy}{imripy}~\cite{Becker:2021ivq} and the dashed part shows the power law extrapolation of $t_f$ for $j_i < 10^{-3}$ with power index equal to $0.46$.}
 \label{fig:tfvsji}
\end{figure}

Now, we define a reference PBH binary with \textbf{static} DM spikes having $m_{i,\mathrm{ref}} = 1 \, \mathrm{M_{\odot}} $, $m_{j,\mathrm{ref}} = 10^{-3} \, \mathrm{M_{\odot}}$ and $a_{i,\mathrm{ref}} = 1.67 \times 10^{2} \, \mathrm{pc}$. This value of $a_{i,\mathrm{ref}}$ is calculated using Eq.~\eqref{eq:initialsemimajoraxishalos} assuming that the comoving separation $x$ of the PBHs is equal to its average value i.e. $x = \bar{x}$.
Then, the final merger time of PBH binaries with \textbf{static} DM spikes calculated by solving Eqs.~\eqref{eq:evolutionofa} and~\eqref{eq:evolutionofe} using the \texttt{imripy} code\footnote{\url{https://github.com/DMGW-Goethe/imripy}}~\cite{Becker:2021ivq} is shown in Fig.~\ref{fig:tfvsji}. In this figure, the value of the final merger time $t_{f}$ is power-law extrapolated for $j_{i} < 10^{-3}$ (i.e.~$t_f \propto j_i{}^\nu$) with power-law index $\nu = 0.46$. For a small range of $j_{i}$, the merger time is approximately constant shown by the flattened part of the curves in Fig.~\ref{fig:tfvsji}.

\begin{figure}[tb!]
\centering
\includegraphics[width=0.49\linewidth]{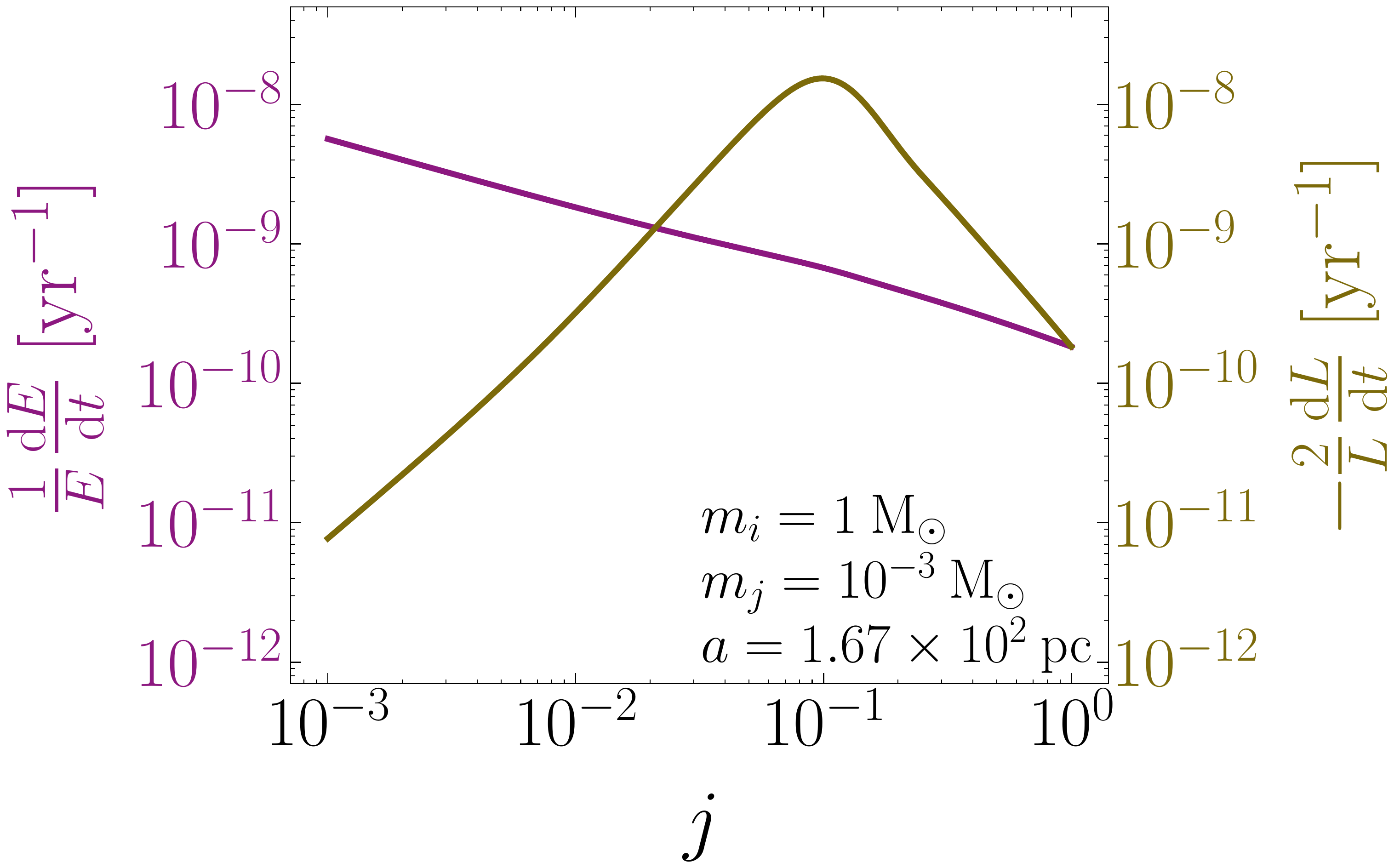} 
\hspace*{\fill}
\includegraphics[width=0.49\linewidth]{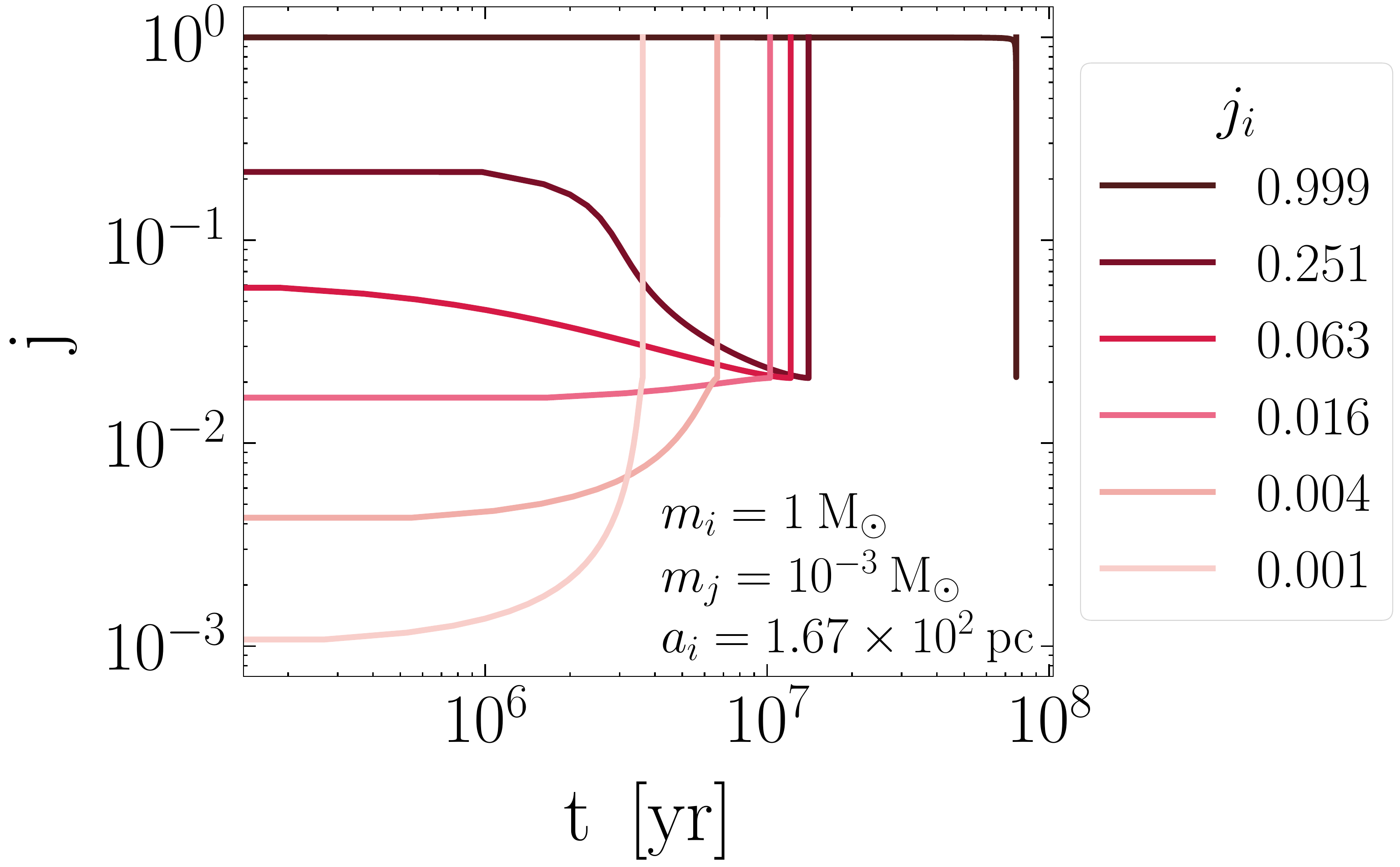}
\caption{The left panel shows the variation of terms $\left(1/E\right)\mathrm{d}E/\mathrm{d}t$ (dark magenta line) and $\left(1/L\right)\mathrm{d}L/\mathrm{d}t$ (olive colored line) with angular momentum $j$, mentioned in the equation of evolution of the eccentricity $e$ given by Eq.~\eqref{eq:evolutionofe}, for a PBH binary with \textbf{static} DM spikes. The values of the total energy $E$ and angular momentum $L$ of the binary orbit is given by Eq.~\eqref{eq:orbitalenergystatic} and Eq.~\eqref{eq:angularmomentumspbhs} respectively. We checked that for the PBH binary mentioned in the left panel, the loss of orbital energy due to gravitational waves $\mathrm{d}E_\mathrm{GW}/\mathrm{d}t$ is $\approx \mathcal{O}(28)$ smaller than the loss of energy due to dynamical friction $\mathrm{d}E_\mathrm{DF}/\mathrm{d}t$. The right panel shows the variation of angular momentum $j$ of the same PBH binary as a function of time $t$ for different values of the initial angular momentum $j_{i}$.}
\label{fig:staticDMbinary}
\end{figure}

This behaviour of PBH binaries with \textbf{static} DM spikes shown in Fig.~\ref{fig:tfvsji} can be explained on the basis of Fig.~\ref{fig:staticDMbinary}. Here, the terms $\left(1/E\right)\mathrm{d}E/\mathrm{d}t$ and $\left(1/L\right)\mathrm{d}L/\mathrm{d}t$, which are opposite in sign, contribute to the evolution of the eccentricity given by Eq.~\eqref{eq:evolutionofe} for PBH binaries with \textbf{static} DM spikes. The rate of energy loss is largest for very eccentric binaries ($j \ll 1$). These therefore merge the fastest, as shown in the right panel of Fig.~\ref{fig:staticDMbinary}. However, in the left panel of Fig.~\ref{fig:staticDMbinary}, we see that for very small values of initial angular momentum $j_{i}$, the orbital energy term dominates over the angular momentum term. This dominance as per Eq.~\eqref{eq:evolutionofe} leads to decrease the eccentricity of the binary with time. This circularization of initially highly eccentric orbits tends to slow their merger compared to the case with a fixed eccentricity. Thus, the merger time grows only slowly with $j$. In the left panel of this figure, we see that for $j_{i} \gtrsim 0.02$, the angular momentum term dominates over the contribution of orbital energy in Eq.~\eqref{eq:evolutionofe} due to which the eccentricity of the PBH binaries increases with time. In general, the merger time of PBH binaries should increase with increase in $j_{i}$. However, due to this eccentrification, the increase in the merger time of PBH binaries is smaller than expected. This can be seen in the right panel of Fig.~\ref{fig:staticDMbinary} and also explains the flattening of the curves in Fig.~\ref{fig:tfvsji} close to $j_{i} \sim 0.02$. Finally, the merger time of PBH binaries increases quickly as $j_{i} \rightarrow 1$, as the process of eccentrification becomes less and less efficient (for $j_i \sim 1$, this eccentrification occurs only very close to the merger, as seen in the top most curve in the right panel of Fig.~\ref{fig:staticDMbinary}).

By studying the variation of the final merger time $t_{f}$ with initial angular momentum $j_{i}$ shown in Fig.~\ref{fig:tfvsji}, and using the parameters of the reference binary, we find the following approximate expression for final merger time $t_{f}$ of PBH binaries with \textbf{static} DM spikes:
\begin{equation} \label{eq:t_final_static_halos}
    t_{f} =  f(j_{i})\cdot C_\mathrm{ref}(a_{i}, m_{i}, m_{j})\,,
\end{equation}
where
\begin{equation} \label{eq:C_ref}
    C_\mathrm{ref}(a_{i}, m_{i}, m_{j}) = \left(\frac{a_{i}}{a_{i, \mathrm{ref}}}\right)^{\alpha} \left(\frac{m_{i}}{m_{i,\mathrm{ref}}}\right)^{\gamma} \left(\frac{m_{j}}{m_{j,\mathrm{ref}}}\right)^{\delta}\,,
\end{equation}
\begin{equation} \label{eq:scalingpowers}
\alpha = 0.75, \quad \gamma = 0.65, \quad \delta = -0.89\,.
\end{equation}
The function $f(j_{i})$ is an interpolation function which gives the value of $j_{i}$ corresponding to a given value of $t_{f}$, following the shape of the curves in Fig.~\ref{fig:tfvsji}. Here, $C_\mathrm{ref}(a_{i}, m_{i}, m_{j})$ is a constant factor which normalizes the final merger time of PBH binaries with \textbf{static} DM spikes with respect to the reference binary defined earlier.  
In Appendix~\ref{sec:scaling}, we provide analytic arguments for the scaling of the merger time with the component masses and initial orbital elements of the PBH binaries with \textbf{static} DM spikes. These scalings derive from the fact that in most of the PBH binaries we are interested in, energy losses due to dynamical friction dominate over gravitational wave emission. We note that for PBH binaries with \textbf{static} DM spikes having very small values of initial semi-major axis and initial angular angular momentum, the loss of orbital energy is dominated by the emission of gravitational waves. So, in such cases, we make the assumption that the merger time of PBH binaries with \textbf{static} DM spikes is the same as that of PBH binaries without DM spikes.

\subsection{Impact of DM spikes on merger time}
\label{subsec:Impact of DM spikes on merger time}
\begin{figure}[tb!] 
\centering
\begin{minipage}[c]{0.48\linewidth}
\centering
\includegraphics[width=\linewidth]{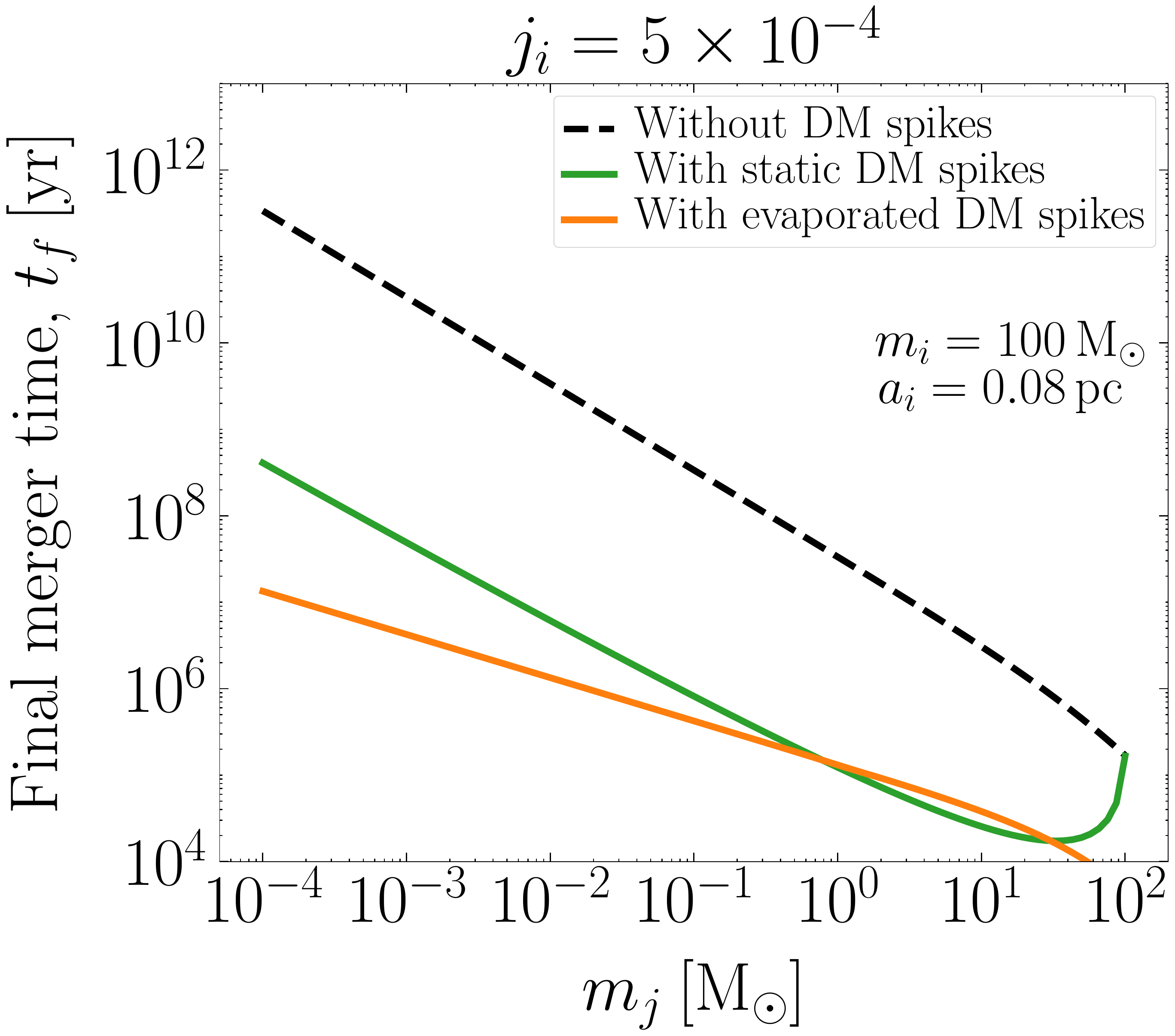}
\end{minipage}
\hfill
\begin{minipage}[c]{0.48\linewidth}
\centering
\includegraphics[width=\linewidth]{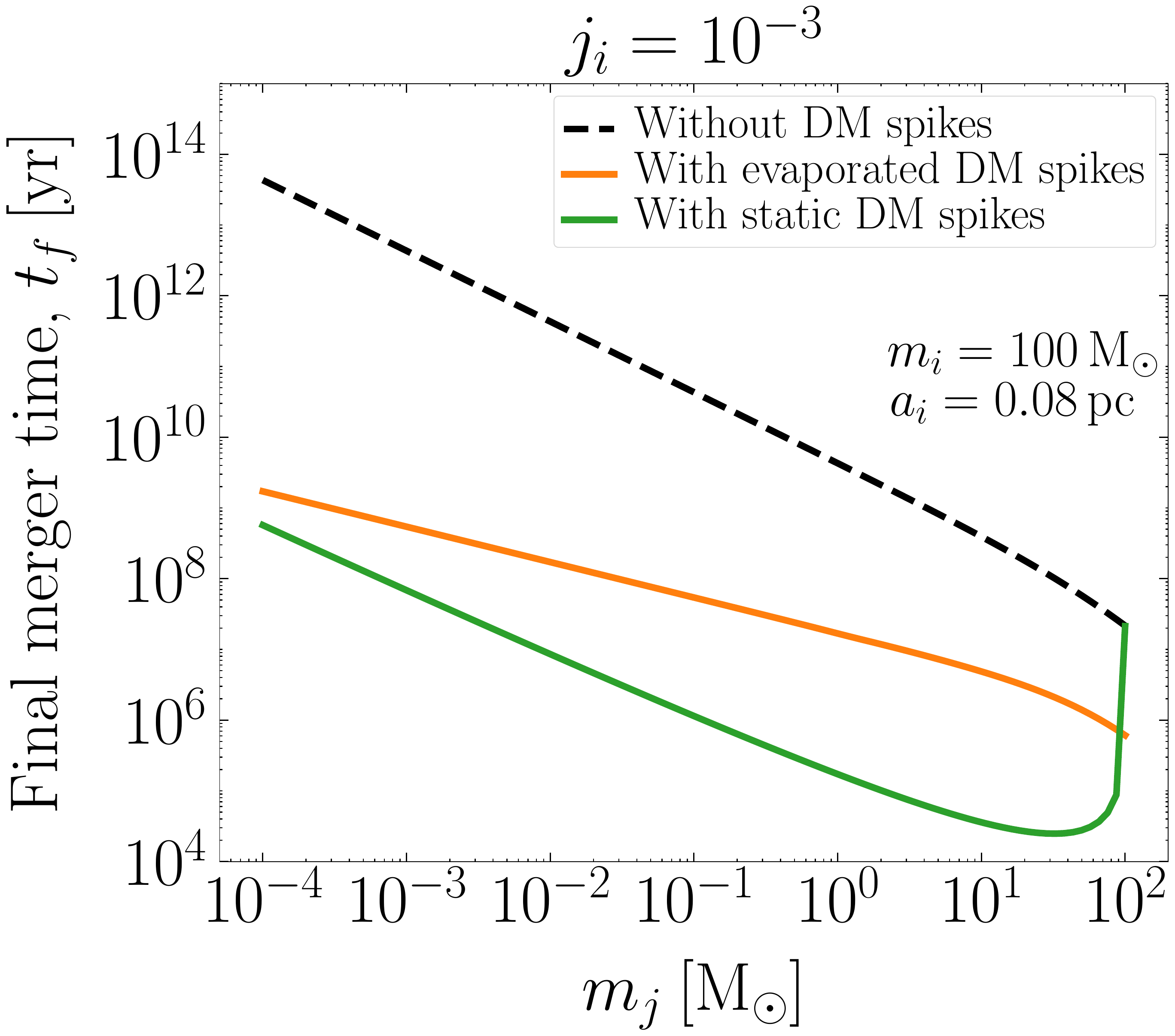}
\end{minipage}
\caption{Variation of final merger time $t_{f}$ as a function of the PBH mass $m_{j}$ for PBH binaries with \textbf{evaporated} DM spikes and \textbf{static} DM spikes. The left and right panels show different values of the initial angular momentum $j_i$ of the binary. Here, we have chosen the values of initial semi-major axis as $a_{i} = 0.08 \,\mathrm{pc}$.}
\label{fig:finalmergertime}
\end{figure}

Now, the variation of final merger time $t_{f}$ of PBH binaries without DM spikes (Eq.~\eqref{eq:mergertime}), with \textbf{evaporated} DM spikes  (Eq.~\eqref{eq:finalmergertime}) and with \textbf{static} DM spikes (Eq.~\eqref{eq:t_final_static_halos}) is shown in Fig.~\ref{fig:finalmergertime}. In both panels of the figure, we see that the PBH binaries without DM spikes take longest time to merge. The merger time of PBH binaries without DM spikes decreases with increase in $m_{j}$ as $t_{f} \approx 1/m_{j}$ for $m_{j} \ll m_{i}$. Comparing the left and right panels, we see that the merger time of PBH binaries without DM spikes increases as per Eq.~\eqref{eq:mergertime} with increase in the initial angular momentum $j_{i}$. 

In the case of PBH binaries with \textbf{evaporated} DM spikes, the final semi-major axis decreases with decrease in $m_{j}$ i.e.\ $\beta \propto m_{j}$ in Eq.~\eqref{eq:a_final}. This implies that for $m_{j} \ll m_{i}$, the loss of energy during the evaporation of DM spikes is much larger than the initial orbital energy of the PBH binaries. Hence, most of the energy of PBH binaries with \textbf{evaporated} DM spikes lies in the form of the binding energy of DM spikes. Then, using $\beta \propto m_{j}$ in  Eq.~\eqref{eq:finalmergertime} we get $t_{f}\propto 1/\sqrt{m_{j}}$ due to which the final merger time of PBH binaries with \textbf{evaporated} DM spikes decreases with increase in $m_{j}$ (but more slowly than the case without spikes).

In the case of PBH binaries with \textbf{static} DM spikes, the final merger time decreases with increase in $m_{j}$ as $t_{f} \propto m_{j}^{-0.89}$ (see Appendix~\ref{sec:scaling}). As the value of the initial eccentricity increases then PBH binaries with \textbf{static} DM spikes merge faster which is already explained as per Fig.~\ref{fig:tfvsji} in the previous section. But, for $m_{i} = m_{j}$, the merger time of PBH binaries with \textbf{static} DM spikes rises rapidly and coincides with the merger time of PBH binaries without DM spikes. This is because for $m_{i} = m_{j}$, the value of Coulomb logarithm $\log \Lambda = \sqrt{m_{i}/m_{j}}$ in dynamical friction becomes zero. This implies that the loss of orbital energy happens via emission of gravitational waves only due to which its dynamics becomes similar to PBH binaries without DM spikes. Hence, the final merger time of equal mass PBH binaries with \textbf{static} DM spikes coincides with the merger time of equal mass PBH binaries without DM spikes.

In Fig.~\ref{fig:finalmergertime} we also see that there are scenarios in which the system with a \textbf{static} DM spike merges faster than the same system assuming an \textbf{evaporated} spike. The loss of energy which tends to accelerate the merger comes from the binding energy of the DM spike. Since all the DM particles get ejected before the merger, so the evaporation of DM spikes can be interpreted as the maximum possible amount of energy which can be extracted from the PBH binaries. Hence, we would expect that PBH binaries with \textbf{evaporated} DM spikes should have the minimum possible merger time. When systems with \textbf{static} spikes merge faster than this, it suggests that more energy is being extracted by dynamical friction than is available in the binding energy of the spike, signalling the breakdown of the \textbf{static} spike formalism described above. As we have noted already, the actual physics of PBH binaries and spikes lies somewhere in between these two  scenarios and it is the ratio of PBHs masses $q = m_{j}/m_{i}$ which plays the key role to decide which description of the dynamics is closer to the truth. In particular, from Fig.~\ref{fig:finalmergertime}, we see that the merger time for binaries with \textbf{static} spikes becomes increasingly smaller than the merger time with \textbf{evaporated} spikes, as $q \rightarrow 1$, matching our intuition that the \textbf{static} spike assumption should be poor for equal-mass binaries. Of course, in the general case, we require a more complete model to describe the dynamics of dressed PBH binaries which includes the feedback of DM spikes in the picture. We address this issue in more detail in Sec.~\ref{sec:conclusions}.

\section{Merger rates of PBH binaries}
\label{sec:Merger rates of PBH binaries}
We assume that the PBHs are uniformly distributed in the Universe and there is no other PBH lying inside the comoving volume $V = \frac{4 \pi}{3} x^{3}$ occupied by the binary with PBHs of masses $m_{i}$ and $m_{j}$.
We note that if $f_\mathrm{pbh} \approx 1$, then the binaries can get trapped in a PBH cluster. The dense environment of the PBH cluster can disrupt the binaries leading to the suppression of their current merger rates~\cite{Raidal:2018bbj,Vaskonen:2019jpv,Eroshenko:2023bbe}.\footnote{We use the term `current merger rate' to refer to the merger rate of PBH binaries at the present time $t = t_0$, with $t_0$ the age of the Universe today.} We neglect this suppression factor in our calculations by avoiding $f_\mathrm{pbh} \approx 1$ because we are focusing on subdominant PBHs. Moreover, an abundance of $f_\mathrm{pbh} \approx 1$ for Solar-mass PBHs is disfavoured by LVK estimates of BBH merger rates. The dimensionless variable $\lambda$ given by Eq.~\eqref{eq:lambda} quantifies the initial separation of the PBHs in the binary. Then, the probability distribution of $\lambda$ for a PBH binary takes the form~\cite{Chen:2018czv}:
\begin{equation}   \label{eq:Plambda1}
 P(\lambda) \equiv \frac{\mathrm{d} P}{\mathrm{d} \lambda} = \left(\frac{f_{b}\left(\Delta_{i} \Delta_{j}\right)^{1/2}}{\mu}\right)\,
e^{-\lambda\, f_{b}\left(\Delta_{i} \Delta_{j}\right)^{1/2} \,\frac{4\pi}{3} \bar{x}^{3}n_{T}}\,, \end{equation}    
with $n_{T} = f\rho_{\mathrm{eq}}\int_{0}^{\infty} \frac{P(m)}{m}\,\mathrm{d}m$ as the total number density of PBHs. Here, $P(\lambda)$ is the probability of finding a pair of nearest neighbouring PBHs of mass $m_i$ and $m_j$ with a dimensionless separation $\lambda \propto x^{3}$, with no other PBHs within the comoving distance $x$.

Now, we assume that the combined probability distribution function $P(\lambda, j_{i})$ is separable i.e.\ $ P(\lambda, j_{i})= P(\lambda)\cdot P(j_{i})$. Then, the combined probability distribution of dimensionless variable $\lambda$ and final merger time $t_{f}$ of PBH binaries is: 
\begin{equation} \label{eq:Plambdatf1}
    P(\lambda, t_{f}) = \frac{\mathrm{d^{2}} P}{\mathrm{d} \lambda\, \mathrm{d}t_{f}} =\frac{\mathrm{d} j_{i}}{\mathrm{d} t_{f}} \,\, \frac{\mathrm{d^{2}} P}{\mathrm{d} \lambda\, \mathrm{d}j_{i}} = \frac{\mathrm{d} j_{i}}{\mathrm{d} t_{f}} \,\,P(\lambda, j_{i}) = \frac{\mathrm{d} j_{i}}{\mathrm{d} t_{f}} \,\,P(\lambda)\cdot P(j_{i})\,.
\end{equation}
Substituting the value of $P(j_{i})$ given by Eq.~\eqref{eq:jipdf} in the above equation, the probability distribution of the final merger time $t_f$ of PBH binaries at a given value of $\lambda$ becomes:
\begin{equation} \label{eq:Ptflambda}
    P(t_f|\lambda) = \left( \frac{\mathrm{d} j_{i}}{\mathrm{d} t_{f}} \right) \cdot \,   \frac{1}{j_{i}} \:  \frac{\gamma_{\lambda}^{2}}{\left(1+\gamma_{\lambda}^{2}\right)^{3/2}}\,,
\end{equation}
with 
\begin{equation} \label{eq:gammaX}
   \gamma_{\lambda} = \frac{j_{i}(t_i; \lambda)}{j_{\lambda}}\,.
\end{equation}
Then, using Eqs.~\eqref{eq:Plambda1} and \ref{eq:Ptflambda}, the combined probability distribution of $(\lambda,t_{f})$ can be written as:
\begin{equation} \label{eq:Plambdatf2}
    P(\lambda,t_{f})= \left( \frac{\mathrm{d} j_{i}}{\mathrm{d} t_{f}}\right) \cdot \, \left(\frac{f_{b}\left(\Delta_{i} \Delta_{j}\right)^{1/2}}{\mu\,j_{i}}\right) \: \frac{\gamma_{\lambda}^{2}}{\left(1+\gamma_{\lambda}^{2}\right)^{3/2}} \quad e^{-\lambda\, f_{b}\left(\Delta_{i} \Delta_{j}\right)^{1/2} \,\frac{4\pi}{3} \bar{x}^{3}n_{T}}\,.
\end{equation}
Using Eq.~\eqref{eq:Plambdatf1}, the distribution of the final merger time can be calculated as:
\begin{equation} \label{eq:Ptf}
  P(t_{f}) \equiv  \frac{\mathrm{d} P}{\mathrm{d} t_{f}} = \int_{0}^{\infty}  \mathrm{d}\lambda \,\, P(\lambda, t_{f}) \,.
\end{equation}
Then, similar to Ref.~\cite{Chen:2018czv}, for PBH binaries having PBHs of masses $m_{i}$ and $m_{j}$, the merger rate at time $t_{f}$ is equal to the number of mergers $N_\mathrm{merger}$ per unit merger time $t_{f}$ per unit comoving volume $V_{m}$, written as: 
\begin{equation}  \label{eq:mergerrate}
    \mathcal{R}_{ij}(t_{f}) = \frac{\mathrm{d} N_{\mathrm{merger}}}{\mathrm{d} t_{f} \, \mathrm{d}V_{m}} = \rho_{\mathrm{m}}^{0} \cdot \mathrm{min}\left(\frac{f_{i}\Delta_{i}}{m_{i}}, \frac{f_{j}\Delta_{j}}{m_{j}}\right) \frac{\mathrm{d} P}{\mathrm{d} t_{f}}\,, 
\end{equation} 
where $\rho_{m}^{0} \simeq 4\times10^{19}\, \mathrm{M_{\odot}} \,\mathrm{Gpc}^{-3}$ is the matter density of the Universe today. The final merger time used in our further calculations is equal to the age of the Universe today i.e.\ $t_{f} = 13.78 \,\mathrm{G yr}$. Here, $\mathcal{R}_{ij}$ is the merger rate of PBH binaries having extended PBH mass functions but for monochromatic PBH mass functions, $\mathcal{R}_{ij}$ can be calculated using $m_{i} = m_{j} = m$ i.e.\ $P (m_{i}) = P (m_{j}) = P (m)/2$.

\subsection{Evaporated DM spikes}
\label{sec:merger rate Ejected DM spikes}
For PBH binaries with \textbf{evaporated} DM spikes, the dependence of angular momentum on merger time is given by Eq.~\eqref{eq:mergertime} which along with Eq.~\eqref{eq:finalangularmomentum} gives:
\begin{equation}
   \left( \frac{\mathrm{d} j_{i}}{\mathrm{d} t_{f}}\right) = \frac{j_i}{ 7\, t_f} \,.
\end{equation}
Substituting the above equation in Eq.~\eqref{eq:Plambdatf2}, the probability distribution of $(\lambda,t_f)$ for binaries with \textbf{evaporated} DM spikes becomes:
\begin{equation} \label{eq:PXtfejected}
    P(\lambda,t_{f})_\mathrm{evaporated} =  \left(\frac{f_{b}\left(\Delta_{i} \Delta_{j}\right)^{1/2}}{7 \,\mu\, t_{f}}\right) \cdot  \frac{\gamma_{\lambda}^{2}}{\left(1+\gamma_{\lambda}^{2}\right)^{3/2}} \quad e^{-\lambda\, f_{b}\left(\Delta_{i} \Delta_{j}\right)^{1/2} \,\frac{4\pi}{3} \bar{x}^{3}n_{T}}\,.
\end{equation}
We note that the functional form of $P(\lambda,t_{f})$ is the same for PBH binaries with \textbf{evaporated} DM spikes and without DM spikes (though in that case the relationship between $j_i$ and $t_f$ is different). Now using Eq.~\eqref{eq:Ptf}, the probability distribution of the final merger time for PBH binaries with \textbf{evaporated} DM spikes is:
\begin{equation} \label{eq:ejecteddistributiontf}
\left(\frac{\mathrm{d} P}{\mathrm{d} t_{f}}\right)_\mathrm{evaporated} = \left(\frac{f_{b}\left(\Delta_{i} \Delta_{j}\right)^{1/2}}{7 \,\mu\, t_{f}}\right) \int_{0}^{\infty}  \mathrm{d} \lambda   \,   \frac{\gamma_{\lambda}^{2}}{\left(1+\gamma_{\lambda}^{2}\right)^{3/2}} \:\: e^{-\lambda\, f_{b}\left(\Delta_{i} \Delta_{j}\right)^{1/2} \,\frac{4\pi}{3} \bar{x}^{3}n_{T}} \,.
\end{equation}
Substituting the value of $\mu$ given by Eq.~\eqref{eq:mu} and probability distribution of final merger time given by Eq.~\eqref{eq:ejecteddistributiontf} in Eq.~\eqref{eq:mergerrate}, the merger rate of PBH binaries with \textbf{evaporated} DM spikes at time $t_{f}$ is:
\begin{equation} \label{eq:ejectedmergerrate}
\begin{aligned}
\mathcal{R}_{ij,\, \mathrm{evaporated}} = \left(\frac{1}{7 \:
t_{f}}\right) \rho_{m}^{0}\: f^{2} \, \left(m_{i} + m_{j}\right) \,  \left(\frac{P(m_{i})\Delta_{i}}{m_{i}} + \frac{P(m_{j})\Delta_{j}}{m_{j}}\right)\\ \quad \quad \cdot \quad \mathrm{min}\left(\frac{P(m_{i})\Delta_{i}}{m_{i}} , \frac{P(m_{j})\Delta_{j}}{m_{j}}\right)\,
 \int_{0}^{\infty}  \mathrm{d}\lambda \: \frac{\gamma_{\lambda}^{2}}{\left(1+\gamma_{\lambda}^{2}\right)^{3/2}} \:\: e^{-\lambda\, f_{b}\left(\Delta_{i} \Delta_{j}\right)^{1/2} \,\frac{4\pi}{3} \bar{x}^{3}n_{T}}\,,
\end{aligned}
\end{equation}
with
\begin{equation}  \label{eq:gammaXspike}
\gamma_{\lambda} = \frac{\left(1+0.0861\,\lambda\right)^{-1/14}}{\left(1+0.0696\,\lambda\right)}\,\, \mathcal{B}_\mathrm{evaporated}\,\,\beta^{-1/14} \,\,m_{b}^{1/2}\,\,\lambda^{-37/21}\,.
\end{equation}
Here, 
\begin{equation}  \label{eq:Bspike}
\mathcal{B}_\mathrm{evaporated}  = 22.34 \left(\frac{G^{3}\rho_\mathrm{eq}^{4/3}}{c^{5}}\right)^{1/7} \left[\frac{\left(m_{i} \, m_{j}\,  t_{f}\right)^{1/7}\left(m_{i} + m_{j}\right)^{-23/42}}{\sqrt{\left(f^{2}+0.52\,\sigma_\mathrm{eq}^{2}\right)}}\right] \,,
\end{equation}
is a dimensionless constant and $c$ is the speed of light. The merger rate $\mathcal{R}_{ij,\,0}$ of PBH binaries without DM spikes~\cite{Ali-Haimoud:2017rtz,Chen:2018czv} at time $t_{f}$ can be calculated by setting the mass of DM spikes equal to zero and $\beta = 1$ in the merger rate of PBH binaries with \textbf{evaporated} DM spikes, given by Eq.~\eqref{eq:ejectedmergerrate}.

\subsection{Static DM spikes}
\label{sec:merger rate Static DM spikes}
 The probability distribution of $(\lambda,t_{f})$ for PBH binaries having \textbf{static} DM spikes can be calculated using Eq.~\eqref{eq:t_final_static_halos} in Eq.~\eqref{eq:Plambdatf2} as:
\begin{equation} \label{eq:PXtfstatic}
    P(\lambda,t_{f})_\mathrm{static} = \left( \frac{f_{b}\left(\Delta_{i} \Delta_{j}\right)^{1/2}}{j_{i} \cdot f^{\prime}(j_{i})} \right) \cdot C_\mathrm{ref} \cdot
    \frac{\gamma_{\lambda}^{2}}{\left(1+\gamma_{\lambda}^{2}\right)^{3/2}} \:\: e^{-\lambda\, f_{b}\left(\Delta_{i} \Delta_{j}\right)^{1/2} \,\frac{4\pi}{3} \bar{x}^{3}n_{T}}\,.
\end{equation}
From Eq.~\eqref{eq:Ptf}, the probability distribution of final time of merger for PBH binaries with \textbf{static} DM spikes can be written as:
\begin{equation}    \label{eq:staticdistributiontf}
\left(\frac{\mathrm{d} P}{\mathrm{d} t_{f}}\right)_\mathrm{static} = \:\: \frac{f_{b}\left(\Delta_{i} \Delta_{j}\right)^{1/2}}{\mu} \int_{0}^{\infty}  \mathrm{d}\lambda  \,   \left( \frac{1}{j_{i} \cdot f^{\prime}(j_{i})} \cdot C_\mathrm{ref}\right) \,
    \frac{\gamma_{\lambda}^{2}}{\left(1+\gamma_{\lambda}^{2}\right)^{3/2}} \: \: e^{-\lambda\, f_{b}\left(\Delta_{i} \Delta_{j}\right)^{1/2} \,\frac{4\pi}{3} \bar{x}^{3}n_{T}} \,.
\end{equation}
Again, substituting the value of $\mu$ given by Eq.~\eqref{eq:mu} and the probability distribution of the final merger time given by Eq.~\eqref{eq:staticdistributiontf} in Eq.~\eqref{eq:mergerrate}, the merger rate of PBH binaries with \textbf{static} DM spikes is given as:
\begin{equation} \label{eq:staticmergerrate}
\begin{aligned}
 \mathcal{R}_{ij,\, \mathrm{static}} = \rho_{m}^{0}\,f^{2}\left(m_{i}+m_{j}\right) \, \left(\frac{P(m_{i})\Delta_{i}}{m_{i}} + \frac{P(m_{j})\Delta_{j}}{m_{j}}\right) \, \cdot \,\,\mathrm{min}\left(\frac{P(m_{i})\Delta_{i}}{m_{j}},\, \frac{P(m_{j})\Delta_{j}}{m_{j}}\right) \,\\  \int_{0}^{\infty}  \mathrm{d} \lambda \, 
      \left( \frac{1}{j_{i} \cdot f^{\prime}(j_{i})} \cdot C_\mathrm{ref}\right) \frac{\gamma_{\lambda}^{2}}{\left(1+\gamma_{\lambda}^{2}\right)^{3/2}} \: \: e^{-\lambda\, f_{b}\left(\Delta_{i} \Delta_{j}\right)^{1/2} \,\frac{4\pi}{3} \bar{x}^{3}n_{T}} 
     \end{aligned}\,.
\end{equation}

\subsection{Component mass dependence of merger rates}
\label{subsec:Component mass dependence of merger rates}
\begin{figure}[tb!] 
\centering
\includegraphics[width=0.65\textwidth]{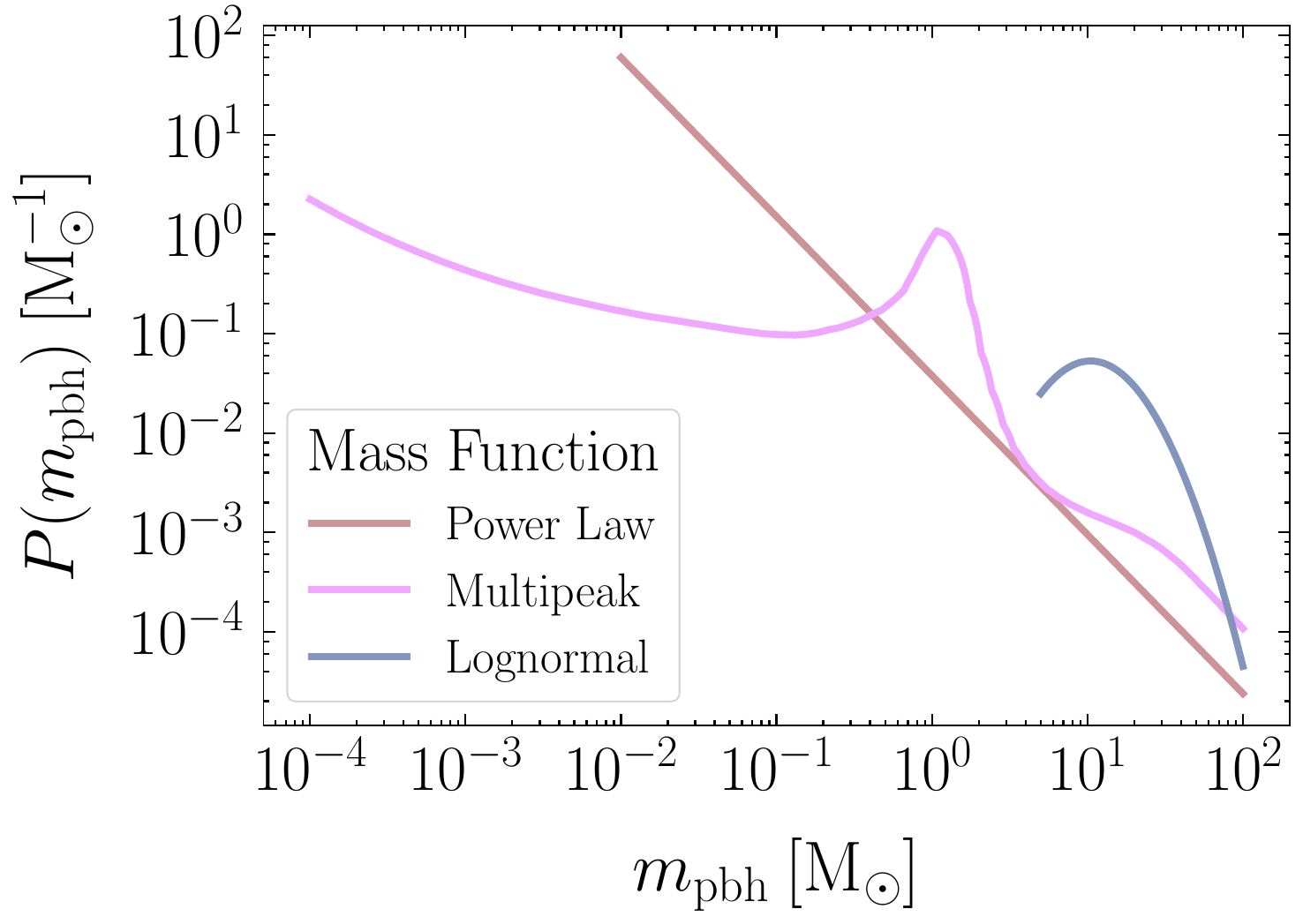}
\caption{Variation of Lognormal, Power Law and Multipeak mass functions with PBH mass $m_\mathrm{pbh}$ in mass range from $m_\mathrm{min}$ to $m_\mathrm{max}$.}
\label{fig:PBHmassPDFs}
\end{figure}

To calculate the merger rates of PBH binaries with \textbf{evaporated} and \textbf{static} DM spikes given by Eq.~\eqref{eq:ejectedmergerrate} and Eq.~\eqref{eq:staticmergerrate} respectively, we consider three specific examples of PBH mass functions (MFs):
Lognormal~\cite{Dolgov:1992pu}, Power-Law~\cite{Chen:2018czv, Carr...201....1C} and Multipeak~\cite{Cole:2022ucw} illustrated in Fig.~\ref{fig:PBHmassPDFs}. The details of these three MFs are as follows:
\begin{itemize}
\item Lognormal,  $ P(m) = \frac{1}{\sqrt{2 \pi} \sigma m} \exp\left(-\frac{\log^{2}(m/m_{c})}{2\, \sigma^{2}}\right)$ with $\sigma = 0.6, \, m_{c} = 15 \: \mathrm{M_{\odot}}$ in the mass range of $5 -  100 \: \mathrm{M_{\odot}}\,$\,.

\item Power Law, $ P(m)  = \frac{\alpha-1}{M} \left(\frac{m}{M}\right)^{-\alpha}$ with $\alpha = 1.6, \, M = 10^{-2} \: \mathrm{M_{\odot}}$ in the mass range of $10^{-2} - 100 \: \mathrm{M_{\odot}}$\,.

\item Multipeak in mass range of $10^{-4} - 100 \: \mathrm{M_{\odot}}$. This MF is extracted from a piece-wise primordial power spectrum (PPS) which is in agreement with all the current constraints on the abundance of PBHs in the Universe~\cite{Cole:2022ucw}. The variation of the threshold for PBH formation with the equation of state in the early Universe leads to the multipeak structure of this MF~\cite{Carr:2019kxo}. 
\end{itemize}
In the rest of the work, we shall denote the lower mass limit of these MFs as $m_\mathrm{min}$ and the upper mass limit as $m_\mathrm{max}$.

For the Lognormal MF, we fix $f_\mathrm{{pbh}} = 1.61 \times 10^{-3}$ which corresponds to a merger rate of $\mathcal{R} =  44 \:\, \mathrm{Gpc}^{-3} \, \mathrm{yr}^{-1}$ for PBH binaries without DM spikes and roughly matches with the merger rate observed by the LVK Collaboration. Then, we use this value of $f_\mathrm{{pbh}}$ to calculate the merger rates of PBH binaries with and without DM spikes having Lognormal MF. For the Power Law and Multipeak MFs, we use $f_\mathrm{pbh} = 2.53\times 10^{-2}$ and $f_\mathrm{pbh} = 0.066$ respectively which are in agreement with all the constraints on PBHs having extended mass functions. To simplify the structure of the paper, we shall focus more on the Power Law and Multipeak MFs in the main text and mention some of the further details about the Lognormal MF in Appendix~\ref{sec:lognormal}. In the following, we will present the merger rates as a function of the component masses $m_i$ and $m_j$. In addition, in Appendix~\ref{sec:Mass ratio Distribution of merging PBHs}, we present the same information as a function of the mass ratio $q = m_j/m_i$. This is particularly instructive for understanding whether mergers with large or small values of $q$ dominate the present day merger rate and therefore which formalism (\textbf{evaporated} or \textbf{static} spikes) might be more applicable, following the discussion in Sec.~\ref{subsec:Impact of DM spikes on merger time}.
\begin{figure}[tb!] 
\centering
\textbf{Current Merger rates without DM spikes }\par\medskip
\begin{minipage}[c]{0.48\linewidth}
\includegraphics[width=\linewidth]{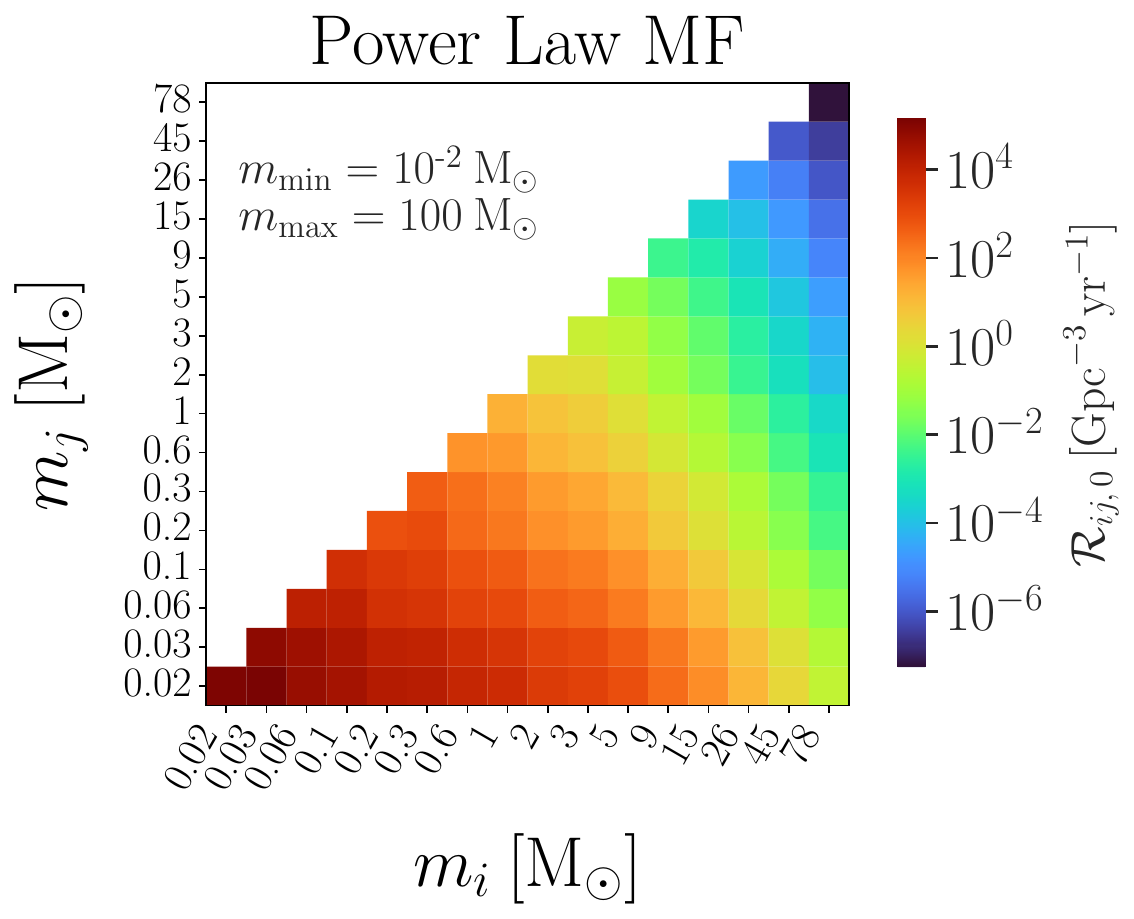}
\end{minipage}
\hspace*{\fill}
\begin{minipage}[c]{0.50\linewidth}
\includegraphics[width=\linewidth]{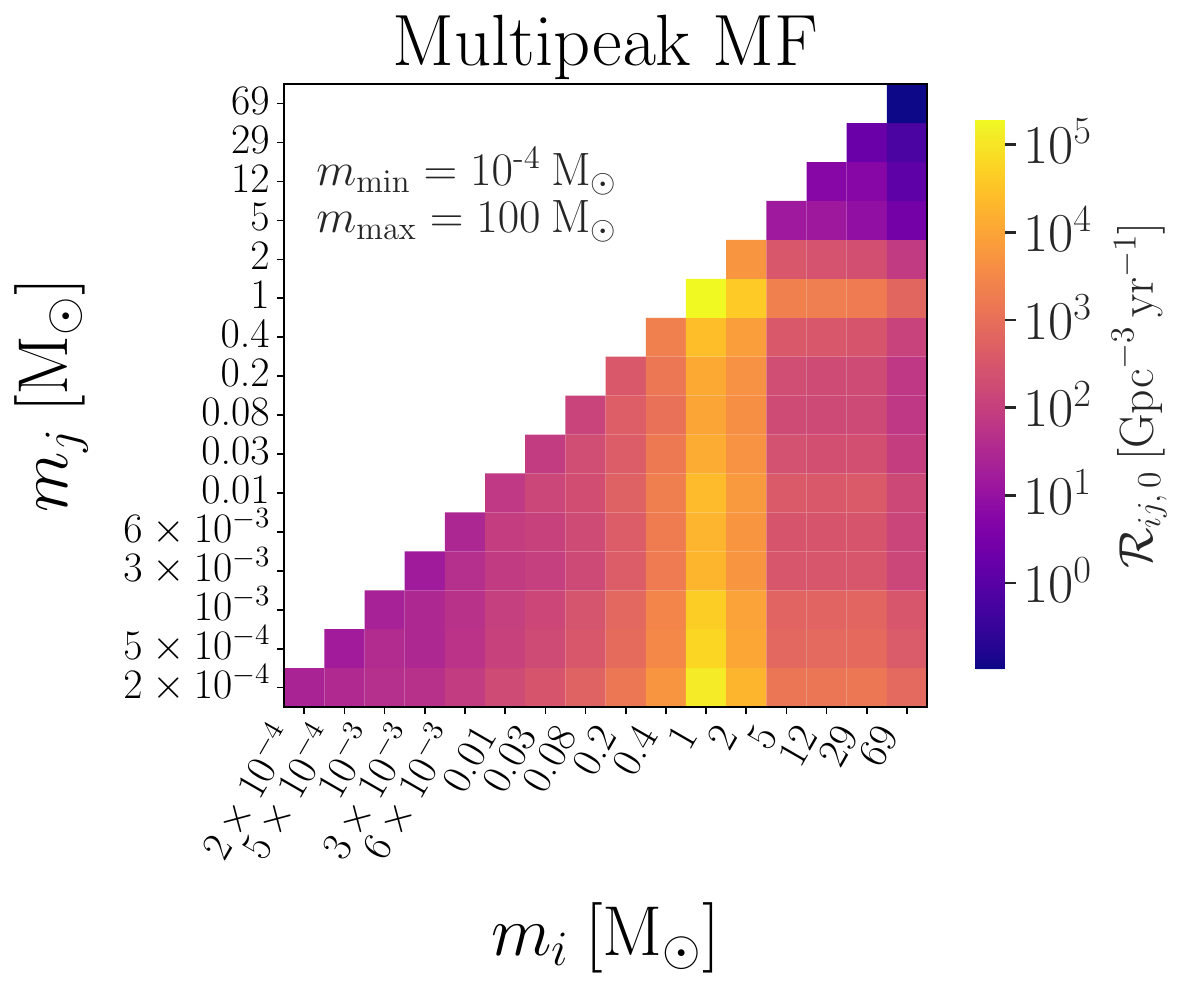}
\end{minipage}
\caption{Current merger rates of PBH binaries without DM spikes $\mathcal{R}_{ij,\,0}$ for Power Law and Multipeak MFs with PBH mass range from $m_\mathrm{{min}}$ to $m_\mathrm{{max}}$. 
}
\label{fig:mergerwithoutDMspikes}
\end{figure}

Now, the present-day merger rates of PBH binaries without DM spikes are shown in Fig.~\ref{fig:mergerwithoutDMspikes}. We see that for both the MFs, the merger rates are largest for binaries composed of the most abundance PBHs (compare with Fig.~\ref{fig:PBHmassPDFs}). If the mass fraction of PBHs $P(m)/m$ were constant with $m$, the merger rate would be expected to decrease roughly proportional to $1/m$~\cite{Kocsis:2017yty}. However, the bin widths $\Delta$ we have used in  Fig.~\ref{fig:mergerwithoutDMspikes} grow with $m$. Thus, the binary population and therefore the binned merger rates simply trace the underlying abundance of PBHs with masses $m_i$ and $m_j$, as expected from Eq.~\eqref{eq:mergerrate}. 

\begin{figure}[tb!] 
\centering
\textbf{Impact of Evaporated DM spikes on current merger rates }\par\medskip
\begin{minipage}[c]{0.48\linewidth}
\includegraphics[width=\linewidth]
{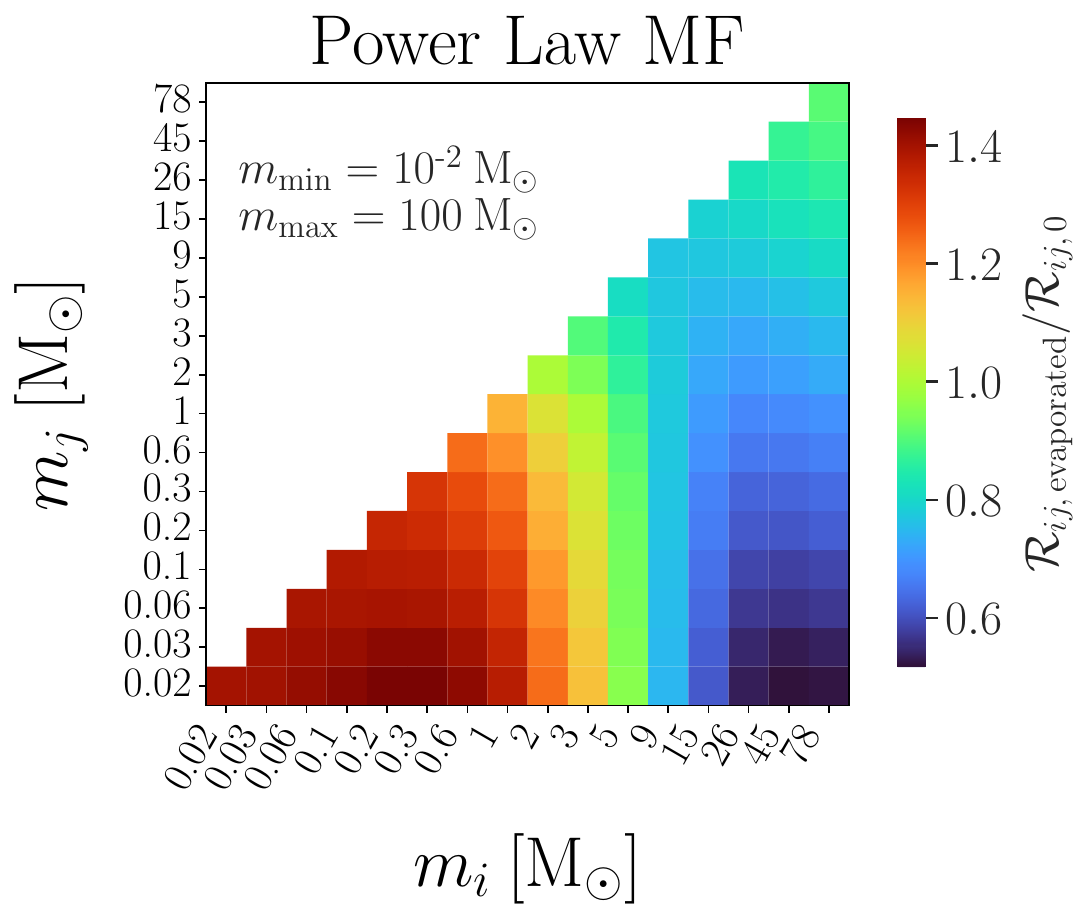}
\end{minipage}
\hspace*{\fill}
\begin{minipage}[c]{0.50\linewidth}
\includegraphics[width=\linewidth]{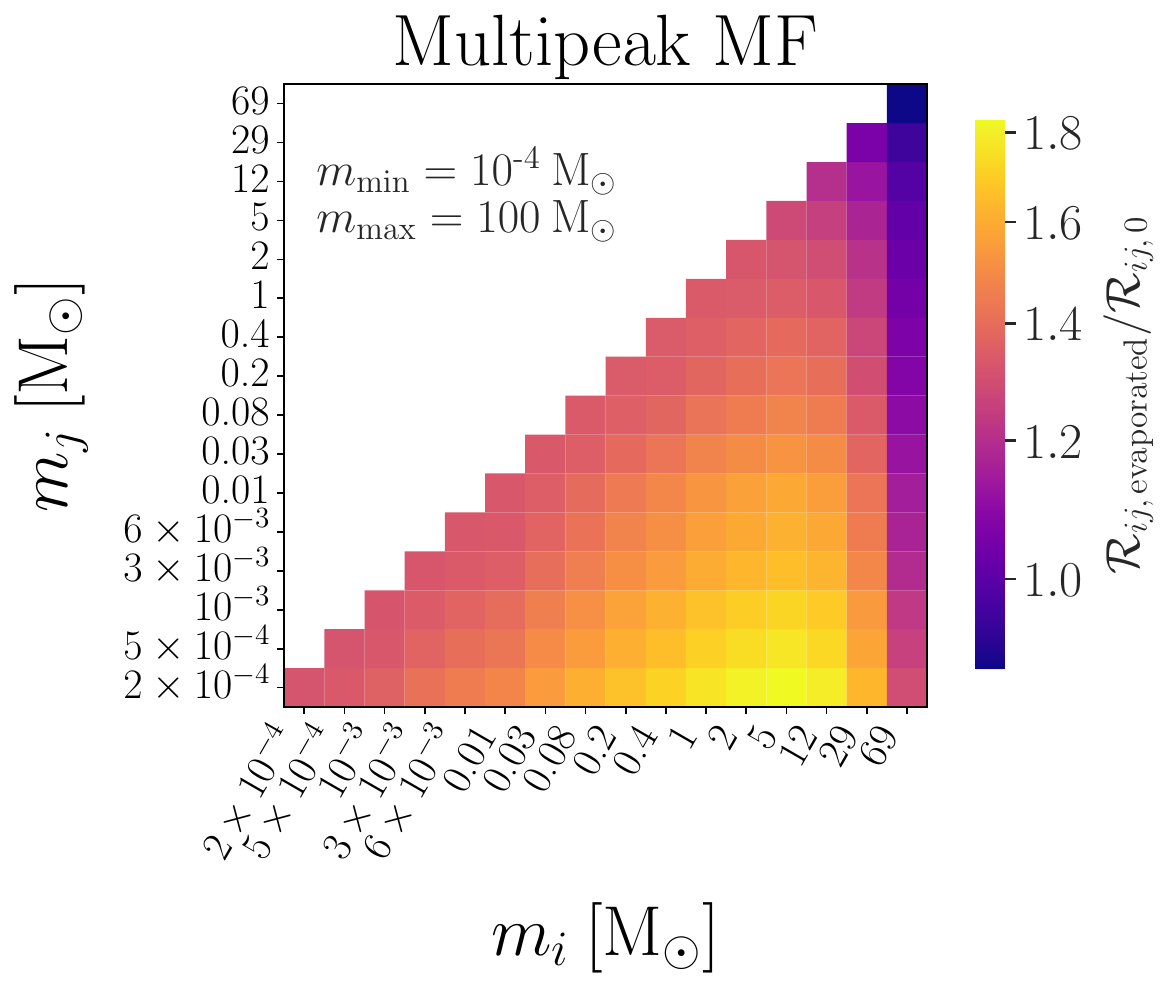}
\end{minipage}
\caption{Ratio of current merger rates of PBH binaries with \textbf{evaporated} DM spikes and without DM spikes having Power Law and Multipeak MF in mass range from $m_\mathrm{{min}}$ to $m_\mathrm{{max}}$. Here, $\mathcal{R}_{ij,\,\mathrm{evaporated}}$ represents the current merger rate of PBH binaries consisting of DM spikes which get completely ejected before the merger and $\mathcal{R}_{ij,\,0}$ signifies the current merger rate of PBH binaries without DM spikes.}
\label{fig:mergerratesDMejected}
\end{figure}

In Fig.~\ref{fig:mergerratesDMejected}, we show the ratio of the present-day merger rates of PBH binaries with \textbf{evaporated} DM spikes given by Eq.~\eqref{eq:ejectedmergerrate} and binaries without DM spikes. In this figure, we see that for our Power Law and Multipeak MF examples, the presence of \textbf{evaporated} DM spikes may enhance or reduce the present-day merger rate by an $\mathcal{O}(1)$ factor, depending on the component masses.  

\begin{figure}[tb!] 
\centering
\begin{minipage}[c]{0.49\linewidth}
\includegraphics[width=\textwidth]{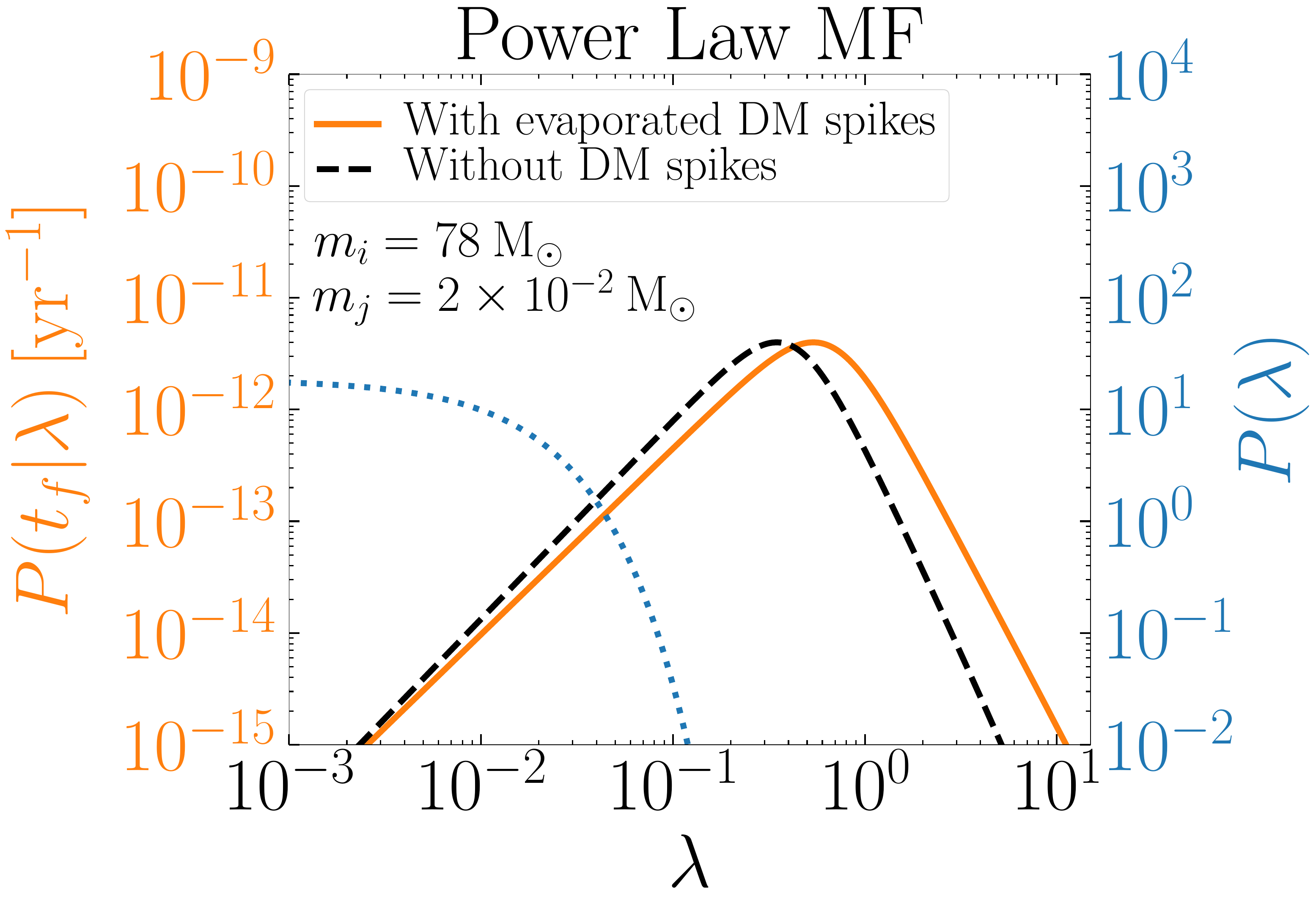}
\end{minipage}
\hspace*{\fill}
\begin{minipage}[c]{0.49\linewidth}
\includegraphics[width=\textwidth]{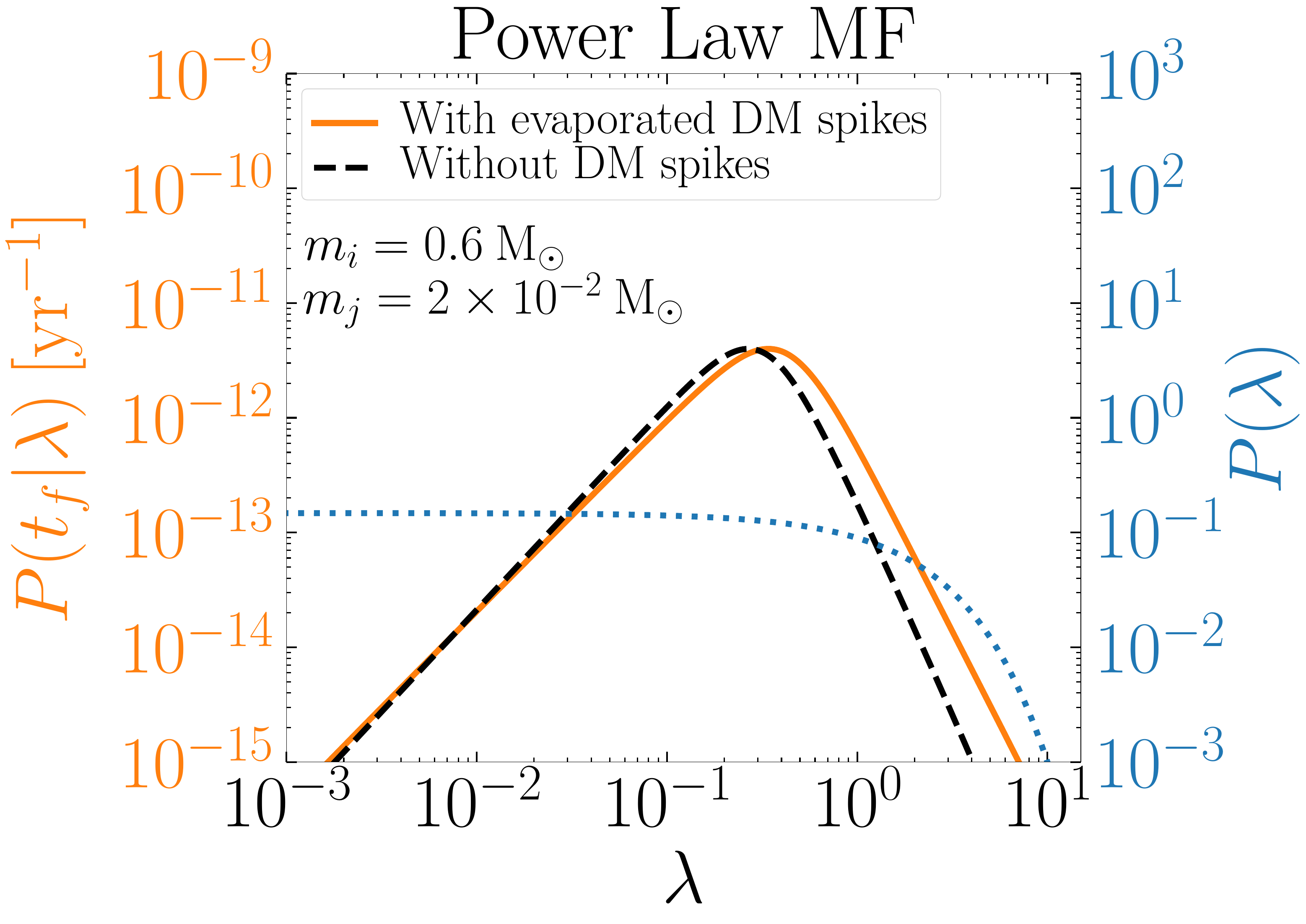}
\end{minipage}
\caption{Variation of $P(\lambda)$ and $P(t_f|\lambda)$ with $\lambda$ for PBH binaries with \textbf{evaporated} DM spikes and without DM spikes. The merger rate is proportional to the integral over the product of these distribution: $\mathcal{R} \propto \int P(t_\mathrm{f}|\lambda) P(\lambda)\,\mathrm{d}\lambda$. For illustration, we assume a Power Law MF, with different component masses in the left and right panels. In the left panel, the merger rate is decreased by the addition of DM spikes, while in the right panel the merger rate is increased.} 
\label{fig:pdfoftf}
\end{figure}

This behaviour can be explained on the basis of how the initial properties of PBH binaries merging today is affected by the presence of the DM spike. In Fig.~\ref{fig:pdfoftf}, we show the probability distribution of the final merger time $t_f$ as a function of the dimensionless PBH separation $\lambda$, i.e.\ $P(t_f|\lambda)$ given by Eq.~\eqref{eq:Ptflambda}, for two different choices of component PBH masses. The black dashed line shows the case without DM spikes. The peak of this distribution represents the value of $\lambda$ for binaries which are most likely to merge today, which we denote $\lambda_\mathrm{peak}$. Both panels of Fig.~\ref{fig:pdfoftf} show that in comparison to the PBH binaries without DM spikes, the value of $\lambda_\mathrm{peak}$ is shifted towards higher values in the presence of \textbf{evaporated} DM spikes (shown by the orange colored curves). This can be explained as per Eq.~\eqref{eq:finalmergertime} which signifies that for fixed values of $a_i \sim \lambda^{4/3}$ and initial angular momentum $j_{i}$, the presence of \textbf{evaporated} DM spikes reduces the final merger time of the PBH binaries. Hence, one needs to go to larger values of $\lambda$ to find PBH binaries with \textbf{evaporated} DM spikes which are merging today. We also show the probability distribution for the dimensionless separation $\lambda$ of PBH binaries in Fig.~\ref{fig:pdfoftf} (blue dotted lines). In the left panel, $P(\lambda)$ decreases rapidly close to $\lambda_\mathrm{peak}$. In this case, the shift in  $\lambda_\mathrm{peak}$ to larger values  means that the binaries which are most likely to merge today are \textit{rarer} in the presence of the DM spikes. This leads to a reduction in the merger rate. Instead, in the right panel, $P(\lambda)$ remains roughly constant close to $\lambda \sim \lambda_\mathrm{peak}$. The shift to larger values of $\lambda$ leads to the availability of more binaries merging today (without much penalty from larger values of $\lambda$ being rarer). In this case, then, the merger rate is increased by the presence of DM spikes.

In order to distinguish these two possibilities, we can rewrite $P(\lambda)$, given by Eq.~\eqref{eq:Plambda1}, as:
\begin{equation} \label{eq:Plambdatyp}
    P(\lambda) = \left(\frac{f_{b}\left(\Delta_{i} \Delta_{j}\right)^{1/2}}{\mu}\right)\,
e^{-\frac{\lambda}{\lambda_\mathrm{typ}}} \,,
\end{equation}
with the typical value of $\lambda$:
\begin{equation} \label{eq:lambdatyp}
    \lambda_\mathrm{typ} = \left[\frac{4\pi}{3}\bar{x}^{3} n_{T}f_{b}\left(\Delta_{i} \Delta_{j}\right)^{1/2}\right]^{-1} \equiv \frac{V_\mathrm{avg}}{V_{ij}}\,.
\end{equation}
Here, $V_\mathrm{avg} = 1/n_{T}$ is the average comoving volume occupied by any PBH in the Universe (regardless of its mass) and $V_{ij} = \frac{4\pi}{3}\bar{x}_{ij}^{3} $ is the average comoving volume occupied by a pair of PBHs of mass $m_i$ and $m_j$. As the mass of the component PBHs increases, their number density decreases and $V_{ij}$ increases, leading to a decrease in $\lambda_\mathrm{typ}$. Then, Eq.~\eqref{eq:Plambdatyp} signifies that the formation of PBH binaries is pushed to smaller values of $\lambda < \lambda_\mathrm{typ}$ (as in the left panel of Fig.~\ref{fig:pdfoftf}). Physically, as $V_{ij}$ becomes larger, the probability that \textit{another} PBH (with mass $m_k \neq m_i, m_j$) will be in the intervening volume increases (preventing the formation of the $m_i-m_j$ binary). This in turn pushes the typical separation of binaries to smaller values. We should then compare $\lambda_\mathrm{typ}$ to $\lambda_\mathrm{peak}$ to understand whether the merger rate increases or decreases with the addition of DM spikes. The value of $\lambda_\mathrm{peak}$ can be estimated by fixing the initial angular momentum of the binary to its most likely value $j_i \sim j_\lambda \propto \lambda$, given by Eq.~\eqref{eq:tidalj_Xwithspikes}, and choosing then $\lambda = \lambda_\mathrm{peak}$ to give the desired merger time. If $\lambda_\mathrm{peak} > \lambda_\mathrm{typ}$, then $P(\lambda)$ decreases rapidly close to $\lambda_\mathrm{peak}$, as in the left panel of Fig.~\ref{fig:pdfoftf}, leading to a decrease in the present-day merger rate in the presence of DM spikes. Instead, If $\lambda_\mathrm{peak} < \lambda_\mathrm{typ}$, then $P(\lambda)$ is roughly constant close to $\lambda_\mathrm{peak}$, as in the right panel of Fig.~\ref{fig:pdfoftf}, and the merger rate increases.

\begin{figure}[tb!] 
\centering
\textbf{Impact of Static DM spikes on current merger rates }\par\medskip
\begin{minipage}[c]{0.47\linewidth}
\includegraphics[width=\linewidth]
{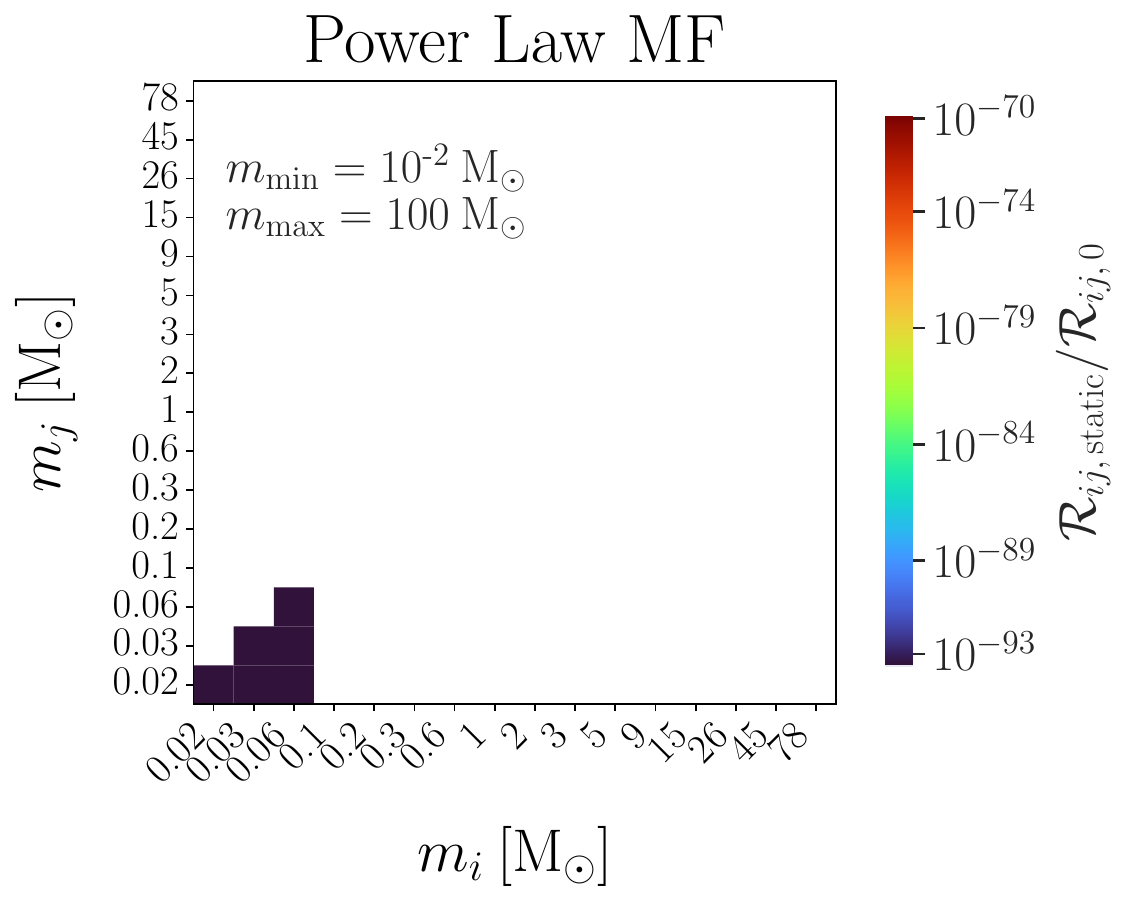}
\end{minipage}
\hspace*{\fill}
\begin{minipage}[c]{0.51\linewidth}
\includegraphics[width=\linewidth]{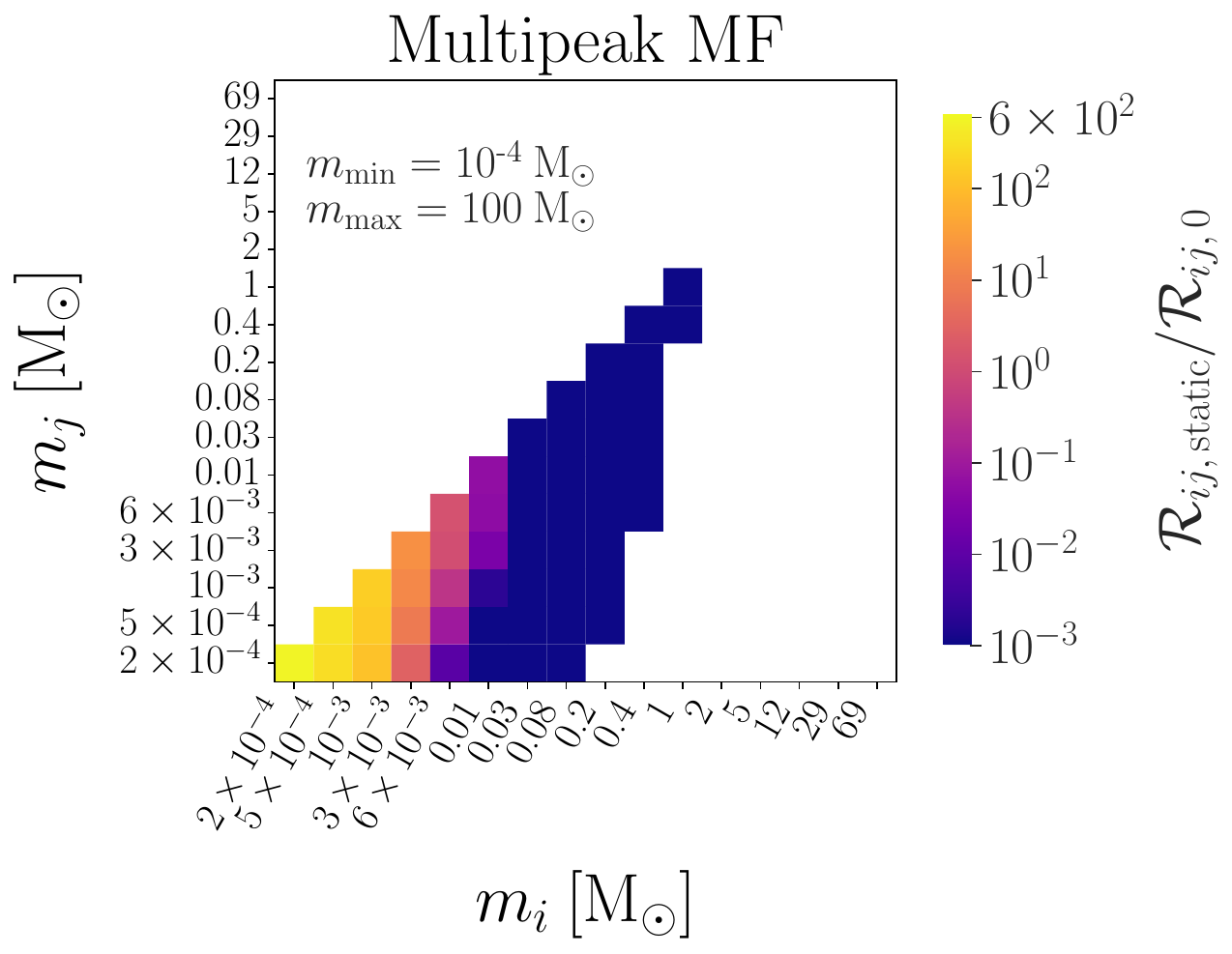}
\end{minipage}
\caption{Ratios of current merger rates of PBH binaries with \textbf{static} DM spikes and without DM spikes having Power Law and Multipeak MF in PBH mass range from $m_\mathrm{{min}}$ to $m_\mathrm{{max}}$. Here, $\mathcal{R}_{ij,\,\mathrm{static}}$ signifies the current merger rate of PBH binaries containing DM spikes which remain static until the merger and $\mathcal{R}_{ij,\,0}$ represents the current merger rate of PBH binaries without DM spikes.}
\label{fig:mergerratesDMstatic}
\end{figure}

In Fig.~\ref{fig:mergerratesDMstatic}, we show the ratios of the current merger rates of PBH binaries with \textbf{static} DM spikes, given by Eq.~\eqref{eq:staticmergerrate}, and binaries without DM spikes. From this figure, we see that the merger rates for most of the PBH binaries with \textbf{static} DM spikes are massively suppressed for Power Law MF. This behaviour can be explained as in the case of \textbf{evaporated} spikes, though the addition of \textbf{static} DM spikes typically reduced the merger time substantially more than \textbf{static} DM spikes (see e.g.~Fig.~\ref{fig:finalmergertime}). This pushes $\lambda_\mathrm{peak}$ to much higher values, meaning that binaries wide enough to be observed merging today are very rare ($\lambda_\mathrm{peak} \gg \lambda_\mathrm{typ}$). This in turn leads to a huge suppression of the merger rate. For the Multipeak MF, we note that the presence of DM spikes enhances the merger rate for PBH binaries composed of the lightest PBHs (bottom left corner). For a fixed mass fraction, very light PBHs have a much larger number density, decreasing the typical volume $V_{ij}$ in between PBHs. In addition, light PBHs are the most abundant in the Multipeak MF, meaning that the probability of finding heavier PBHs in the volume $V_{ij}$ is small. This means that the typical dimensionless separation $\lambda_\mathrm{typ}$ is large ($\lambda_\mathrm{peak} < \lambda_\mathrm{typ}$). The shift in $\lambda_\mathrm{peak}$ to larger values increases the availability of binaries merging today, which in turn can enhance the merger rates for these light, abundance PBH binaries. 

We note that if the DM density profile is shallower (for example, $\rho_\mathrm{sp}(r) \propto r^{-3/2}$) the behaviour of the PBH binaries is expected to be similar to the case with $\rho_\mathrm{sp}(r) \propto r^{-9/4}$. That is, DM spikes of density profile of $\rho_\mathrm{sp}(r) \propto r^{-3/2}$ should also lead to an increase in the merger rate of some PBH binaries and a decrease in others. The shrinking of the binaries (due to DM spikes) leads to a shift in the peak separation of binaries merging today $\lambda_\mathrm{peak}$. But because as per the right panel of Figure~\ref{fig:semimajoraxis}, the shrink in the final separation of the binary is expected to be less in the case of $\rho_\mathrm{sp}(r) \propto r^{-3/2}$, the corresponding shift in $P(t_m|\lambda)$ to larger values of $\lambda$ is expected to be smaller than for $\rho_\mathrm{sp}(r) \propto r^{-9/4}$. We thus expect the merger rate in that case to be closer to the rate without DM spikes. That is, when the presence of DM spikes is expected to reduce the merger rate (left panel of Figure~\ref{fig:pdfoftf}), we expect this reduction to be smaller for $\rho_\mathrm{sp}(r) \propto r^{-3/2}$ than for $\rho_\mathrm{sp}(r) \propto r^{-9/4}$. Similarly, when the merger rates increase (right panel of Figure~\ref{fig:pdfoftf}) we expect a smaller increase for $\rho_\mathrm{sp}(r) \propto r^{-3/2}$ in comparison with $\rho_\mathrm{sp}(r) \propto r^{-9/4}$. Ultimately, we expect that DM spikes having a density profile of $\rho_\mathrm{sp}(r) \propto r^{-9/4}$ will have a more drastic impact on the merger rates than when the density profile is shallower.

\subsection{Effect of $f_\mathrm{pbh}$ on merger rates}
\label{subsec:Effect of fpbh on merger rates}
\begin{figure}[tb!]
\centering
\begin{minipage}[c]{0.48\linewidth}
\centering
\includegraphics[width=\linewidth]{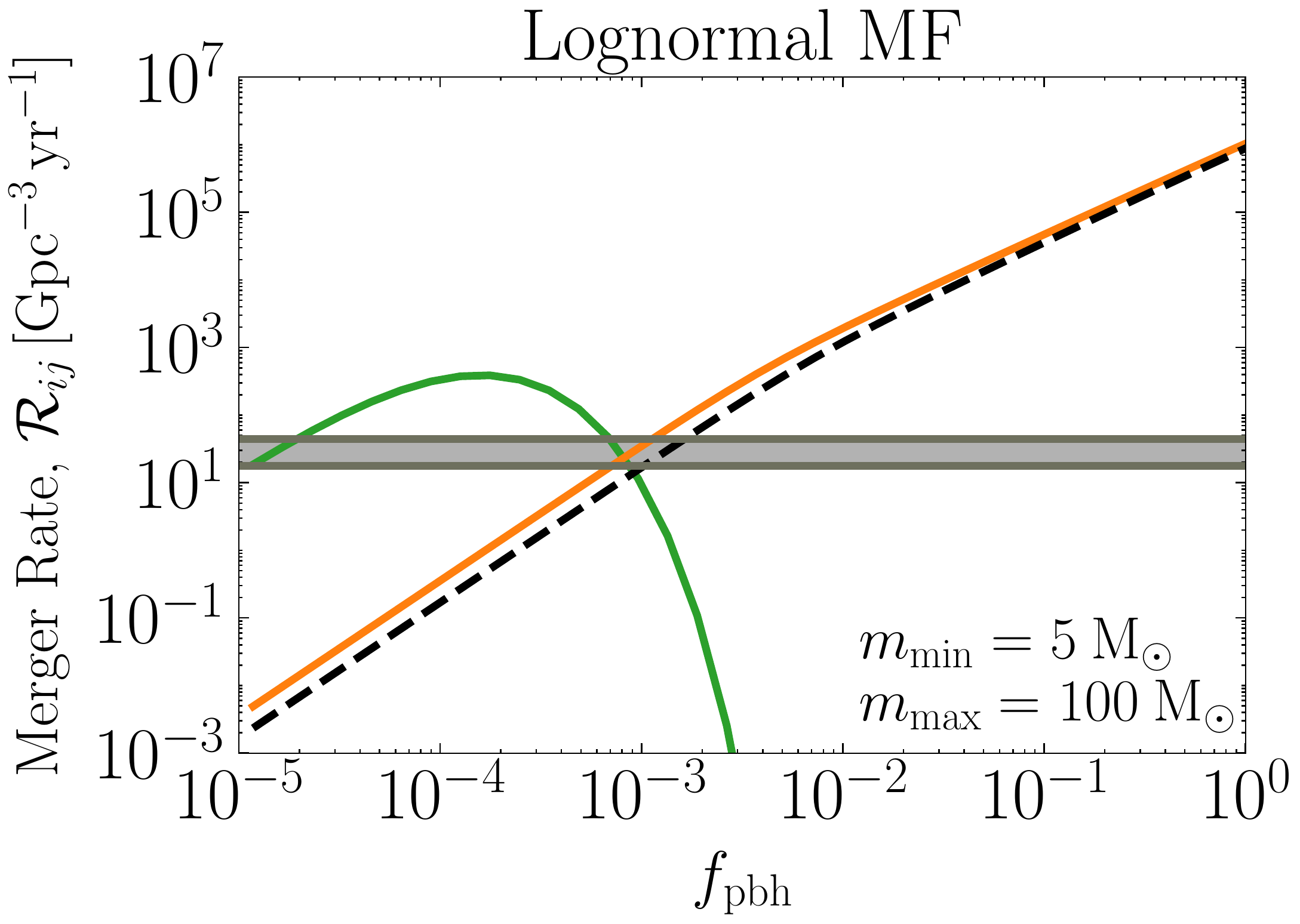}
\end{minipage}
\hspace*{\fill}
\begin{minipage}[c]{0.48\linewidth}
\centering
\includegraphics[width=\linewidth]{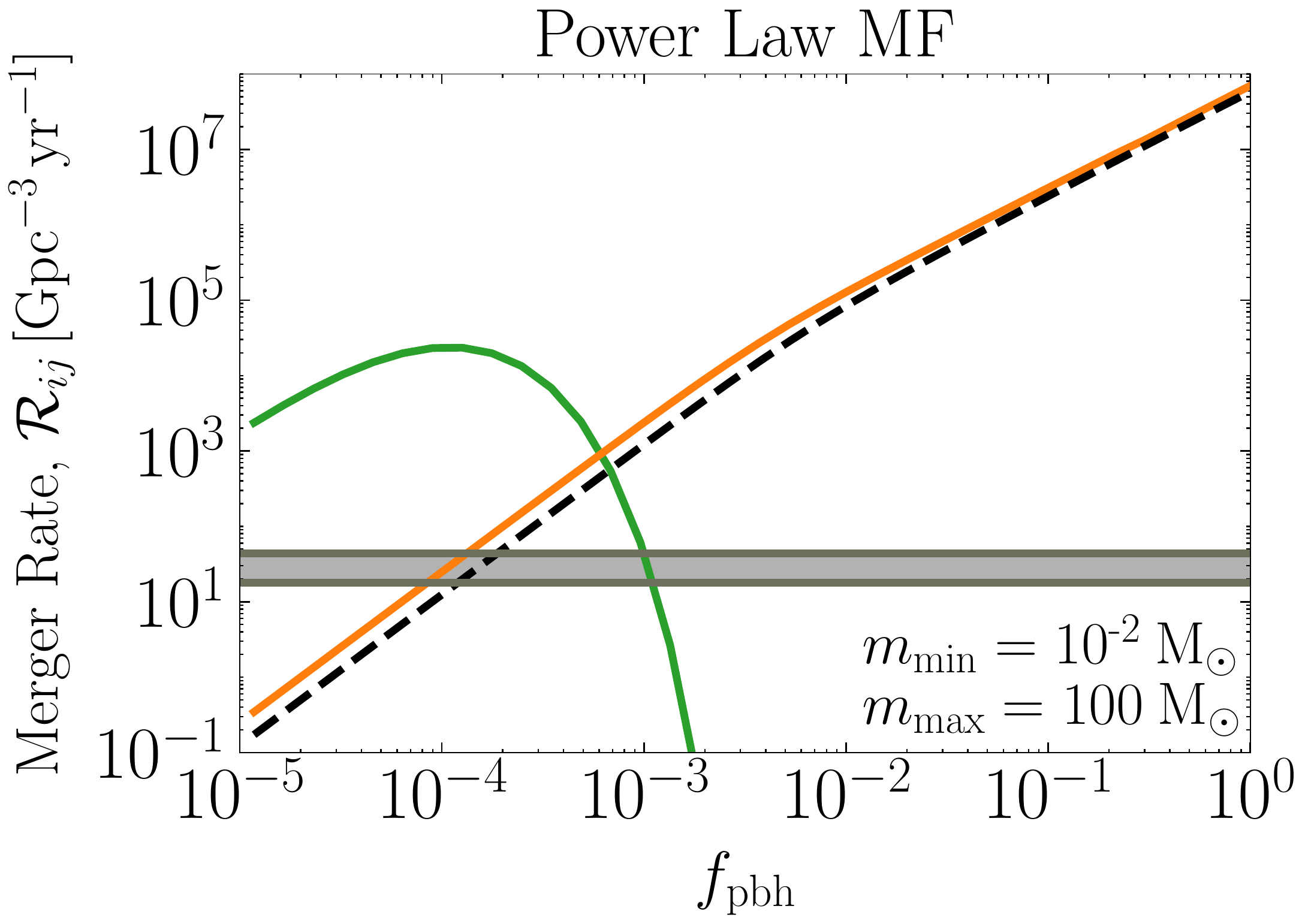}
\end{minipage}
\vspace*{\fill}
\begin{minipage}[c]{0.47\linewidth}
\centering
\includegraphics[width=\linewidth]{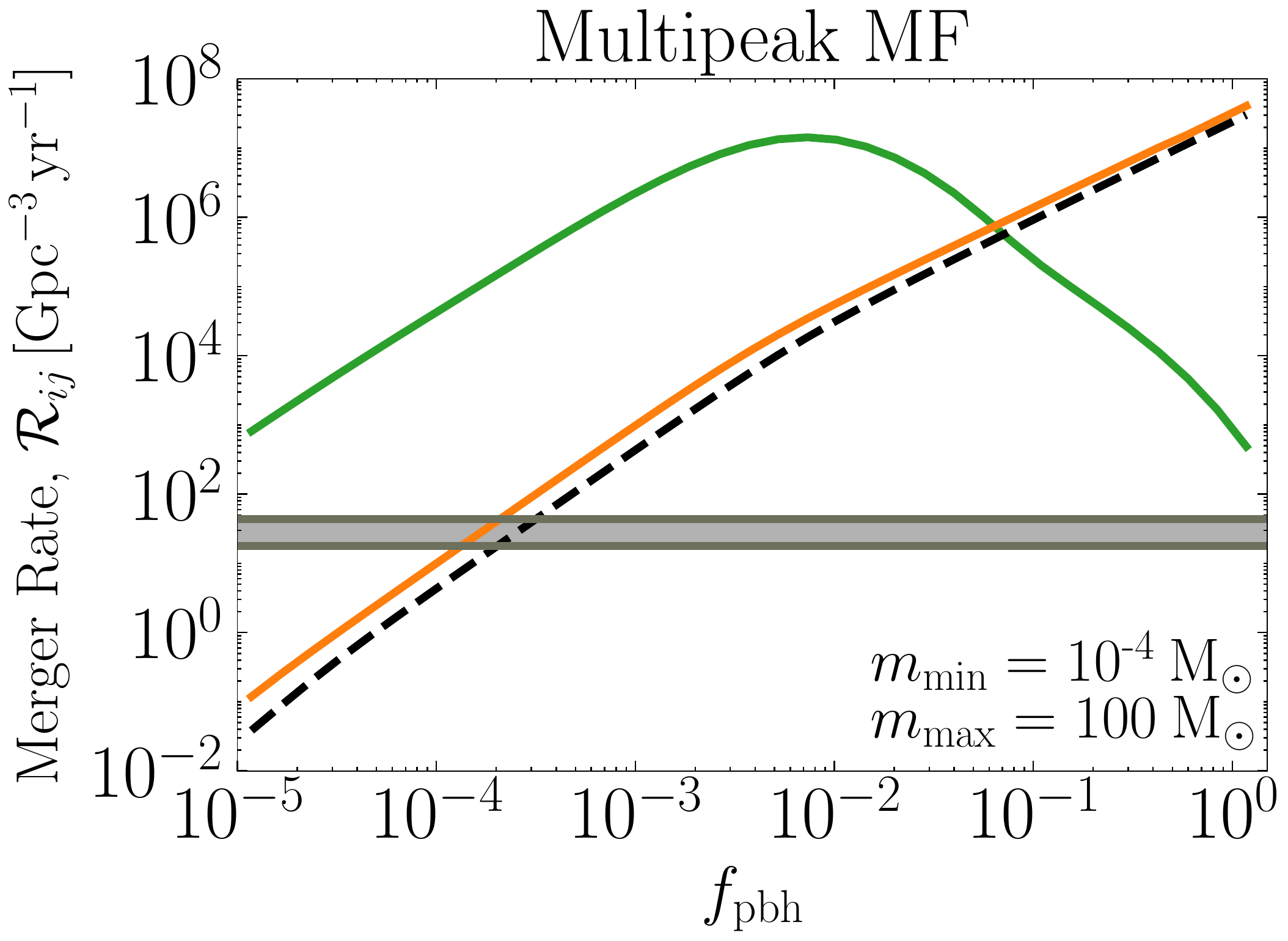}
\end{minipage}
\hspace*{\fill}
\begin{minipage}[c]{0.41\linewidth}
\centering
\includegraphics[width=\linewidth]{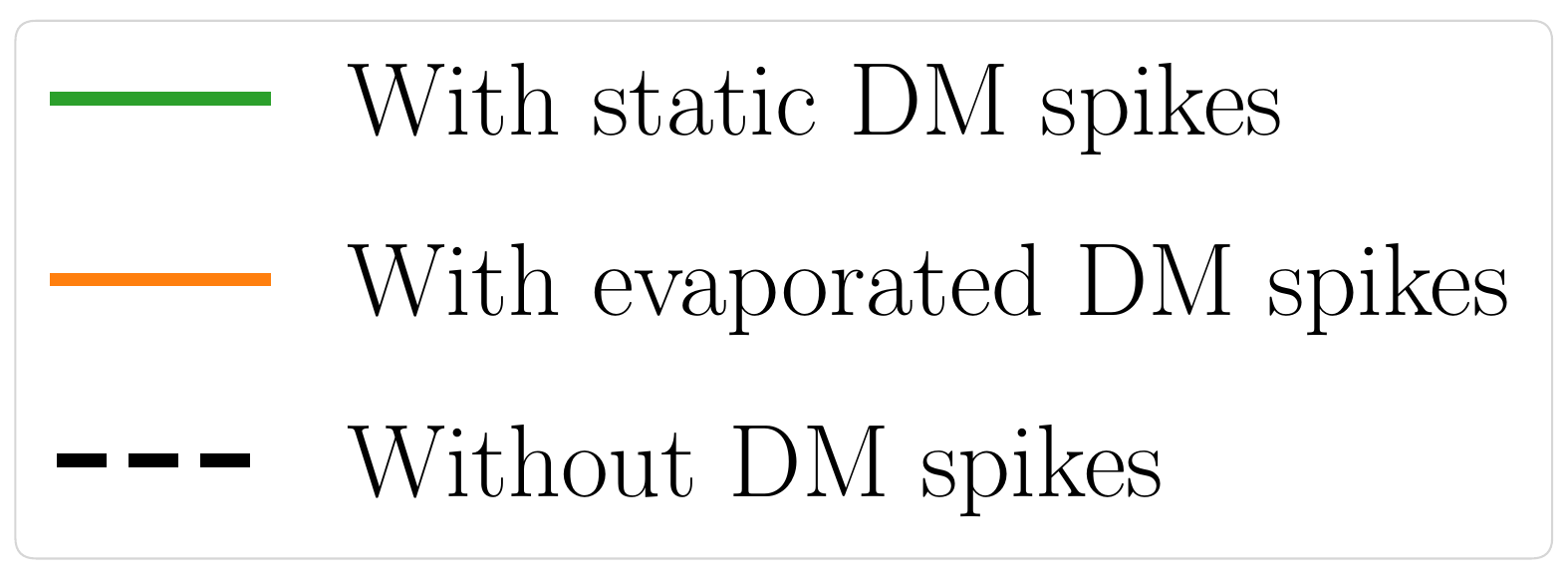}
\end{minipage}
\caption{Variation of current merger rates of PBH binaries with and without DM spikes as a function of $f_\mathrm{pbh}$. Here, $f_\mathrm{pbh} = f/0.85$ is the fractional abundance of PBHs in cold dark matter. The shaded region corresponds to the latest merger range of $\left(17.9  - 44 \right) \:\, \mathrm{Gpc}^{-3} \, \mathrm{yr}^{-1}$ seen by LVK Collaboration for BBHs mergers in mass range of $ 5-100 \: \mathrm{M_{\odot}}$ such that $m_{i} \geq m_{j}$.}
\label{fig:LVKmerger}
\end{figure}

To see how the PBH abundance affects the current merger rates of PBH binaries, we calculated the merger rates of PBH binaries as a function of $f_\mathrm{pbh}$ as shown in Fig.~\ref{fig:LVKmerger}. In this figure, we show the total merger rate of binaries with component masses in the rate $5 - 100 \, \mathrm{M_\odot}$. We see that as $f_\mathrm{pbh}$ decreases, the impact of the presence of the DM spikes on the merger rates of PBH binaries becomes more prominent. This means that as $f_\mathrm{pbh}$ decreases, the merger rate in the presence of \textbf{evaporated} and \textbf{static} DM spikes becomes larger (relative to the case of without DM spikes). From Fig.~\ref{fig:LVKmerger}, we also see that as $f_\mathrm{pbh}$ decreases, the ratio of the merger rates with and without DM spikes tends to a constant value. This can be explained as per the variations of probability distributions $P(t_f|\lambda)$ and $P(\lambda)$ with $f_\mathrm{pbh}$ in the presence of DM spikes.

\begin{table}[tb]
\centering
\caption{Values of $f_\mathrm{pbh}$ constrained by latest LVK merger data as per the current merger rates of PBH binaries with and without DM spikes shown in Fig.~\ref{fig:LVKmerger}.}
\label{tbl:table}
\medskip
\begin{tabularx}{\linewidth}{X X X}
    \multicolumn{3}{c}{PBH Binaries without DM spikes} 
    \\ \hline 
    Mass function & PBH mass range $[\mathrm{M_{\odot}}]$ & $f_\mathrm{pbh}$
    \\\hline \addlinespace[2.5pt]
    Lognormal &$5-100$  &$1.02\times 10^{-3} - 1.63\times 10^{-3}$ \\
    Power Law &\,$10^{-2}-100 \: $ & $1.19\times10^{-4} - 1.90\times 10^{-4}$\\
    Multipeak &\,$10^{-4}-100 \: $ & $2.03\times10^{-4} - 3.18\times 10^{-4}$\\ \addlinespace[2.5pt]
    \hline
\end{tabularx}

\bigskip

\begin{tabularx}{\linewidth}{X X X}
    \multicolumn{3}{c}{PBH binaries with \textbf{evaporated} DM spikes} 
    \\\hline
    Mass function & PBH mass range $[\mathrm{M_{\odot}}]$ & $f_\mathrm{pbh}$
    \\\hline \addlinespace[2.5pt]
     Lognormal &$5-100$    &$7.10\times 10^{-4} - 1.14\times 10^{-3}$\\
     Power Law &\,$10^{-2}-100 \: $ & $8.40\times10^{-5} - 1.33\times 10^{-4}$\\
     Multipeak &\,$10^{-4}-100 \: $ & $1.33\times10^{-4} - 2.10\times 10^{-4}$\\ \addlinespace[2.5pt]
    \hline
\end{tabularx}

\bigskip

\begin{tabularx}{\linewidth}{X X X}
    \multicolumn{3}{c}{PBH binaries with \textbf{static} DM spikes}
    \\\hline
    Mass function & PBH mass range $[\mathrm{M_{\odot}}]$ & $f_\mathrm{pbh}$
   \\\hline \addlinespace[2.5pt]
    Lognormal &$5-100 $ & $1.17 \times 10^{-5} - 1.95 \times 10^{-5}$ $6.90 \times 10^{-4} - 8.60\times 10^{-4}$\\
    Power Law &\,$10^{-2}-100 \: $ & $9.90\times 10^{-4} - 1.10\times 10^{-3}$ \\ \addlinespace[2.5pt]
    \hline
\end{tabularx}
\end{table}
The typical value of $\lambda$ given by Eq.~\eqref{eq:lambdatyp} decreases as we increase $f_\mathrm{pbh}$.\footnote{The total number density $n_T$ grows proportional to $f_\mathrm{pbh}$, while the mean volume  $\bar{x}^3$ decreases proportional to $f_\mathrm{pbh}$. The abundance of the binary components grows as $f_b \propto f_\mathrm{pbh}$, such that $\lambda_\mathrm{typ} \propto f_\mathrm{pbh}^{-1}$.} However, as we increase $f_\mathrm{pbh}$, the most probable value of $\lambda_\mathrm{peak}$ for binaries merging today also decreases. This can be seen by noting that the distribution $P(t_f|\lambda)$, given by Eq.~\eqref{eq:Plambda1}, peaks at $\gamma_{\lambda} = \sqrt{2}$. The typical initial angular momentum of PBH binaries scales as $j_\lambda \sim \lambda (f^2 + \sigma_\mathrm{eq}^2)^{1/2}$, so setting $j_i \sim \sqrt{2} j_\lambda$ and fixing $\lambda_\mathrm{peak}$ choosing the final merger time, we obtain the estimate:
\begin{equation} \label{eq:lambdam}
  \lambda_\mathrm{peak} \sim  \left(f^{2} + \sigma_\mathrm{eq}^{2}\right)^{\frac{-1}{2\left(\kappa + 1\right)}}\,, 
\end{equation}
where we have neglected some $\mathcal{O}(1)$ factors inside the brackets.
Here, the value of the exponent $\kappa$ for different PBH binaries can be calculated as:
\begin{equation} \label{eq:kappa}
 \kappa =
\begin{cases}
     \frac{16}{21} &\quad\text{for no  DM spikes,}\\
      \frac{14}{21} &\quad\text{for \textbf{evaporated} DM spikes,}\\
      2 &\quad\text{for \textbf{static} DM spikes.}
    \end{cases}
\end{equation}
 The different values of $\kappa$ arise because of the different scaling of the merger time with $j_i$ and $\lambda$ in each scenario. The presence of \textbf{evaporated} spikes causes large binaries to shrink as they eject DM, meaning that the merger time scales more slowly with the initial semi-major axis (and therefore $\lambda$) than in the absence of DM spikes. This translates to a smaller value of $\kappa$. In contrast, in the \textbf{static} spike case, the inspiral is driven by dynamical friction. The merger time then scales much more slowly with $j_i$ than in the absence of DM spikes. This translates to a larger value of $\kappa$. Thus, $\lambda_\mathrm{peak}$ decreases as we increase $f_\mathrm{pbh}$, but the precise scaling with $f_\mathrm{pbh}$ varies depending on our assumptions about the presence and behaviour of DM spikes. 

We see from the above discussion that as we decrease $f_\mathrm{pbh}$, the value of $\lambda_\mathrm{peak}$ grows more slowly for binaries without DM spikes than for binaries with \textbf{evaporated} DM spikes. This means that the relative shift in $\lambda_\mathrm{peak}$ induced by \textbf{evaporated} spikes grows as we decrease $f_\mathrm{pbh}$. In the mass range of interest in Fig.~\ref{fig:LVKmerger}, we are typically in the scenario where $\lambda_\mathrm{typ} > \lambda_\mathrm{peak}$, leading to an increase in the merger rate as we decrease $f_\mathrm{pbh}$. However, the impact of \textbf{evaporated} DM spikes on the merger rate grows only slowly (the two values of $\kappa$ are similar in Eq.~\eqref{eq:kappa}) and eventually saturates as $f_\mathrm{pbh} \ll \sigma_\mathrm{eq}$. In contrast, as we decrease $f_\mathrm{pbh}$, the value of $\lambda_\mathrm{peak}$ grows more quickly for binaries without DM spikes than for binaries with \textbf{static} DM spikes. At large values of $f_\mathrm{pbh}$, we find that $\lambda_{\mathrm{peak, static}} \ll \lambda_\mathrm{typ}$, strongly suppressing the merger rate. However, as we decrease $f_\mathrm{pbh}$, $\lambda_\mathrm{peak}$ in the static case becomes closer to the case without DM spikes. Eventually, for $f_\mathrm{pbh} \ll \sigma_\mathrm{eq}$, we reach the point where $\lambda_{\mathrm{peak, static}} > \lambda_\mathrm{typ}$ and the merger rate can be substantially enhanced. 

In Fig.~\ref{fig:LVKmerger}, we also highlight with a grey band the merger rates of BH binaries detected by the LVK Collaboration in mass range of $5 - 100\, \mathrm{M_{\odot}}$: $\mathcal{R} = (17.9 - 44) \:\: \mathrm{Gpc}^{-3} \:\mathrm{yr}^{-1}$, as reported in GWTC-$3$~\cite{LIGOScientific:2021psn}. We note that these rates can be well explained by the mergers of PBH binaries with and without DM spikes having either Lognormal or Power Law MF (though of course the contribution of astrophysical binaries should also be taken into account). To emphasise the effect that DM spikes would have on an inference of the PBH abundance, we list the values of $f_\mathrm{pbh}$ which would be compatible with the reported LVK merger rates in Table~\ref{tbl:table}.

\section{Conclusions}
\label{sec:conclusions}
In this work, we have reviewed the accretion of dark matter (DM) spikes around isolated primordial black holes (PBHs). Then we studied the effect of the growth of DM spikes on the initial orbital parameters of PBH binaries formed in the early Universe. To probe the merger dynamics of PBH binaries, we made assumptions about the ways in which these spikes can impact the evolution of PBH binaries. We have calculated the current merger rates of PBH binaries with DM spikes for three different extended PBH mass functions under the assumption that either the DM spikes are completely evaporated before merger (``evaporated" spikes) or they remain static up to the merger (``static" spikes).

Our calculations verify that for the density profile of $\rho(r)\propto r^{-9/4}$, the mass of the DM spike being accreted around an isolated PBH is directly proportional to the mass of the PBH itself. We have demonstrated for the first time that the \textit{impact} of DM spikes on the formation of PBH binaries is independent of the masses of the component PBHs. We find that the effect of DM spikes on the orbital parameters increases with the scale factor at which the formation of the binary takes place, as this allows more time for the DM spike to grow. The impact of DM spikes on the initial semi-major axis can be substantial for binaries forming around matter-radiation equality or later, but the impact on the initial binary eccentricity is typically negligible. 

Our results for the merger rates of PBH binaries suggest a subtle behaviour for extended mass functions. We found that in the scenario of evaporated DM spikes, the merger rates can be increased or decreased by around a factor of two. This is due to the fact that the presence of DM spikes speeds up the merger of PBH binaries. This in turn means that in order to find a PBH binary which merges today, the initial PBH separation required is shifted to larger values compared to the case without DM spikes. If this required separation is larger than the typical PBH separation, this corresponds to a shift to increasingly rare PBH binaries, leading to a reduction in the merger rate. Instead, if this required separation is smaller than the typical separation, this shift leads to an increase in the availability of binaries merging today, leading to an increase in the merger rate. The same logic applies when the DM spikes are considered static until merger, though in that case the merger rates of PBH binaries can be reduced by many orders of magnitude. This is especially true for very asymmetric binaries, which potentially has implications for searches for dressed PBH IMRIs and EMRIs. These results strongly disagree with~\cite{Kavanagh:2018ggo,Pilipenko:2022emp} in the sense that the presence of DM spikes around PBHs might \textit{only} leads to increase the merger rates of PBH binaries. We would also like to point out that the focus of our numerical results was on the present-day merger rates of PBH binaries. However, our formalism still holds for PBH binaries merging at arbitrary redshift, and can be used to study the merger rates over cosmic time.

Our calculations suggest that the current data of the LVK Collaborations could include a contribution from the mergers of PBH binaries having DM spikes. As shown in Table~\ref{tbl:table}, this constrains the fraction of PBHs in CDM $f_\mathrm{pbh}\leq \mathcal{O}(10^{-5} - 10^{-3})$, depending on the PBH mass function, consistent with previous results. We have also highlighted how the gap between the merger rates of PBH binaries with and without DM spikes increases with  decrease in $f_\mathrm{pbh}$. This is also an important aspect which has not been considered so far and must be taken into account in order to derive accurate limits from future data of the LVK Collaboration, as existing constraints are pushed to smaller and smaller values of $f_\mathrm{pbh}$. In addition, these results may also be relevant for studying the rapid and enhanced mergers of PBH binaries in the early Universe due to the presence of DM spikes. Such PBHs may provide the seeds for the growth of supermassive black holes and early-forming galaxies~\cite{1984MNRAS.206..801C,Bean:2002kx}.

In our description of DM spikes around PBHs, we have ignored the annihilation, mutual gravitational attraction and growth of DM spikes after decoupling of PBH binaries. We have also ignored processes such as the capture of binaries in PBH clusters or the interaction of PBHs lying very close to the binaries. We have argued that for sufficiently small $f_\mathrm{pbh}$ some of these should be negligible (such as the formation of PBH clusters). However, a more complete picture of these factors should be taken into account for careful analysis of the accretion of DM spikes around PBHs. 

In this work, we have considered two extreme scenarios for the behaviour of DM spikes around PBHs binaries. We expect our description of evaporated spikes to be accurate for binaries with close-to-equal mass ratios, in which the injection of energy leads to the rapid unbinding of the spike. Instead, the static spike scenario may be more accurate when the mass ratio is very large and the lighter PBH does not substantially perturb the spike of the larger companion. Of course, the true dynamics will be somewhere in between these extremes. This true dynamics can be very complex, including the co-evolution of the binaries and the DM spike. To accurately probe the merger rates of PBH binaries dressed with DM spikes, the feedback of the PBHs on DM spikes also need to be considered in a way similar to Ref.~\cite{Kavanagh:2020cfn}.

Even so, our results highlight that the presence of DM spikes around PBH binaries may enhance or suppress their merger rates, depending strongly on their abundance and mass functions. In some cases, our results suggest that this suppression can be extreme. The development of a prescription for describing the feedback of DM spikes in highly eccentric systems should therefore be considered a priority. With this, it will be possible to develop a more general and complete approach which can probe the real merger dynamics of the PBH binaries dressed with DM spikes.

\begin{acknowledgments}
 The authors thank Pippa Cole for providing a tabulated version of the mass function described in Ref.~\cite{Cole:2022ucw}. We also appreciate the efforts of Zu-Cheng Chen and Qing-Guo Huang for providing technical explanations regarding their work published in Ref.~\cite{Chen:2018czv}. The authors also thank the `Dark Collaboration at IFCA' working group and Gianfranco Bertone for useful discussions and comments.

 Pratibha Jangra acknowledges support of project PGC2018-101814-B-100 (MCIU/AEI/
 MINECO/FEDER, UE) Ministerio de Ciencia, Investigaci\'on y Universidades. This project was funded by the Agencia Estatal de Investigaci\'on, Unidad de Excelencia Mar\'ia de Maeztu, ref. MDM-2017-0765.  BJK also acknowledges funding from the Ram\'on y Cajal Grant RYC2021-034757-I, financed by MCIN/AEI/10.13039/501100011033 and by the European Union ``NextGenerationEU"/PRTR.
\end{acknowledgments}

\appendix

\section{Scaling of merger time for static DM spikes}
\label{sec:scaling}
In this appendix, we aim to provide a physical justification for the scaling of the PBH merger time as a function of $m_i, m_j, a_i$ assuming \textbf{static} DM spikes, as described in Sec.~\ref{subsec:merger time Static DM spikes}. In that section, numerical calculations using \texttt{imripy}~\cite{Becker:2021ivq} suggested that the final merger time $t_f$ should scale with the initial semi-major axis, $a_i$ and the component masses, $m_i$ and $m_j$ as:
\begin{equation}
\label{eq:scalings}
t_f \sim (a_i)^\alpha \, (m_i)^\gamma \, (m_j)^\delta\,,
\end{equation}
with $\alpha = 0.75$, $\gamma = 0.65$ and   $\delta = -0.89$.

We begin by noting that for large enough initial separations, gravitational wave energy losses are typically subdominant to dynamical friction energy losses (see e.g.\ Ref.~\cite{Coogan:2021uqv}). Then, we can write:
\begin{equation}
\frac{\mathrm{d}a}{\mathrm{d}t} \approx \frac{\mathrm{d}E}{\mathrm{d}t} \frac{\mathrm{d}a}{\mathrm{d}E}\,,
\end{equation}
where the orbital energy is $E = - G \,m_i\, m_j/2\, a$. The loss of total energy $E$ of the binary orbit occurs via the emission of gravitational waves (GW) and due to dynamical friction (DF), given as:
\begin{equation}
\frac{\mathrm{d}E}{\mathrm{d}t} = \left(\frac{\mathrm{d}E_\mathrm{GW}}{\mathrm{d}t} + \frac{\mathrm{d}E_\mathrm{DF}}{\mathrm{d}t}\right)\,.
\end{equation}
Similar to Ref.~\cite{Maggiore:2007ulw}, the rate of loss of orbital energy of the binary due to the emission of gravitational waves is given as:
\begin{equation} \label{eq:lossofGW}
\frac{\mathrm{d}E_\mathrm{GW}}{\mathrm{d}t} = -\frac{32}{5} \frac{\mu_{0}^{2} \left(m_i + m_j\right)^{3} \left( 1 + \frac{73}{25}e^{2} + \frac{37}{96}e^{4}\right) }{a^{5} \left( 1 - e^{2}\right)^{7/2}}\,.
\end{equation}
with $\mu_{0} = m_{i} m_{j}/\left(m_{i} + m_{j}\right)$ as the reduced mass of the PBH binary in the absence of DM spikes. The rate of loss of energy of the PBH binary due to the dynamical friction is~\cite{Becker:2021ivq}:
\begin{equation}   \label{eq:lossofDF}
\frac{\mathrm{d}E_\mathrm{DF}}{\mathrm{d}t}(r, v) = 4 \pi\left(G m_j\right)^2 \rho_{\mathrm{sp}}(r) v^{-1} \log \Lambda\,,
\end{equation}
such that Coulomb logarithm $\log \Lambda = \log \sqrt{m_i/m_j}$. Counting powers of the masses and semi-major axis, we find that $\rho_{\mathrm{sp}} \sim m_i^{3/4} a^{-9/4}$ and $v \sim (m_i/a)^{1/2}$. We thus find that:
\begin{equation}  \label{eq:lossofDFmimj}
\frac{\mathrm{d}E_\mathrm{DF}}{\mathrm{d}t} \sim m_i^{1/4} m_j^2 \,a^{-7/4}\,.
\end{equation}
We can now estimate the scaling of the typical final merger time as:
\begin{align}
t_f \sim \int \left(\frac{\mathrm{d}a}{\mathrm{d}t}\right)^{-1}\,\mathrm{d}a \sim m_i^{3/4} m_j^{-1} \int a^{-1/4}\,\mathrm{d}a \sim a^{3/4}\,m_i^{3/4} m_j^{-1}\,.
\end{align}
This roughly matches the empirical scaling observed in Eq.~\eqref{eq:scalings}, with small differences coming from the dependence of $\log\Lambda$ on $m_i$ and $m_j$ and on the effects of eccentricity in the orbit.

\section{Merger rates for Lognormal Mass function}
\label{sec:Merger rates for Lognormal Mass function}
\begin{figure}[tbh!] 
\centering
\textbf{Lognormal mass function}\par\medskip
\begin{minipage}[c]{0.48\linewidth}
\includegraphics[width=\linewidth]{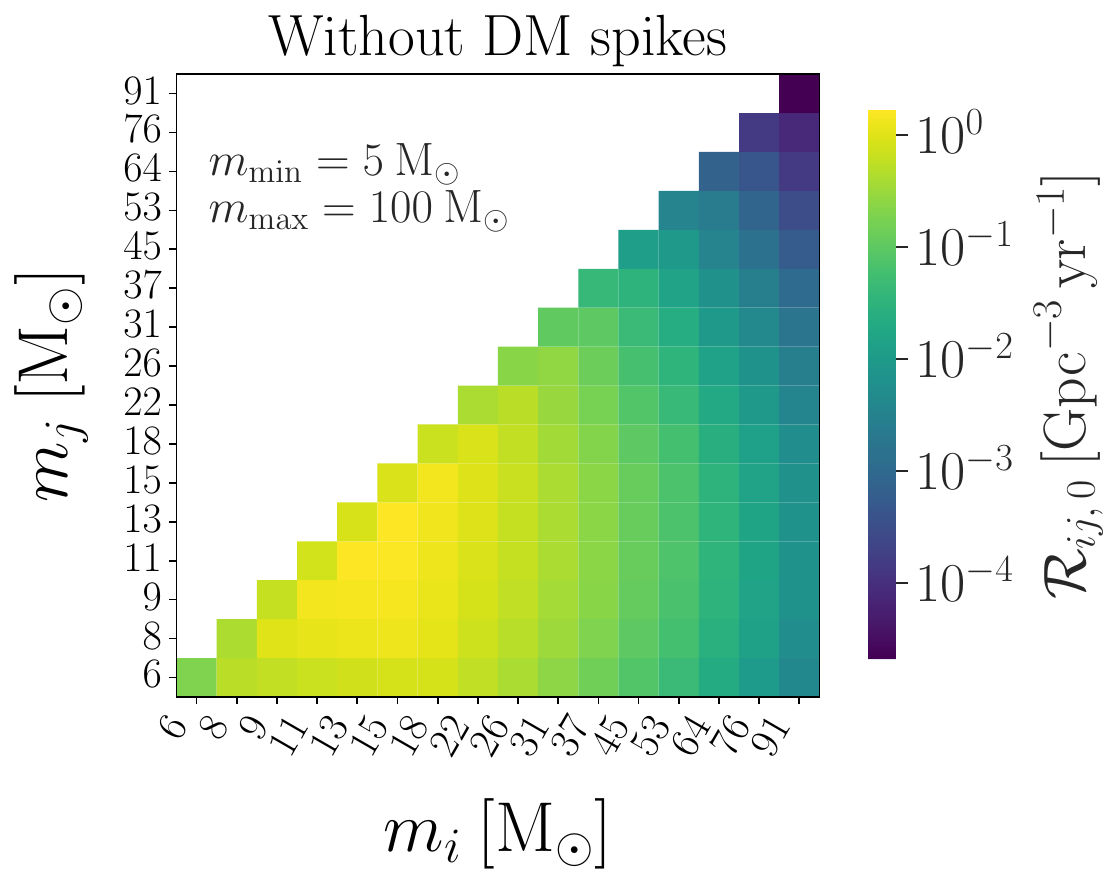}
\end{minipage}
\hspace*{\fill}
\begin{minipage}[c]{0.48\linewidth}
\includegraphics[width=\linewidth]{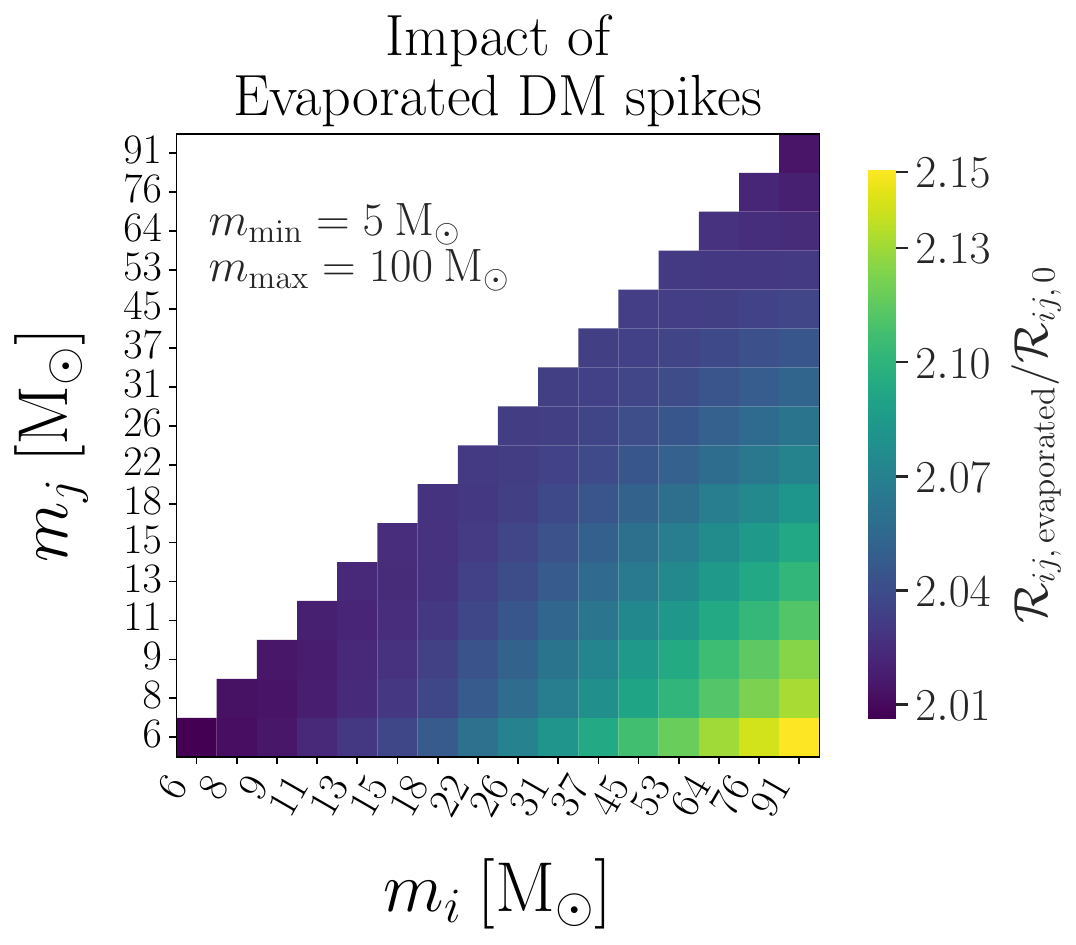}
\end{minipage}
\vspace*{\fill}
\begin{minipage}[c]{0.48\linewidth}
\includegraphics[width=\linewidth]{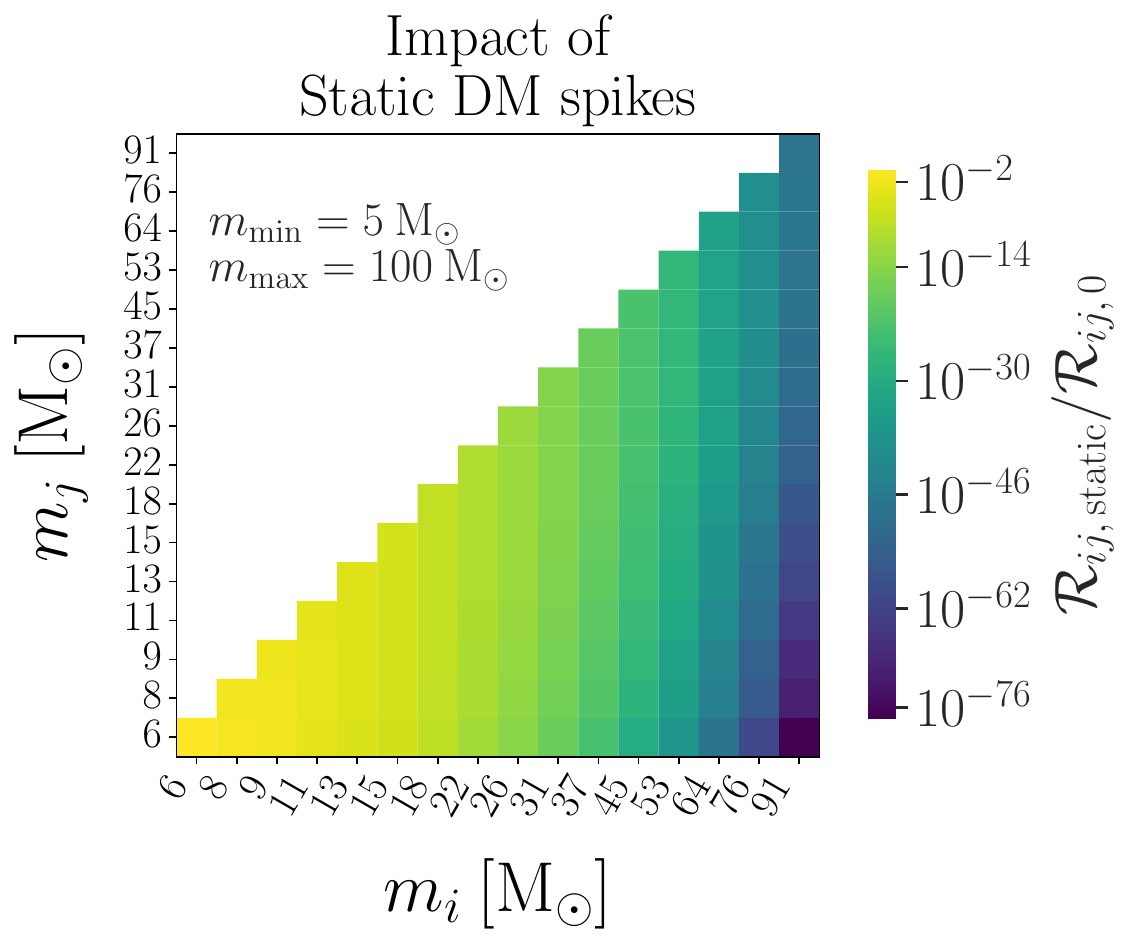}
\end{minipage}
\caption{Current merger rates of PBH binaries with and without DM spikes having Lognormal mass function with PBH mass range from $m_\mathrm{{min}}$ to $m_\mathrm{{max}}$. Here, $\mathcal{R}_{ij,\,0}$ represents the current merger rate of PBH binaries having no DM spikes around the PBHs. $\mathcal{R}_{ij,\,\mathrm{evaporated}}$ denotes the current merger rate of PBH binaries consisting of DM spikes which get completely evaporated before merger. And $\mathcal{R}_{ij,\,\mathrm{static}}$ signifies the current merger rate of PBH binaries containing DM spikes which remain static until the merger.}
\label{fig:mergerlognormal}
\end{figure}

Figure~\ref{fig:mergerlognormal} shows the current merger rates of PBH binaries with and without DM spikes in the mass range of $5-100\,\mathrm{M_{\odot}}$ for our Lognormal benchmark mass function. The top left panel signifies that in the absence of DM spikes the merger rates are for PBH binaries which consist of the most abundant PBHs as shown in Fig.~\ref{fig:PBHmassPDFs}. The top right panel shows that the presence of \textbf{evaporated} DM spikes increases the merger rate of PBH binaries by roughly a factor of $2$ for all PBH masses. The bottom panel shows that the merger rates of PBH binaries with \textbf{static} DM spikes are very much suppressed in comparison to the binaries without DM spikes. This behaviour is similar to the dynamics of PBH binaries with \textbf{static} DM spikes having Power Law and Multipeak MF shown in Fig.~\ref{fig:mergerratesDMstatic}. The merger rates of PBH binaries with DM spikes shown in Figure.~\ref{fig:mergerlognormal} can be explained on the basis of the probability distributions of PBHs separation and final merger time, which has been already addressed in Sec.~\ref{subsec:Component mass dependence of merger rates} in Fig.~\ref{fig:pdfoftf}. Since the presence of DM spikes might affects these parameters in two possible ways hence depending on the response of DM spikes, the merger rates get enhanced for some PBH binaries and reduced in other cases. 
\label{sec:lognormal}

\section{Mass ratio Distribution of merging PBHs}
\label{sec:Mass ratio Distribution of merging PBHs}
\begin{figure}[tbh!] 
\centering
\begin{minipage}[c]{0.483\linewidth}
\includegraphics[width=\linewidth]{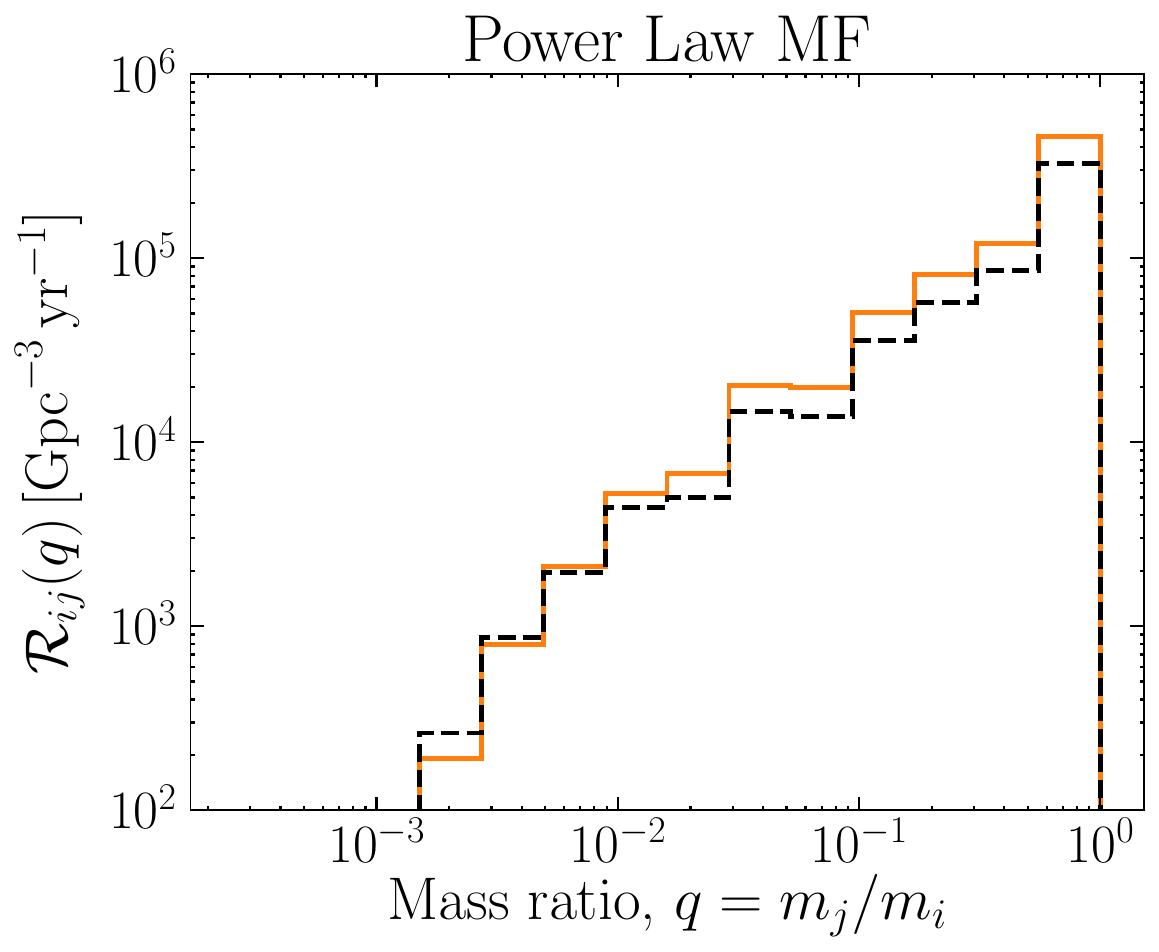}
\end{minipage}
\hspace*{\fill}
\begin{minipage}[c]{0.47\linewidth}
\includegraphics[width=\linewidth]{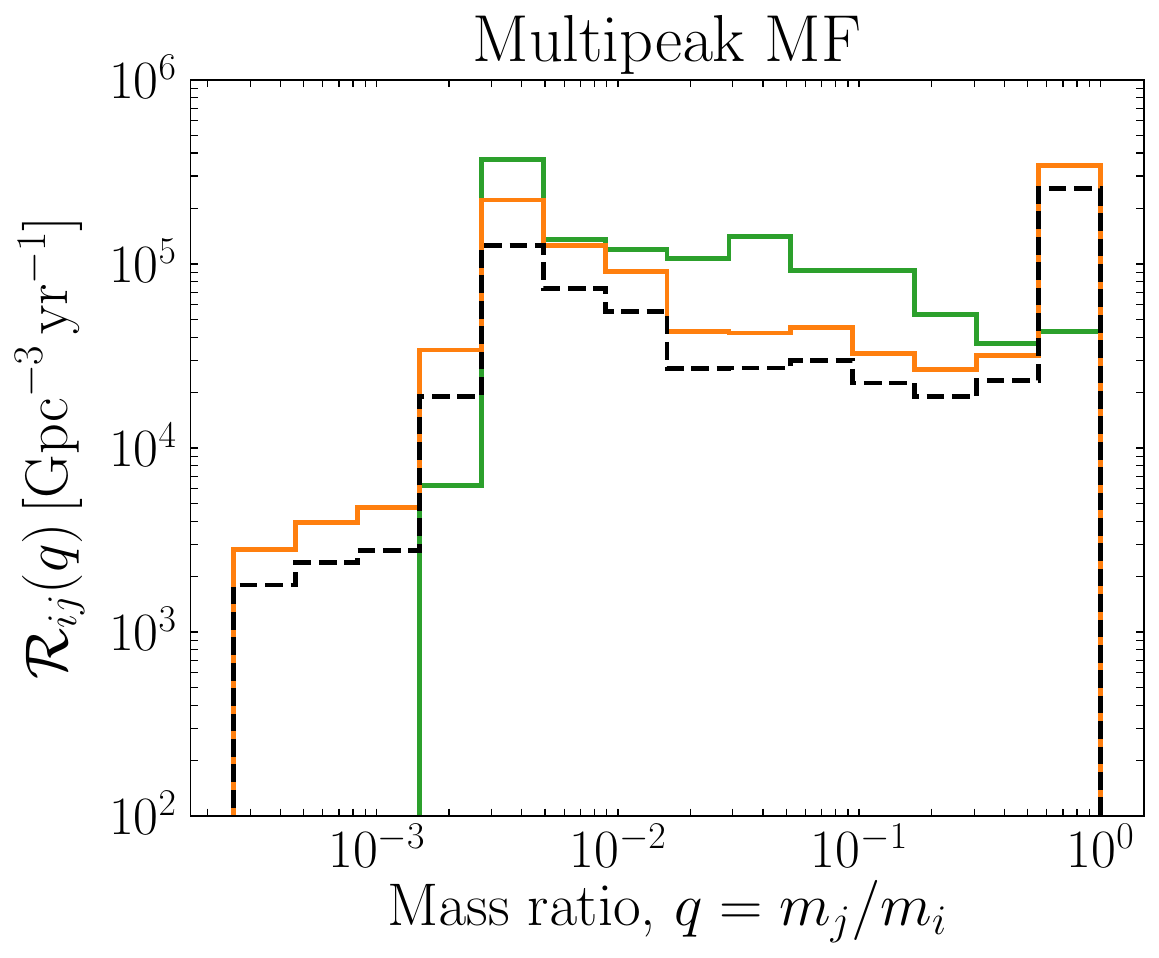}
\end{minipage}
\vspace*{\fill}
\begin{minipage}[c]{0.48\linewidth}
\includegraphics[width=\linewidth]{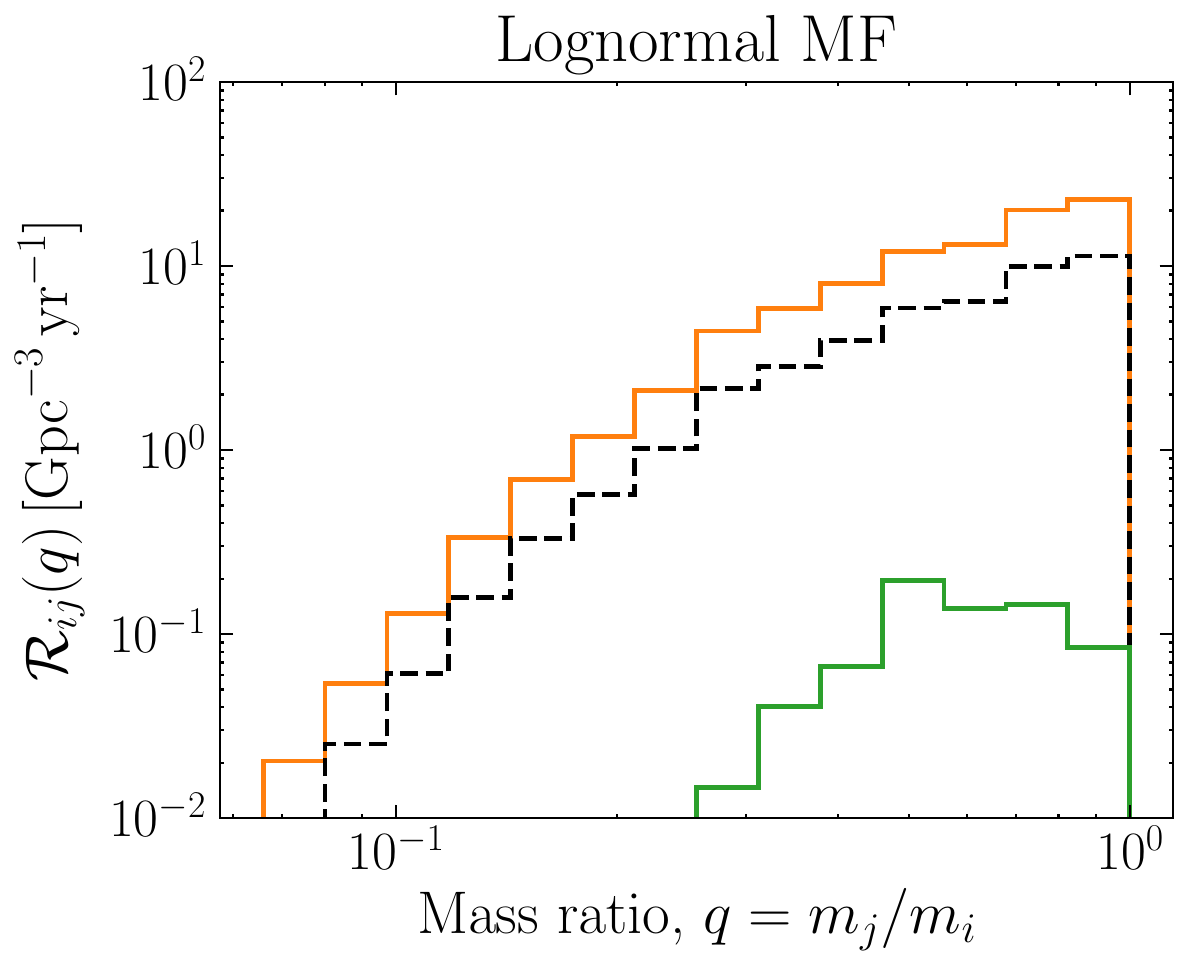}
\end{minipage}
\hspace*{\fill}
\begin{minipage}[c]{0.41\linewidth}
\includegraphics[width=\linewidth]{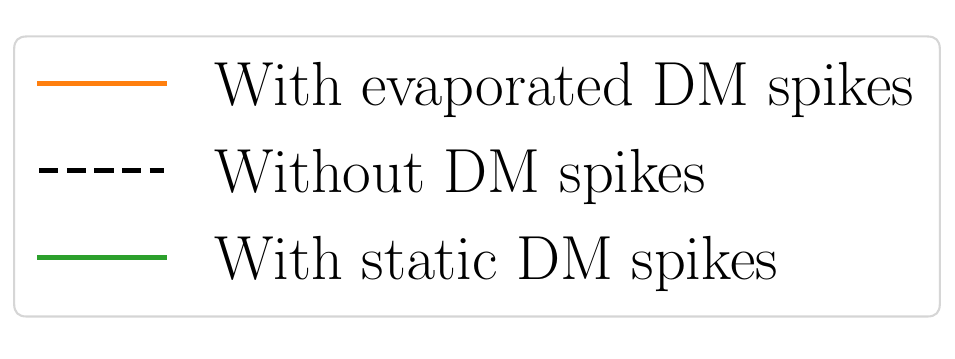}
\end{minipage}
\caption{Mass ratio distributions of PBHs merging currently as per the merger rates shown in Figures~\ref{fig:mergerwithoutDMspikes},~\ref{fig:mergerratesDMejected},~\ref{fig:mergerratesDMstatic} and~\ref{fig:mergerlognormal}. Here, $q = m_j/m_i$ is the mass ratio of two PBHs in the binaries merging today with and without the presence of DM spikes around isolated PBHs.}
\label{fig:PBHmassratiodistribution}
\end{figure}

In this section, we estimate the mass ratio $q = m_j/m_i$ distribution of the PBHs merging today. As per the merger rates of PBH binaries with and without DM spikes shown in Figures~\ref{fig:mergerwithoutDMspikes},~\ref{fig:mergerratesDMejected},~\ref{fig:mergerratesDMstatic} and~\ref{fig:mergerlognormal}, the distribution of PBHs mass ratios for different PBH mass functions discussed in Sec.~\ref{subsec:Component mass dependence of merger rates} is shown in Figure~\ref{fig:PBHmassratiodistribution}. In this figure, we see that overall the merger rates for the three specified mass function are dominated by larger value of $q$ which is discussed in details in Sec~\ref{subsec:Component mass dependence of merger rates}. This figure also clearly depicts that for a given mass function whether the merger rates of PBH binaries are dominated via \textbf{evaporated} or \textbf{static} nature of DM spikes which is something that has not been addressed in the previous literature. 

\clearpage
\bibliographystyle{JHEP}
\bibliography{PBHspike}
\end{document}